\documentclass[onecolumn]{aastex62}

\usepackage{amsmath}

\accepted{May 19, 2024}
\submitjournal{ApJ}

\shorttitle{AASTeX v6.3.1 Sample article}
\shortauthors{\c{S}ahin and Antolin}
\graphicspath{{./}{figures/}}
\def\ion#1#2{#1$\;${\small\rm\@{#2}}\relax}
\newcommand{\uat}[2]{\href{http://astrothesaurus.org/uat/#2}{#1 (#2)}}

\begin{document}

\title{From Chromospheric Evaporation to Coronal Rain: An Investigation of the Mass and Energy Cycle of a Flare}

\correspondingauthor{Seray \c{S}ahin}
\email{seraysahin93@gmail.com}

\author[0000-0003-3469-236X]{Seray \c{S}ahin}
\author[0000-0003-1529-4681]{Patrick Antolin}
\affiliation{Department of Mathematics, Physics and Electrical Engineering, Northumbria University, Newcastle Upon Tyne, NE1 8ST, UK}

\begin{abstract}

Chromospheric evaporation (CE) and coronal rain (CR) represent two crucial phenomena encompassing the circulation of mass and energy during solar flares. While CE marks the start of the hot inflow into the flaring loop, CR marks the end, indicating the outflow in the form of cool and dense condensations. With \textit{IRIS} and \textit{AIA/SDO}, we examine and compare the evolution, dynamics, morphology, and energetics of the CR and CE during a C2.1 flare. The CE is directly observed in imaging and spectra in the \ion{Fe}{XXI} line with \textit{IRIS} and in the \ion{Fe}{XVIII} line of AIA, with upward average total speeds of $138\pm[35]~$km~s$^{-1}$ and a temperature of $[9.03\pm3.28]\times10^{6}$~K. An explosive to gentle CE transition is observed, with an apparent reduction in turbulence. From quiescent to gradual flare phase, the amount and density of CR increases by a factor of $\approx4.4$ and 6, respectively. The rain's velocity increases by a 1.4, in agreement with gas pressure drag. In contrast, the clump widths variation is negligible. The location and morphology of CE match closely those of the rain showers, with similar CE sub-structure to the rain strands, reflecting fundamental scales of mass and energy transport. We obtain a CR outflow mass three times larger than the CE inflow mass, suggesting the presence of unresolved CE, perhaps at higher temperatures. The CR energy corresponds to half that of the CE. These results suggest an essential role of coronal rain in the mass-energy cycle of a flare.
\end{abstract}

\keywords{Coronal rain; \uat{Solar prominences}{1519}; \uat{Solar chromosphere}{1479}; \uat{Solar transition region}{1532}; \uat{Solar coronal heating}{1989}, chromospheric evaporation, mass-energy cycle}

\section{Introduction} \label{sec:intro}

Solar flares are one of the most fascinating and, at the same time, the most energetic phenomena in the universe that can be spatially and temporally resolved. These events involve the reconfiguration of the magnetic field, which is known as magnetic reconnection \citep[e.g.,][]{Sweet1958IAUS....6..123S, Shibata2011LRSP....8....6S}. During this reconnection, a solar flare suddenly releases energy (10\textsuperscript{28}-10\textsuperscript{32} erg) stored in the magnetic field \citep{Fletcher2011SSRv..159...19F} within a typical timescale of tens of minutes. 
%This released energy travels mostly in the form of accelerated particles along newly reconnected magnetic field lines and is deposited in the chromosphere at the footpoints. 
Solar flares have three distinct phases \citep{Golub2009soco.bookG} in their temporal evolution known as pre-flare (or precursor), impulsive, and gradual phases. In the pre-flare phase, a series of small brightenings occur before the onset of a flare, playing a crucial role as early indicators of a flare trigger. These triggers encompass various phenomena, such as the emergence of new flux or activation of filaments. As the system undergoes destabilization, the interaction between magnetic fluxes occurs, primarily through reconnection, releasing large amounts of free magnetic energy and subsequently initiating the flare and starting the impulsive phase. This results in the visibility of coronal plasma in soft X-ray and extreme ultraviolet (EUV) lines. During the impulsive phase, a substantial acceleration of electron and ion beams and a reconfiguration of the magnetic field also occur. The energy released through magnetic field dissipation and accelerated particles is readily observed in flare lightcurves derived from various sources, including hard and soft X-rays (HXR and SXR, respectively), $\gamma$-rays, EUV, and, in certain instances, white light emission \citep{Fletcher2011SSRv..159...19F}. The accelerated electron beams propagate downward along the loop legs and heat the chromospheric plasma rapidly. At the same time, the reconfigured magnetic field following reconnection does work through the Lorentz force on the plasma, and magnetic energy is converted to kinetic energy in the loops. Energy is transported downwards through thermal conduction, thereby also heating the chromosphere. The increase of temperature in the chromosphere increases the pressure, which drives the chromospheric material upward, filling up the flare loop in a process commonly known as chromospheric evaporation (CE) \citep{Fisher1985ApJ...289..414F, Dudik2016ApJ...823...41D}. The onset of chromospheric evaporation is marked by the appearance of flare ribbons in chromospheric lines (like H$\alpha$). Following the impulsive phase, the gradual phase of a solar flare ensues, spanning several hours or even extending to a day or longer. This phase is characterized by cooling processes and a decline in the intensity of X-ray and EUV emissions. The cooling process within this phase can be further divided into three distinct stages \citep{scullion2016}. Thermal conduction is initially the primary loss mechanism, driven by elevated temperatures. As the temperature gradually decreases, radiative cooling becomes more effective, aided by the increasing radiative loss function at lower temperatures \citep{Culhane1970SoPh...15..394C}. Ultimately, thermal instability (TI) is thought to set in, accelerating the cooling and producing dense and cool condensations radiating strongly in chromospheric lines that subsequently fall down as coronal rain. The flare loop is thought to largely evacuate and return to thermal equilibrium. Flare-driven coronal rain seems to be present in most flares \citep{Mason2019ApJ...874L..33M}, but a large-scale statistical investigation awaits. 

\begin{figure*}[ht!]
\includegraphics[width=\textwidth]{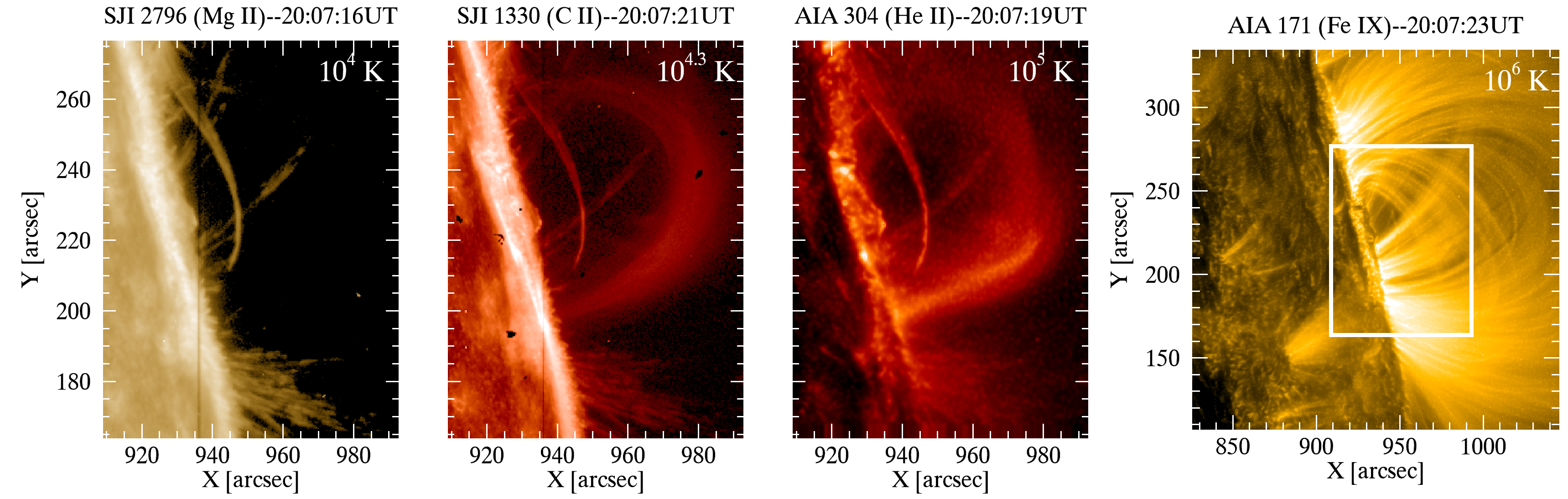}
\includegraphics[width=\textwidth]{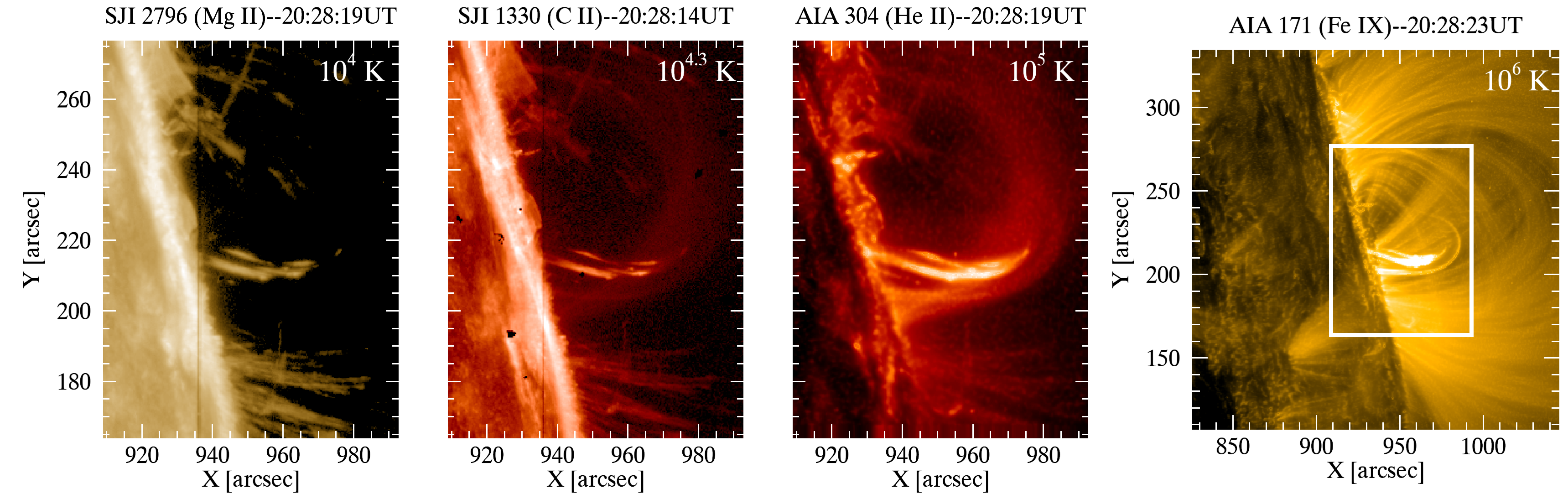}
\caption{The studied active region (AR NOAA 12158) at the West limb of the Sun,  observed by IRIS/SJI 2796~\AA (left), SJI 1330~\AA (second from left), and SDO/AIA 304~\AA~(third from left) taken at 20:07:16~UT (top) and 20:28:19~UT (bottom) on 2014 September 17 for the C-type flare. The separated panel on the right shows a wider FOV with SDO/AIA 171~\AA over the same region,  with the white rectangle outlining the FOV shown on the left panels. The top panels show a snapshot during the CE, while the bottom panels show a snapshot during the flare-driven rain. A movie of these rows is available. The movie shows off-limb coronal rain seen in the SJI 2796~\AA, SJI 1330~\AA, and AIA 304~\AA~falling towards the solar surface along coronal loop structures and chromospheric evaporation seen in the SJI 1330~\AA\ and AIA 304~\AA~.}
\label{figure1}
\end{figure*}

The mass and energy cycle of a flare loop is, therefore, bounded by two processes: chromospheric evaporation at the start and coronal rain at the end. From this perspective, it is interesting to compare the morphology, dynamics and energetics of these two processes since, in the flare model, mass conservation already suggests a direct link between both. A first step in this direction has been taken by \citet{Jing2016NatSR...624319J}, in which the sizes of the ribbons in H$\alpha$ were observed to match the widths of the rain, thus suggesting a fundamental scale of the energy transport. Direct observations of chromospheric evaporation are, however, scarce. This is because of the large expected upward velocities of the heated material and the lack of resolution at very high temperatures. Previous studies and observations (most of them spectroscopic) showed that the speed of chromospheric evaporation in EUV and SXR wavelengths ranges from 50 to 800 km s$^{-1}$ \citep{Doschek1994ApJ...431..888D, Milligan2009ApJ...699..968M, Young2015ApJ...799..218Y, Tian2018ApJ...856...34T, Li2022ApJ...926..216L}.

Coronal rain is another captivating phenomenon that is observed in the solar atmosphere. It is cool and dense partially ionised plasma that forms in a few minutes and falls toward the solar surface along loop-like trajectories \citep{antolin2012}. Under quiescent conditions, its temperature and density varies between 10$^{3}$–$10^{5}$~K and $\approx 10^{10}$–$10^{12}~$cm$^{-3}$, respectively. Coronal rain is best-observed off-limb, where it stands out clearly against the dark background. It can be seen in emission in both chromospheric \citep[such as H$\alpha$, \ion{Ca}{II} K $\&$ H and \ion{He}{I};][]{DeGroof2004, Schad2018ApJ...865...31S, froment2020} and transition region lines \citep[such as \ion{Si}{IV}~1402~\AA, or/and in \ion{He}{II}~304~\AA;][]{Vashalomidze_2015AA...577A.136V,antolin2015}. It is also observed generally in absorption on disk, where it often corresponds to supersonic downflows into sunspot umbrae \citep{Kleint2014ApJ...789L..42K, Ahn2014SoPh..289.4117A, Chitta2016A&A...587A..20C, Nelson2020A&A...636A..35N, ChenH_2022AA...659A.107C}. However, it is also accompanied by corona and upper transition region emission due to the Condensation Corona Transition Region (CCTR) \citep{antolin2020}. In quiescent active region conditions, thermal instability (TI) within a coronal loop in a state of thermal non-equilibrium (TNE), known as the TNE-TI scenario \citep{antolin2020, Antolin_Froment_10.3389/fspas.2022.820116} is the leading cause of coronal rain generation. During TNE-TI, coronal heating concentrated towards the loop footpoints leads to chromospheric evaporation and, therefore, a density increase of the loop. If the heating is frequent enough, this increased density results in excessive radiative losses in the loop, surpassing the energy input from the heating mechanism, particularly around the loop apex \citep{muller2003,muller2004}. Consequently, TI arises, producing coronal rain. The exact role of TI remains a subject of ongoing debate \citep{Klimchuk_2019SoPh..294..173K}. However, conditions leading to the onset of TNE are typically obtained in coronal loops characterized by strong stratification and high-frequency heating \citep{Antiochos_1999ApJ...512..985A, Klimchuk_2019ApJ...884...68K}. The loop is subsequently drained, and the heating (evaporation) and cooling (condensation) restart, marking a TNE-TI cycle. 

%Numerical simulations allow to study the parameter space of TNE-TI and coronal rain formation. The results can then be compared with observations, which can then constrain the parameter space of the heating \citep{fang2013,fang2015,Froment2018ApJ...855...52F, Klimchuk_2019ApJ...884...68K, johnston2019}. The specific spatio-temporal heating distribution requirements to generate coronal rain, therefore, constitute stringent proxies of coronal heating. 

Coronal rain is typically observed in three different forms in the solar atmosphere \citep{Antolin_Froment_10.3389/fspas.2022.820116}. The most frequently observed form is called quiescent coronal rain, which occurs in coronal loops in active regions. The second form, flare-driven coronal rain, usually during the gradual phase of a solar flare \citep{Foukal1978ApJ...223.1046F, scullion2016}, is one of the most mesmerizing yet least understood phenomena observed in the solar atmosphere. The last form is called hybrid prominence/coronal rain complexes, which are formed in magnetic dips over null point topologies \citep{Liu_2016SPD....47.0402L, LiL_2018ApJ...864L...4L, Mason2019ApJ...874L..33M, ChenH_2022AA...659A.107C}.

Coronal rain morphology (lengths and widths) varies greatly according to the temperature of formation of the observed wavelength (due to differences in line opacities) and the spatial resolution of the instrument. At the smallest scales, observations in H$\alpha$ of quiescent coronal rain report widths of $\approx$200-300 km with the Solar Swedish Telescope (SST) \citep{antolin2012, froment2020}, $\approx$500~km from EUV absorption in the HRI$_{EUV}$ passband (forming at $\approx10^{5.9}$) with Solar Orbiter \citep{Antolin2023A&A...676A.112A} and ranges between 400~km and 900~km in SJI~2796 and 1400 passbands of IRIS \citep{antolin2015, Sahin2023ApJ...950..171S}. For a given wavelength, there is minimal difference in rain widths across various regions, as shown in \citet{Sahin2023ApJ...950..171S}, which suggests a fundamental MHD mechanism. The widths may be determined by magnetic topology and heating length scales \citep{Antolin2022A} or may be a characteristic of the cooling (for instance, set by TI, e.g. \citet{VanderLinden_1991SoPh..134..247V}). The lengths, on the other hand, can vary significantly, from a few Mm up to tens of Mm according to \citet{Sahin2023ApJ...950..171S}, and it is unclear if there is a connection to temperature \citep{antolin2012, antolin2015}. Regarding the coronal rain dynamics, the average velocities vary between 40-70 km s$^{-1}$, with a long tail of 200-220~km~s$^{-1}$ \citep{schrijver2001, antolin2012, Kleint2014ApJ...789L..42K, Schad2016ApJ...833....5S, Sahin2023ApJ...950..171S}. The velocity of coronal rain is characterized by being lower than free-fall and is typically attributed to the presence of gas pressure forces \citep{Oliver2014}. Other candidates (such as ponderomotive force from transverse MHD waves) have also been proposed to explain this lower than free-fall acceleration. However, the observed wave amplitudes are typically insufficient to explain the rain dynamics \citep{Verwichte2017A&A...598A..57V}.

Coronal rain shows spatial and temporal coherence. Neighbouring rain clumps follow similar trajectories and occur closely in time, thus defining a rain shower \citep{antolin2012}. These rain showers indicate the direction of the magnetic field and have a transverse length scale of a few Mm, and lengths of $\approx$27~Mm on average \citep{Sahin2022ApJ...931L..27S}. \citet{Sahin2022ApJ...931L..27S} use rain showers to properly identify coronal loops (as coronal plasma volume with similar thermodynamic evolution) and quantify the TNE-TI volume over an active region. They have estimated that at least 50\% of the AR is subject to TNE. TI is thought to be responsible for the synchronisation of rain clumps within a rain shower \citep{fang2013, fang2015}. In the TNE-TI scenario, the rain needs strongly stratified heating ($Q_{apex}$/$Q_{footpoint}$$<$0.1), sufficiently long-lasting and frequent enough ($\Delta_{heating}<\tau_{rad}$) and not too asymmetric across both footpoints (asymmetry in heating lower than 3) for TNE-TI (and thus coronal rain) to take place \citep{Klimchuk_2019ApJ...884...68K}. This also seems incompatible with the flaring scenario since very strong but short heating is expected at the footpoints of flaring loops. The problem of the flare-driven coronal rain origin is further investigated by \citet{Reep2020ApJ...890..100R} based on a 1D simulation study, in which they show that with the electron beam alone in a 1D setup, it is not possible to form any condensation.

For quiescent coronal rain, the observed downward mass flux of material is consistent with the estimated mass of the loop that is initially present \citep{antolin2015, Kohutova2019A&A...630A.123K}. However, this does not hold for any coronal structure that exhibits coronal rain. For instance, prominence/coronal rain hybrids show a much larger downward mass flux as coronal rain, with continuous downflows lasting over an hour when only 10 min or so would lead to a full depletion of the loop \citep{Chitta2016A&A...587A..20C}. This apparent mismatch has been explained by the fact that these structures have magnetic dips at the top that act as large mass reservoirs \citep{ChenH_2022AA...659A.107C}. Here, we address this question for the flare context and investigate the flare-driven mass flux relative to that observed from chromospheric evaporation. 

To investigate the role of flare-driven rain in the mass and energy circulation during a flare, it is important to investigate the similarities and differences between quiescent and flare-driven rain. Studies indicate that flare-driven rain can exhibit much large densities of 10$^{13}$~$cm^{-3}$ in the clump core emitting in H$\alpha$ \citep{Heinzel2018ApJ...859..143H}, an order of magnitude larger than the upper limit for quiescent rain. \citet{Jing2016NatSR...624319J}, in H$\alpha$ with the Goode Solar Telescope (GST), have shown rain widths down to $\approx$120 km, half the width of the quiescent kind, which may be due to the increase in spatial resolution relative to the SST. 

%(Heinzel & Shibata 2018, perhaps also Martínez-Oliveros et al. 2016? Check also Koza et al. 2019) 

There have been no studies to our knowledge that directly compare the morphologies, dynamics and energetics between quiescent and flare-driven coronal rain. In this paper, we present the first high-resolution statistical study comparing these two types of coronal rain over a flaring AR. Furthermore, we also study the chromospheric evaporation observed by IRIS and AIA instruments and compare its mass and energy flux to that obtained from the rain. We present the data and method in Section \ref{sec:data} and Section \ref{sec:method}, respectively. Then, we show detailed observational results in Section \ref{sec:result}. Finally, in Section \ref{sec:sum}, we summarise and discuss the studied events.  

\section{Observations} \label{sec:data}

We studied the active region (AR) of NOAA 12158, located at the west limb, which is shown in Figure~\ref{figure1}. The data were taken by the Interface Region Imaging Spectrograph \citep[IRIS;][]{depontieu2014} and the Atmospheric Imaging Assembly \citep[AIA;][]{lemen2012} on board the Solar Dynamics Observatory \citep[SDO;][]{pesnell2012} on 2014 September 17 between 18:18:34 and 22:02:51 UT. During this period, a C2.1-class solar flare occurred in 19:29:11-20:24:26 UT time range. 

\begin{deluxetable*}{cccccc}[ht!]
\tablenum{1}
\tablecaption{Observational time range with total average detected rain pixel per image for the SJI~2796, SJI~1330, and Cool~AIA~304 for the pre-flare and gradual phases. \label{tab:phases}}
\tablewidth{0pt}
\tablehead{\colhead{Instruments} & \colhead{pre-flare phase} & \colhead{N$_{pixel}$/N$_{image}$} & \colhead{gradual phase} & \colhead{N$_{pixel}$/N$_{image}$} & \colhead{$\Delta$N}}
\startdata
SJI 2796 & 18:18:40-19:29:06 & 585 & 20:24:42-22:02:56 & 1809 & 1224 \\
SJI 1330 & 18:18:34-19:29:01 & 216 & 20:24:37-22:02:51 & 1326 & 1110 \\
AIA 304 & 18:18:31-19:28:55 & 449 & 20:24:43-22:02:55 & 1746 & 1297 \\
%SJI 2796 & 18:18:40-19:29:06 & 955 & 20:24:42-22:02:56 & 2180 & 1225 \\
%SJI 1330 & 18:18:34-19:29:01 & 460 & 20:24:37-22:02:51 & 1548 & 1088 \\
%AIA 304 & 18:18:31-19:28:55 & 718 & 20:24:43-22:02:55 & 2003 & 1285 \\
\enddata
\end{deluxetable*}

We use IRIS level 2 slit-jaw imager (SJI) 2796 and 1330 data, which are retrieved from the instrument website\footnote{\url{https://www.lmsal.com/hek/hcr?cmd=view-event\&event-id=ivo\%3A\%2F\%2Fsot.lmsal.com\%2FVOEvent\%23VOEvent_IRIS_20140917_181834_3860107353_2014-09-17T18\%3A18\%3A342014-09-17T18\%3A18\%3A34.xml}}. The SJI~1330 passband is dominated by the \ion{C}{II}~1335.5~\AA\ in the upper chromosphere/lower transition region forming at $10^{4.3}$~K, but emission from the \ion{Fe}{XXI}~1354.08~\AA\ coronal line forming at $10^{7.0}$~K can also be observed during flares. The emission from both can be easily differentiated based on its morphology: the hot emission is diffuse while the cool emission (e.g. from rain) appears clumpy. The SJI~2796 passband is dominated by 2796~\AA\ emission from \ion{Mg}{II} 2796.35~\AA\ chromospheric line forming at 10$^{4}$~K. The IRIS SJI obtained 10.36 s cadence images in the 1330~\AA\ and 2796~\AA\ over an area of 120\arcsec$\times$119$\arcsec$ centred at [936$\arcsec$,221$\arcsec$] with an image scale of 0\arcsec.3327 pixel$^{-1}$. In this study, we also used the IRIS spectrograph (SG) instrument, which has a cadence of 5.18~s (half of the SJI cadence) and an exposure time of 4~s, with a slit pixel size of 0$\arcsec$.33 and length of 120$\arcsec$ in sit-and-stare mode. For the SG data we focused on the \ion{Fe}{XXI}~1354.08~\AA\, line, the \ion{Mg}{II}~k~2796.02~\AA\ line, and the \ion{Si}{IV}~1402.77~\AA\, line, with a formation temperature for the latter of 10$^{4.8}$~K. 

The SDO/AIA data also analyzed in this study are level 2 downloaded through the same instrument website as IRIS with a 12~s cadence. The AIA observations contain seven broad passbands (94~\AA, 131~\AA, 171~\AA, 193~\AA, 211~\AA, 304~\AA, and 335~\AA). However, we focused on the AIA~304~\AA\ channel dominated by \ion{He}{II}~303.8~\AA\ line (forming at $\approx10^{5}$ K) for the rain detection and its analysis. The pixel size of AIA is 0$\arcsec$.6 pixel$^{-1}$; however, for the purpose of co-alignment, we have rebinned the AIA data to match the SJI plate scale. 

Based on the C2.1-class solar flare, we have divided our analysis into three parts: the pre-flare, impulsive and gradual phases. In the result section, we first present the analysis of the pre-flare and gradual phases and we focus on the impulsive phase. We provide the observational time range for the pre-flare and gradual phases in Table \ref{tab:phases} for each instrument.

%Npre SJI = 409
%Nimp SJI = 321
%Ngrd SJI = 571

%Npre AIA = 353
%Nimp AIA = 278
%Ngrd AIA = 488

\section{Methodology} \label{sec:method}
\subsection{Data preparation}
We first resample the AIA images to match the IRIS SJIs and co-align them by matching the solar limb and using several characteristic features (such as bright points and filament patterns) on the disc and off-limb. We computed spatial shifts ($x$ and $y$) at various time instances using the images of AIA~304~\AA\ and SJI~1330~\AA. These shifts were then applied to all other AIA channels to align images with respect to the SJI images.

The AIA~304~\AA~channel exhibits a temperature response peak at $\approx10^5$ K, resulting from \ion{He}{II}~304~\AA~emission. However, the bandpass also includes an additional secondary peak at $\approx10^{6.2}$ K, which can be attributed to many other spectral lines but in particular to \ion{Si}{XI}~303.32~\AA. When observing coronal rain off-limb, both components can have similar intensity. This similarity is due to the surrounding diffuse hot corona, being more extended along the line-of-sight (LOS) than the He II emission, which comes from the smaller, clumpy coronal rain \citep{antolin2015,froment2020}. To remove the hot emission from the AIA~304 passband, we follow the procedure by Antolin et al. (manuscript in preparation). We first fit the response function of AIA~304 over the hot temperature range with the other EUV passbands of AIA (which offer good coverage over the temperature range of 10$^{5.5}-$10$^{7.2}$~K), as shown in Equation \ref{coolaia}. Here, we take the response function using the \textit{``aia\_get\_response"} command in \textit{Interactive Data Language (IDL)} with the \textit{chiantifix} and \textit{eve} keywords, and specifying the time of the observation with the help of the CHIANTI (version 10) atomic database \citep{Landi2012ApJ...744...99L, Del_Zanna_2021}. For further details, see \citet{Antolin2024}.

\begin{equation}
\label{coolaia}
\begin{split}
    R_{304,hot} & = c_{1}R_{94} + c_{2}R_{131} + c_{3}R_{171}+c_{4}R_{193} + c_{5}R_{211} + c_{6}R_{335} \\
    R_{304,cool} & = R_{304,original} - R_{304,hot}
\end{split}
\end{equation}

Here, c$_{1}$...c$_{6}$ are the coefficients obtained from least squares minimisation, for which the coefficients are c1$=$0.02982845, c2$=$0.02947056, c3$=$0.00012495, c4$=$0.00165538, c5$=$0.00952938, c6$=$-0.01288442. We will refer to those output images as Cool~AIA~304~\AA. 

\begin{figure*}[ht!]
\includegraphics[width=\textwidth]{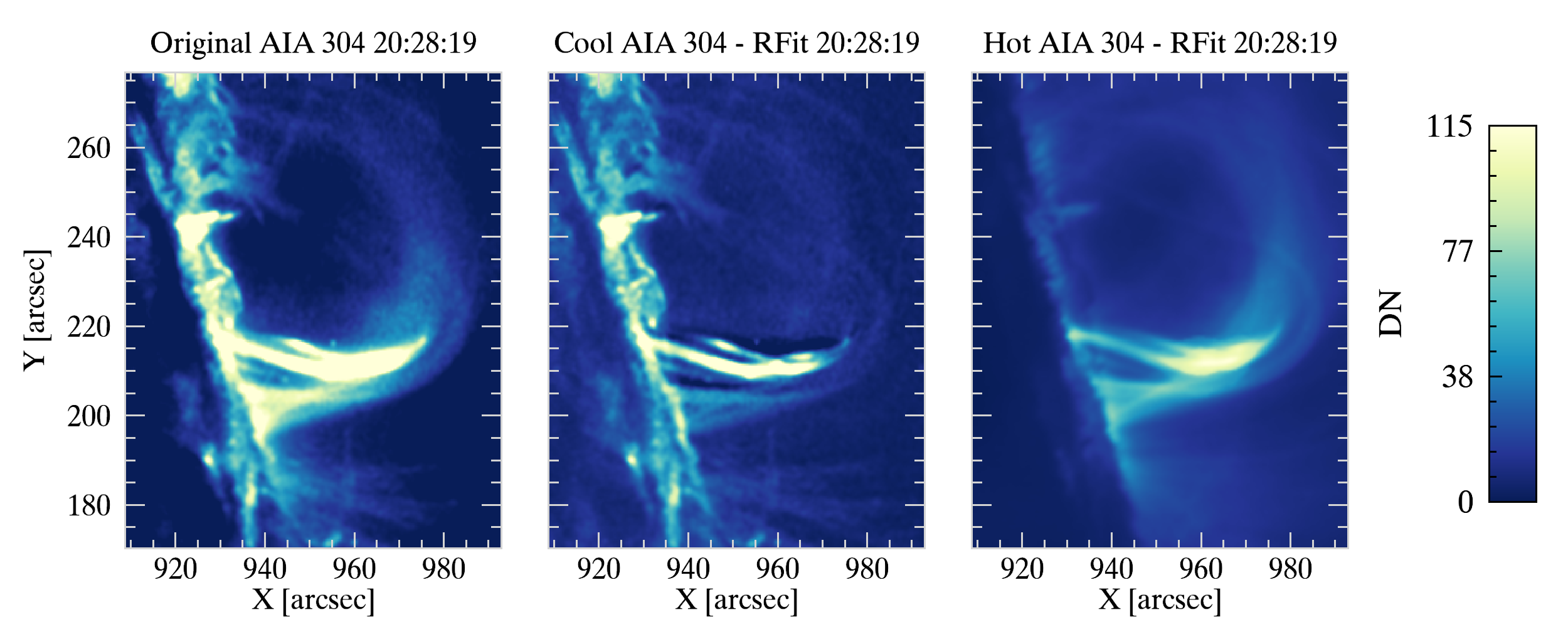}
\caption{Comparison between original AIA~304~\AA\ (left), cool~AIA~304~\AA\ (middle), and hot AIA~304~\AA\ (right). \label{cool_rfit}}
\end{figure*}

Figure \ref{cool_rfit} shows the resulting maps obtained using the Equation \ref{coolaia}. The convolution of the new response function with the EUV passbands, therefore, provides the hot AIA~304 emission (right panel). The cool~AIA~304 (middle panel) emission is then obtained by subtracting this hot component from the original AIA~304 (left panel) intensity. Further details on this procedure can be found in Antolin et al. (manuscript in preparation). 

Similarly to the AIA~304~\AA~channel, the AIA~94~\AA\ channel is blended with three temperature emissions from \ion{Fe}{XV}, \ion{Fe}{XIV}, and \ion{Fe}{XVIII}. Here, in this study, we separated the hot \ion{Fe}{XVIII}~93.96~\AA~line emission with a peak at $\approx10^{6.85}$ K from the warm component in the AIA~94~\AA\ channel since it is particularly important to study heating events \citep{Testa2012ApJ...750L..10T, Ugarte-Urra2014ApJ...783...12U}. For this, we use the empirical method by \citet{DelZanna2013A&A...558A..73D} to compute the hot contribution from a weighted combination of emissions from the AIA~94~\AA, AIA~211~\AA\ and AIA~171~\AA\ channels (see Equation \ref{fexviii}).

\begin{equation}
\label{fexviii}
I_{Fe XVIII}= I_{94} - \frac{I_{211}}{120} - \frac{I_{171}}{450}
\end{equation}

From now on, we will refer to those images as \ion{Fe}{XVIII}, as opposed to the original AIA~94~\AA. 

\begin{figure*}[ht!]
\includegraphics[width=\textwidth]{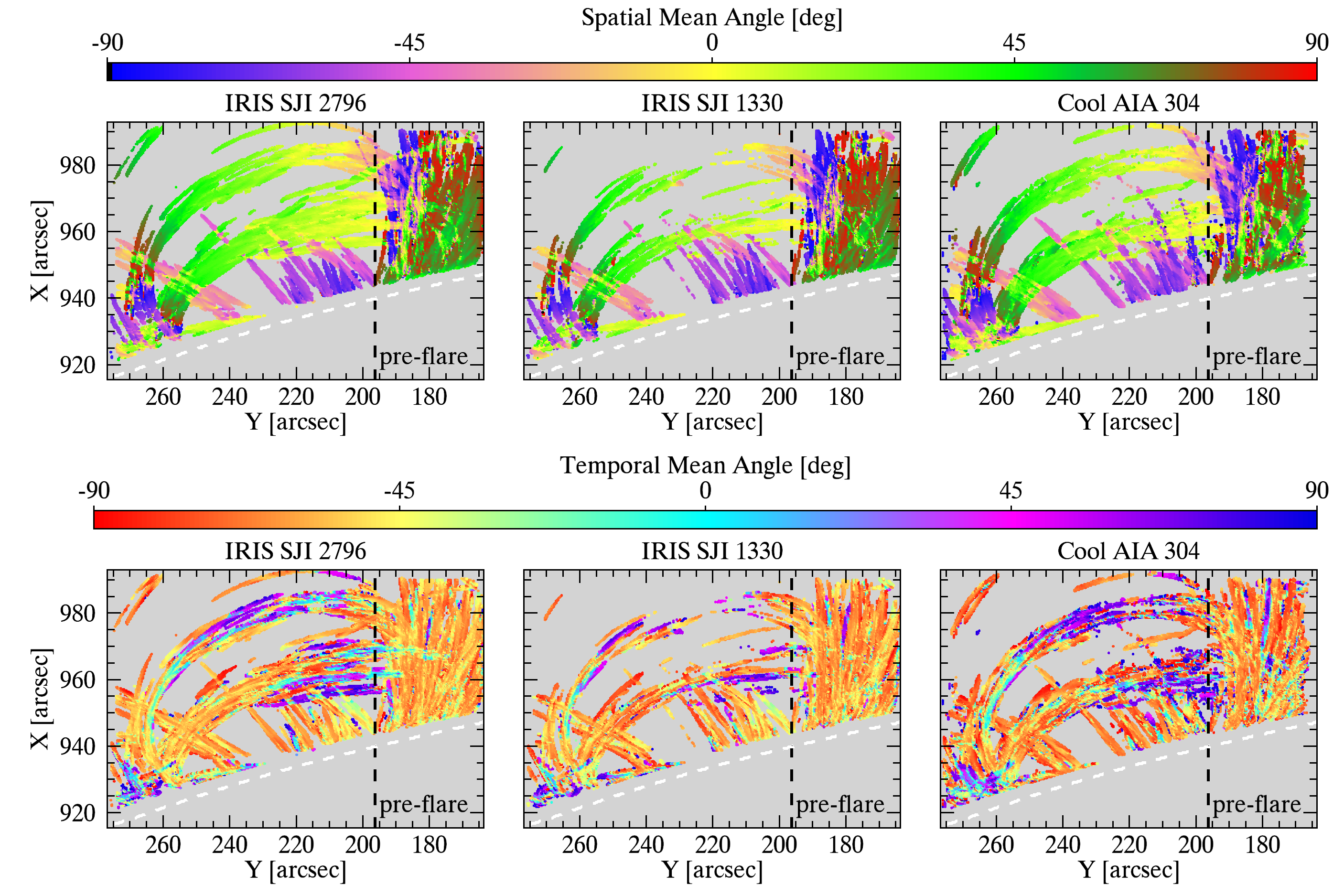}
\caption{Average spatial mean angle (top) and temporal mean angle (bottom) maps over the entire time sequence of the pre-flare phase as derived using the Rolling Hough Transform routine. Average temporal mean angle maps show both downward (negative values) and upward (positive values) motions. The spatial mean angle maps indicate the inclination of the rain with respect to the horizontal direction, while temporal mean angle maps indicate the dynamical change along a trajectory. The dashed white lines indicate the solar limb. The right side of the vertical black dashed line represents the quiescent coronal rain that is observed at all times. \label{pre_mean_angles}}
\end{figure*}

\subsection{Coronal rain detection with imaging instruments}

Before using the semi-automatic coronal rain detection routine, we exclude the region surrounding the solar limb to avoid the significant brightness contrast. We exclude the spicular region above the limb by setting a minimum height of 6 Mm for off-limb observations. This eliminates most of the spicules and other cool chromospheric features that are typically seen above the solar surface. After that, to quantify the coronal rain material, we apply an automatic detection routine called the Rolling Hough Transform \citep[RHT;][]{schad2017} based on the Hough Transform \citep{hough1962method} developed by \citet{schad2017} and further tested for coronal rain detection by \citet{Sahin2023ApJ...950..171S}. First, the images are processed with a high-pass filter and converted to a binary scale before finally being processed with the RHT routine. Some parameters of the routine should be determined before execution for each channel (SJI~2796~\AA\, SJI~1330~\AA\ and AIA~304~\AA). The first parameter is the running mean filter and is determined according to the observed speeds in coronal rain. In this study, we used an 18-step ($\approx$ 3.1 min) for the SJI~2796~\AA, SJI~1330~\AA, and a 16-step running mean filter ($\approx$ 3.2 min) for the AIA~304~\AA. We also use an 18-step bidirectional difference filter along the temporal axis for flow segmentation. Then, we chose the circular RHT kernel width ($D_w$) as 31 pixels for all channels according to the observed minimum length of the rain. This kernel width ($D_w$) is centred on each image pixel to be evaluated at one time. The last parameter that must be specified is the standard deviation regarding the background noise $\sigma_{noise}$. We use 1.1 DN for the SJI~2796~\AA, SJI~1330~\AA\ and 1.4 DN for the AIA~304~\AA. The influence of these parameters and further discussion can be found in \citet{Sahin2023ApJ...950..171S}. Using these parameters, the RHT algorithm produces spatial and temporal information on the mean axial direction, the resultant length, the error in the mean axial direction and the peak of the RHT function. All these outputs have three dimensions in terms of positions (x and y) and time (t). Please refer to \citet{schad2017} and \citet{Sahin2023ApJ...950..171S} for details on these parameters. In this study, we only include the pixels for which $\overline{R}_{xy}\geq0.85$, $max[H_{xy}(\theta)]\geq0.8$, $\overline{R}_{t}\geq0.85$, $max[H_{t}(\theta)]\geq0.8$ for all channels, and $\overline{|\theta}_{t}|\le85^\circ$ for the AIA 304~\AA~and $\overline{|\theta}_{t}|\le84^\circ$ for both SJI channels, since these values ensure relatively small errors for projected velocities.

\subsection{Chromospheric evaporation and coronal rain detection with IRIS/SG}

For this sit-and-stare observation the IRIS slit is located at the footpoint of the flaring loop, which allows to see the dynamics and temperature of the CE episodes we investigate. However, due to the West limb location and the later times of the rain showers compared to CE, we could only capture one of the analysed the rain shower episodes.

For the CE we select the \ion{Fe}{XXI}~1354.08~\AA\, and \ion{Si}{IV}~1402.77~\AA\, lines. For the rain shower, we consider the \ion{Si}{IV} and \ion{Mg}{II}~k lines but focus on the former since \ion{Mg}{II}~k can be optically thick at the low TR heights where the slit is located. For each time step, we select all points along the slit whose spectra are above 5~DN and 8~DN, respectively for the \ion{Fe}{XXI} and \ion{Si}{IV} lines. The number of consecutive points in the wavelength dimension above this threshold needs to be larger than 5. Due to the low count rate of the \ion{Fe}{XXI} line, we only fit a single Gaussian for this line, but we allow either a single or a double Gaussian for the \ion{Si}{IV} line. We select the best fit taking the lowest $\sigma$ error from the fit and also setting a maximum error threshold for each line. We further avoid extremely thin profiles by setting a lower threshold based on the peak temperature formation of the line (we choose 0.13~\AA\, and 0.018~\AA\, for the \ion{Fe}{XXI} and \ion{Si}{IV} lines, respectively). We also allow for the occurrence of additional spectral components in complex profiles. For this, we subtract the result of the fitting (single or double Gaussian) to the original profile and fit the resulting profile subject to the same conditions as for the original profile. In total, a complex line profile is allowed to have 4 distinct spectral components (for example, a main profile fitted with a single or double Gaussian, and 2 residuals to the left and right of the spectrum).

We calculate the spectral quantities from the Gaussian fitting in the usual way, with the position of the peak of the fitting equal to the Doppler velocity, and the FWHM of the line ($\sigma2\sqrt{2\log2}$, with $\sigma$ the Gaussian width) equal to the line width. The errors from the fit provide the corresponding errors in the measurement. The nonthermal velocity $v_{nth}$ is calculated as follows:
\begin{equation}\label{eq:vnth}
v_{nth}=\sqrt{\left(\frac{\sigma c}{\lambda_0}\right)^2-\frac{k_BT}{m_{A}}-\left(\frac{W_{inst}c}{\lambda_02\sqrt{\log2}}\right)^2},
\end{equation}
where $\lambda_0$ is the rest wavelength of the line, $k_B$ is the Boltzmann constant, $c$ is the speed of ight, $m_A$ is the atomic mass of the element and $W_{inst}$ is the instrumental broadening (equal to the IRIS FUV broadening of 26~\AA).

\section{Results} \label{sec:result}

\subsection{Quiescent and Flare-driven Coronal Rain}
\subsubsection{Spatial and Temporal Perspective}

\begin{figure*}[ht!]
\includegraphics[width=\textwidth]{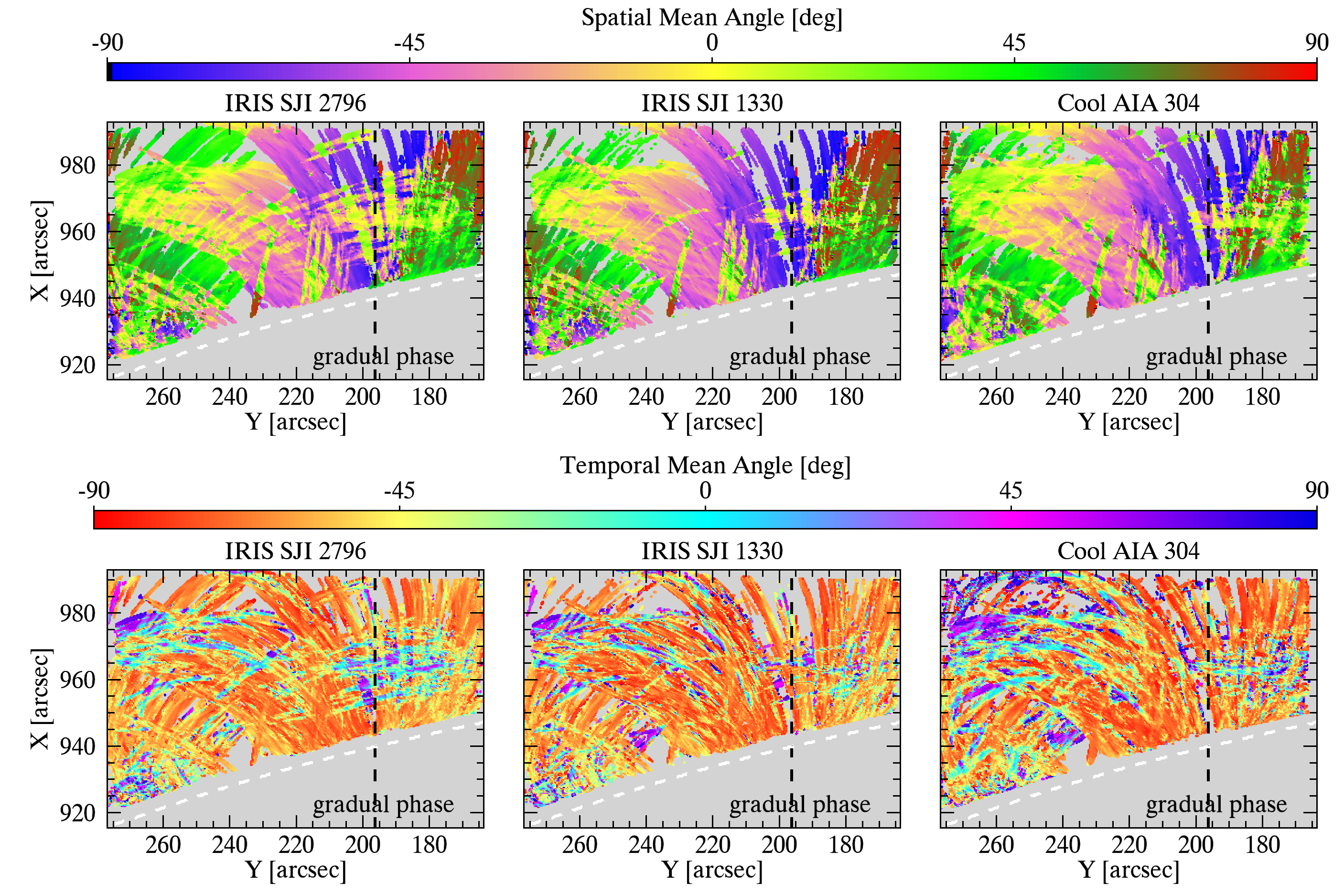}
\caption{The same maps as shown in Figure \ref{pre_mean_angles} but for the gradual phase.} \label{grd_mean_angles}
\end{figure*}

We first examine the pre-flare and gradual phases, where we observe the quiescent and flare-driven coronal rain, respectively. For this purpose, we utilize data from the SJI~2796~\AA, SJI~1330~\AA, and AIA~304~\AA\ channels. Figure \ref{pre_mean_angles} presents the average spatial (top panels) and temporal (bottom panels) mean angle maps obtained from the RHT algorithm during the pre-flare phase, while Figure \ref{grd_mean_angles} shows the same information but for the gradual phase. The spatial mean angle is associated with the inclination of the coronal rain observed in the plane-of-the-sky (POS), whereas the temporal mean angle indicates the dynamic variations occurring along the trajectory of the coronal rain. The coloured pixels on these maps represent coronal rain, and the negative values in the temporal mean angle maps (bottom panels in Figure \ref{pre_mean_angles} and Figure \ref{grd_mean_angles}) indicate the downward motion of coronal rain. As can be seen in these maps, downward motions are dominant due to gravity's force. In these figures, the dashed white lines indicate the solar limb. We also divide these maps into two regions with dashed black lines. The rain in the region to the right is observed during the entire observation time (i.e. from pre-flare to the gradual phases, including the impulsive phase), with apparently little variation, indicating that it is not part of the flare loop. We refer to it as neighbouring coronal rain, and it is not included in our analysis (subject of future work).

\begin{figure*}[ht!]
\includegraphics[width=\textwidth]{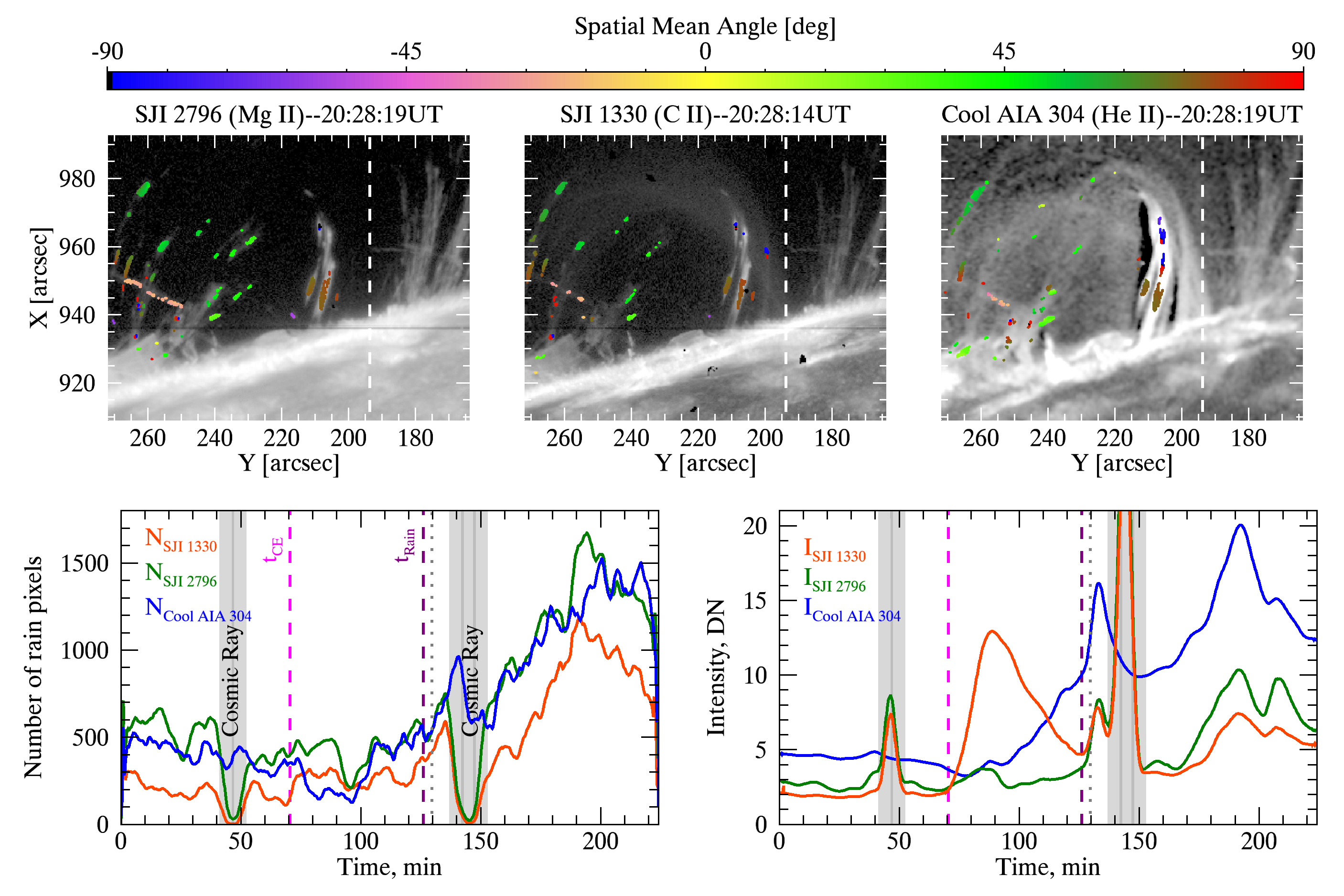}
\caption{Top: Detected rain pixels at 20:28 UT in SJI~2796~\AA\ (left), SJI~1330~\AA\ (middle) and Cool~AIA~304~\AA\ (right). Coloured pixels correspond to the spatial mean angle. Bottom: rain pixel number (left) and rain intensity (right) variation over the entire observational time for the SJI~2796~\AA~(green), SJI~1330~\AA~(red), and cool AIA~304~\AA~(blue). Here, the grey-shaded areas depict where the cosmic ray happens in the SJI channels. The pink and purple dashed lines represent the start time of the CE and gradual phase (rain), respectively. The vertical dotted lines indicate the time of the top panels. An accompanying movie is also available and shows the detected rain pixels in terms of spatial mean angle during the observation time. Note that all curves are 20-point smoothed versions of the averaged number of rain pixels and intensity.} \label{sp_int_variation}
\end{figure*}

Table \ref{tab:phases} provides information about the observational time for the pre-flare and gradual phases and the number of average and maximum detected rain pixels per image in the right column of each phase time. It is worth noting that in the pre-flare phase, we have considered the average number of detected rain pixels, while in the gradual phase, we took the maximum detected rain pixels per image. This can be justified given the relatively constant and increasing rain quantity in the pre-flare and gradual phases, respectively. The detected average rain pixel per image is 585, 216, and 449 for the pre-flare phase in the SJI~2796~\AA, SJI~1330~\AA\ and Cool~AIA~304~\AA, respectively, while the maximum detected rain pixel per image is 1809, 1326, and 1746 for the gradual phase. We would like to highlight that these average numbers are based on the left side of the vertical dashed black line shown in Figure \ref{pre_mean_angles} and Figure \ref{grd_mean_angles}. Even though some horizontal loops (on the left side) have footpoints on the right-hand side of the vertical dashed line, we capture most of these loops (and the respective rain) on the left-hand side. As shown on the bottom-left plot in Figure \ref{sp_int_variation}, the quantity of rain in SJI~2796~\AA, SJI~1330~\AA, and Cool~AIA~304~\AA\ seems to change by roughly the same amount from pre-flare to the gradual phase. However, this is not reflected by the percentage because the initial amounts are different. In order to measure the amount of flare-driven rain, we subtract the average values found in the pre-flare phase, and the results of this subtraction ($\Delta$N) are provided in the last column. It can be seen that these values are very similar to each other in terms of numbers, with similar trends. Remarkably, the SJI~1330~\AA\ displays the lowest values in contrast to SJI~2796~\AA\ and Cool~AIA~304~\AA, possibly due to lower opacity in SJI~1330~\AA\ \citep{Leenaarts2013ApJ...772...89L, Rathore2015ApJ...811...80R}, and thereby making the rain more difficult to detect.

\begin{deluxetable*}{cccc}[ht!]
\tablenum{2}
\tablecaption{Increase factor in intensity and quantity from pre-flare to gradual phases.\label{tab:int_quantity_rain}}
\tablewidth{0pt}
\tablehead{\colhead{} & \colhead{SJI 2796} & \colhead{SJI 1330} & \colhead{AIA 304}}
\startdata
%$\%$Intensity & 214 & 195 & 335  \\
%$\%$Quantity & 209 & 515 & 289  \\
Intensity & 3.14 & 2.95 & 4.35  \\
Quantity & 3.09 & 6.15 & 3.89  \\
\enddata
\end{deluxetable*}

We also examine the relationship between the detected rain pixel number and coronal rain intensity over the solar limb (roughly 8 Mm to avoid eruption occurring in the limb). The top panels in Figure \ref{sp_int_variation} show the detected rain pixels at one particular time (20:36 UT) for the SJI 2796~\AA, SJI 1330~\AA, and Cool AIA 304~\AA. In these panels, the coloured pixels correspond to the spatial mean angle. In the bottom panels, we provide information on the quantity of these detected rain pixels (left) and the average intensity variation (right) above the limb over the observation time. The increase in intensity shows a similar trend as the increase in quantity. The average intensity above the limb exhibits a corresponding increment by a factor of 3.14, 2.95, and 4.35 from pre-flare to the gradual phases for the SJI~2796~\AA, SJI~1330~\AA, and Cool~AIA~304~\AA, respectively. Cool~AIA~304~\AA\ has higher intensity than in the SJI~2796~\AA\ and SJI~1330~\AA. One potential explanation could be associated with opacity, where non-equilibrium ionisation effects greatly enhance the \ion{He}{II} emission \citep{Golding2017A&A...597A.102G}. Additionally, it is conceivable that the observed variations are influenced by the morphology of rain in these channels, with the rain appearing larger in the AIA~304 channel compared to the SJI channels, possibly attributed to differences in resolution (0.87~Mm for the AIA and 0.48~Mm for the SJI channels) \citep{Sahin2023ApJ...950..171S}. Although coronal rain occupies a large POS area on the AR in both phases (see also Figures \ref{pre_mean_angles} and \ref{grd_mean_angles}), it is clear that the rain quantity also increases. Quantifying this variation, we have an increase by a factor of 3.09, 6.15, and 3.89 from pre-flare to the gradual phases for the SJI~2796~\AA, SJI~1330~\AA, and Cool~AIA~304~\AA, respectively. Information on this increase in intensity and quantity is given in Table \ref{tab:int_quantity_rain}. On the bottom-right panel, the big increase in SJI~1330~\AA\ (between two vertical dashed lines) corresponds to chromospheric evaporation (CE) and is not included in the calculation of rain quantity and intensity since it is not part of the gradual phase. The starting time for the gradual phase is given by the purple vertical dashed lines (t$_{Rain}$).

% DISCUSSION --> We found that the intensity increase is larger in the AIA~304\AA\ than the other two channels. %This may be due to the fact that the rains became more visible due to the higher density. Spatial resolution may also affect because the detected blobs are seen thicker in AIA 304 and that means we have more rain pixels. 

\subsubsection{Dynamics of Rain Clumps}

\begin{figure*}[ht!]
\includegraphics[width=\textwidth]{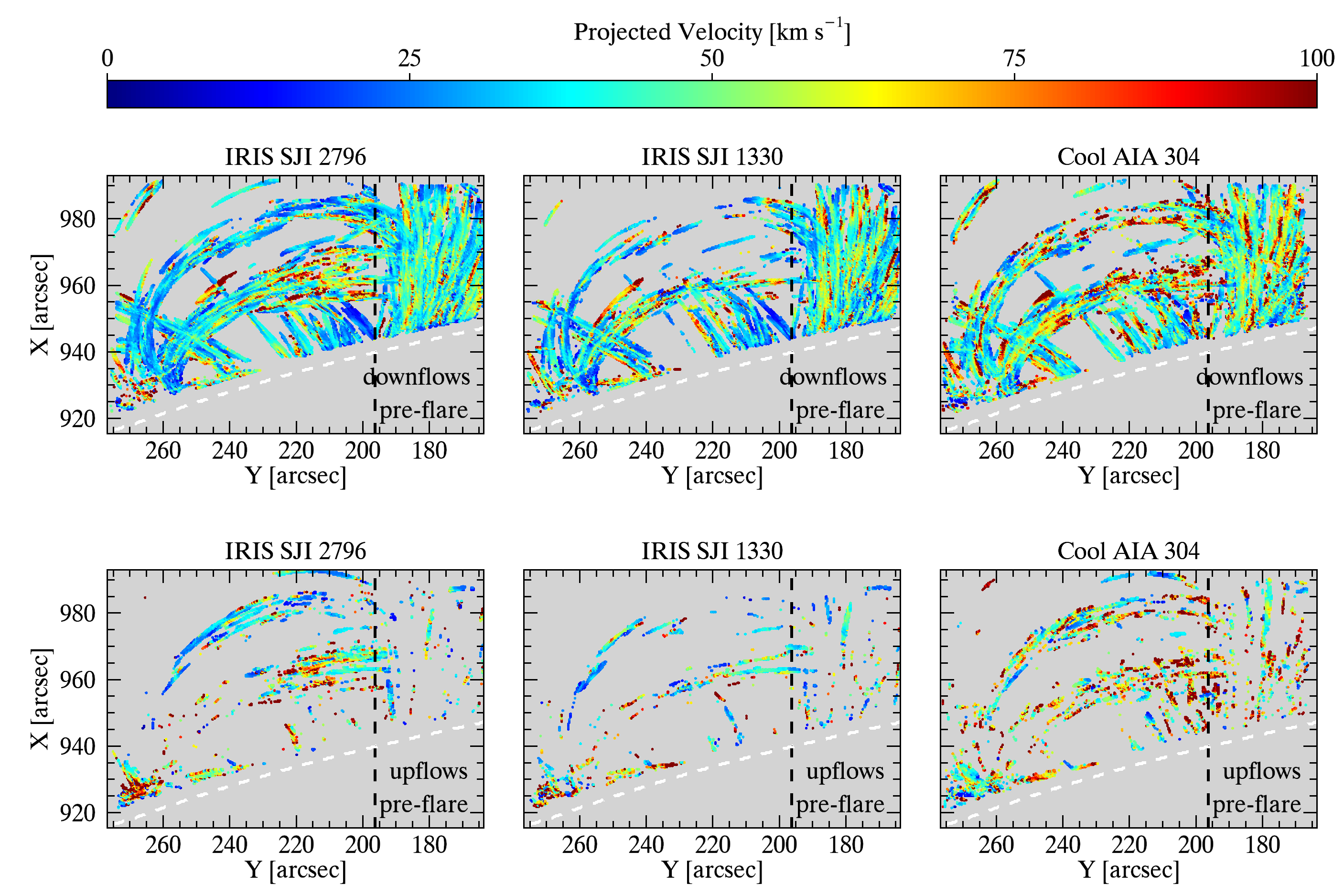}
\caption{Average downflow (top panels) and upflow (bottom panels) projected velocities in SJI~2796~\AA\ (left), SJI~1330~\AA\ (middle), and AIA~304~\AA\ (right) over the pre-flare phase. The dashed white lines over the plots indicate the solar limb. The right side of the vertical black dashed lines represents the quiescent coronal rain that is observed at all times.} \label{pre_rain_dynamics}
\end{figure*}

\begin{figure*}[ht!]
\includegraphics[width=\textwidth]{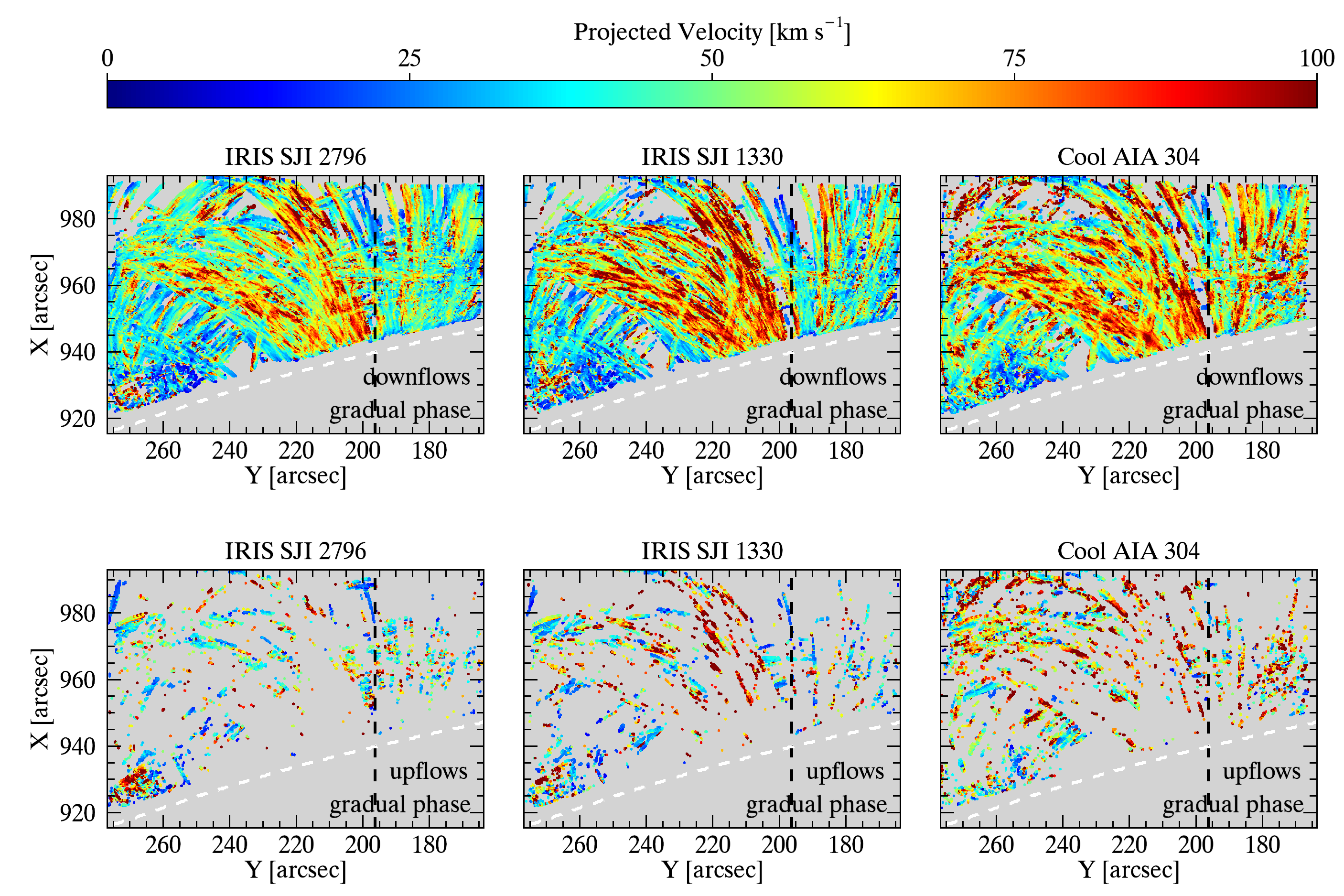}
\caption{The same maps as shown in Figure \ref{pre_rain_dynamics} but for the gradual phase} \label{grd_rain_dynamics}
\end{figure*}

We are now shifting our focus to the analysis of rain clump dynamics. We first show the projected velocity of each rain clump over the pre-flare and gradual phases in Figure \ref{pre_rain_dynamics} and Figure \ref{grd_rain_dynamics}, respectively. Using the Equation \ref{vparalel} \citep{schad2017}, we obtained the velocity for each curved path.

\begin{equation}
\label{vparalel}
v_{||}=\tan{\overline{\theta}_{t}}(\frac{\delta x}{\delta t}).
\end{equation}

Here, $\delta t$ and $\delta x$ represent the average cadence and spatial sampling on the given date, respectively. $\delta x$ is 242.51~km for the SJI channels and  242.44 for the AIA channel. $\delta t$, on the other hand, is 10.36~s and 12~s for the SJI and AIA channels, respectively. Then, the horizontal and vertical velocities are calculated using Equation \ref{vhorz} and Equation \ref{vvert}, respectively, through the RHT spatial mean angle ($\overline{\theta}_{xy}$). 

\begin{equation}
\label{vhorz}
v_{x}= v_{||} \cos \overline{\theta}_{xy}
\end{equation}

\begin{equation}
\label{vvert}
v_{y}= v_{||} \sin \overline{\theta}_{xy}
\end{equation}

Using Equation \ref{vhorz} and Equation \ref{vvert}, tangential (Equation \ref{eq:tan}) and radial (\ref{eq:rad}) velocity components are derived:

\begin{equation}
\label{eq:tan}
v_{tan}= -\cos(\alpha)v_{x}+\sin(\alpha)V_{y}
\end{equation}

\begin{equation}
\label{eq:rad}
v_{rad}= \sin(\alpha)v_{x}+\cos(\alpha)V_{y}
\end{equation}

where $\alpha$ represents the angle between the radial and horizontal directions in radians. These equations provide the projected velocity as shown in Equation \ref{prov}. 

%Using Equation \ref{vhorz} and Equation \ref{vvert}, tangential (Equation \ref{eq:tan}) and radial (\ref{eq:rad}) velocity components are derived, which in turn provide the projected velocity as shown in Equation \ref{prov}. 

\begin{equation}
\label{prov}
 v_p=|v_{||}|=\sqrt{v_{tan}^2 + v_{rad}^2}
\end{equation}

%The projected velocity can also be decomposed in terms of the tangential (Equation \ref{tan}) and radial (Equation \ref{rad}) velocity components:

%\begin{equation}
%\label{tan}
%v_{tan}=v_{x}(-\cos{(\alpha))}+v_{y}(-\sin{(\alpha))} \\
%\end{equation}

%\begin{equation}
%\label{rad}
%v_{rad}=v_{x}(-\sin{(\alpha)})+v_{y}\cos{(\alpha)}, \\
%\end{equation}
%where $\alpha$ denotes the angle between the radial and horizontal directions in radians. The projected velocity is then:

\begin{figure*}[ht!]
\includegraphics[width=\textwidth]{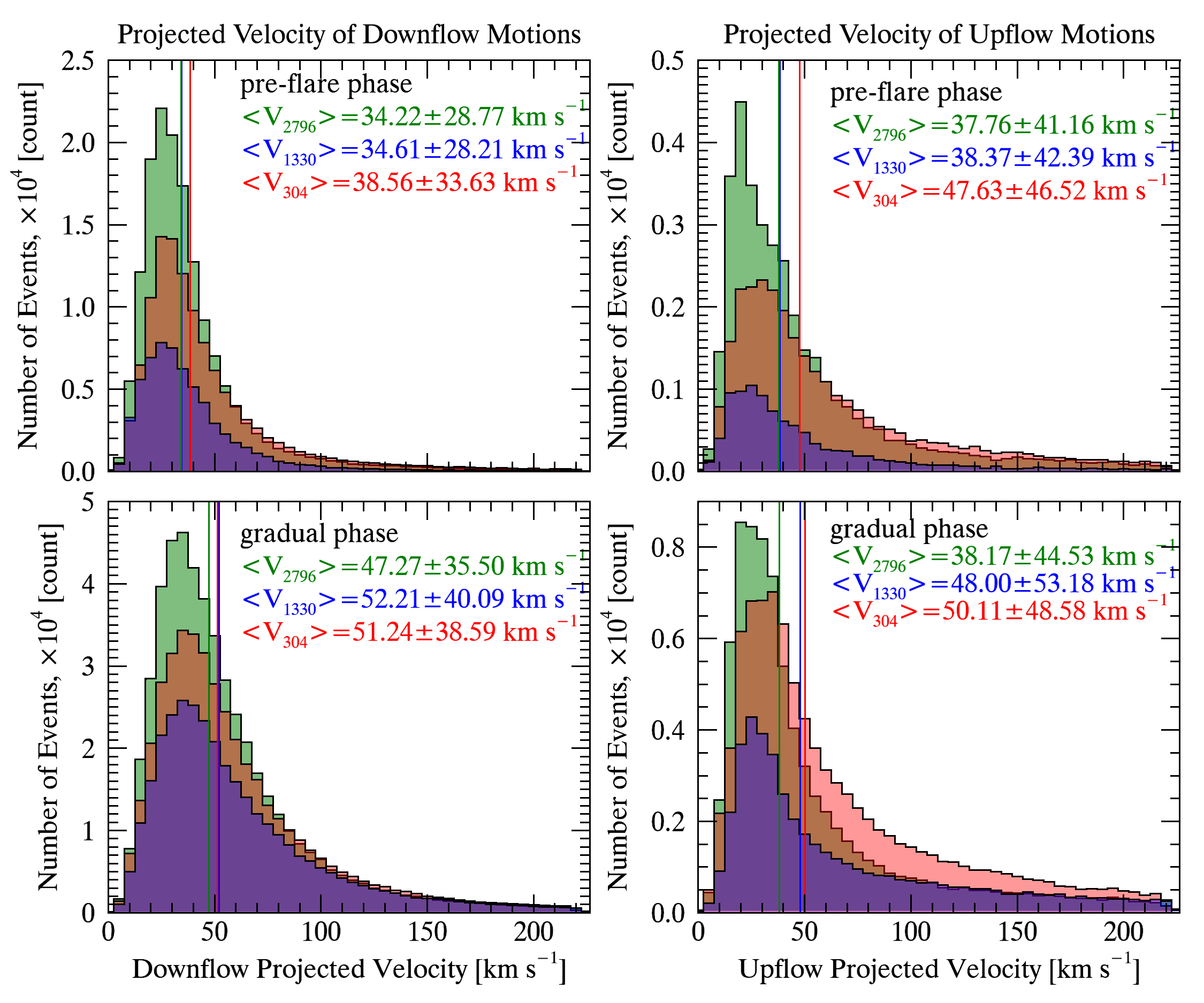} 
\caption{Top: 1D histogram distribution of the projected velocities during the pre-flare phase for the downflow (left) and upflow (right) motions in SJI~2796\AA\ (green), SJI~1330\AA\ (blue), and AIA~304\AA\ (red), with the corresponding mean and standard deviation in the inner caption. Bottom: Same as the top panels, but for the gradual phase.}  \label{1d_histo_pre_grd}
\end{figure*}

We show the average projected velocity maps for downflow and upflow motions in Figure \ref{pre_rain_dynamics} and in Figure \ref{grd_rain_dynamics}, corresponding to the pre-flare and gradual phases, respectively, across the SJI~2796~\AA, SJI~1330~\AA, and Cool~AIA~304~\AA~channels. The majority of coronal rain clumps exhibit increased downward speeds at lower heights, suggesting an acceleration while falling. This is particularly evident during the gradual phase. We also present 1D histogram plots of projected velocity in Figure \ref{1d_histo_pre_grd} for downflows (left) and upflows (right) for the pre-flare (top) and the gradual phases (bottom). Note that we obtained these plots focusing on the left side of the vertical dashed lines in Figures \ref{pre_rain_dynamics} and \ref{grd_rain_dynamics}. It is apparent that both downflow and upflow speeds range from a few km~s$^{-1}$ up to 220 km~s$^{-1}$, and the overall shape of these velocity distributions is very similar across the channels for each phase. The median downflow velocities are found 34$\pm28$~km~s~$^{-1}$, 34$\pm28$~km~s$^{-1}$, and 38$\pm27$~km~s$^{-1}$ during the pre-flare phase and 47$\pm35$~km~s$^{-1}$, 52$\pm40$~km~s$^{-1}$, and 51$\pm38$~km~s$^{-1}$ for the gradual phase in SJI~2796~\AA\ (green), SJI~1330~\AA\ (blue), and Cool~AIA~304~\AA\ (red), respectively. On the other hand, these median velocities are 37$\pm41$~km~s$^{-1}$, 38$\pm42$~km~s$^{-1}$, and 47$\pm46$~km~s$^{-1}$ for the upflow projected velocity during the pre-flare phase and 38$\pm44$~km~s$^{-1}$, 48$\pm53$~km~s$^{-1}$, and 50$\pm48$~km~s$^{-1}$ for the gradual phase in the SJI~2796~\AA\ (green), SJI~1330~\AA\ (blue), and AIA~304~\AA\ (red), respectively. Cool~AIA~304~\AA\ has consistently higher speeds than the SJI channels in the pre-flare phase in both up and downflow motions. The downflow velocity results indicate a factor of 1.38, 1.51, and 1.33 increase in velocity from the pre-flare to the gradual phases for the SJI~2796~\AA, SJI~1330~\AA, and Cool~AIA~304~\AA, respectively. The upflow velocities show almost no change (factors of 1.01 and 1.05 increase for the SJI~2796~\AA\ and Cool~AIA~304~\AA, respectively), or very little change (factor of 1.25 increase for SJI~1330~\AA) from the pre-flare to the gradual phase (see Table \ref{tab:up_down_vp}). We suspect that this discrepancy is due to the very local and sporadic nature of the upflows (further explanation is given below), for which spatial resolution becomes more important, occurring at temperatures higher than the chromospheric range of SJI~2796~\AA.

%This discrepancy could be attributed to a change in the opacity of coronal rain, a consequence of its cooling as it falls and results in an increase in opacity. Hence, some upflow motions can be attributed to the apparent effects due to this opacity change during the falls, resulting in the appearance of rapid upward motion.

\begin{deluxetable*}{cccc}
\tablenum{3}
\tablecaption{Increase factor in downflow and upflow projected velocity from pre-flare to gradual phases.\label{tab:up_down_vp}}
\tablewidth{0pt}
\tablehead{\colhead{} & \colhead{SJI 2796} & \colhead{SJI 1330} & \colhead{AIA 304}}
\startdata
%$\%$ Downflow & 38.13 & 50.85 & 32.88 \\
%$\%$ Upflow & 1.08 & 25.09 & 5.21  \\
Downflow & 1.38 & 1.51 & 1.33 \\
Upflow & 1.01 & 1.25 & 1.05  \\
\enddata
\end{deluxetable*}

\begin{figure*}[ht!]
\includegraphics[width=\textwidth]{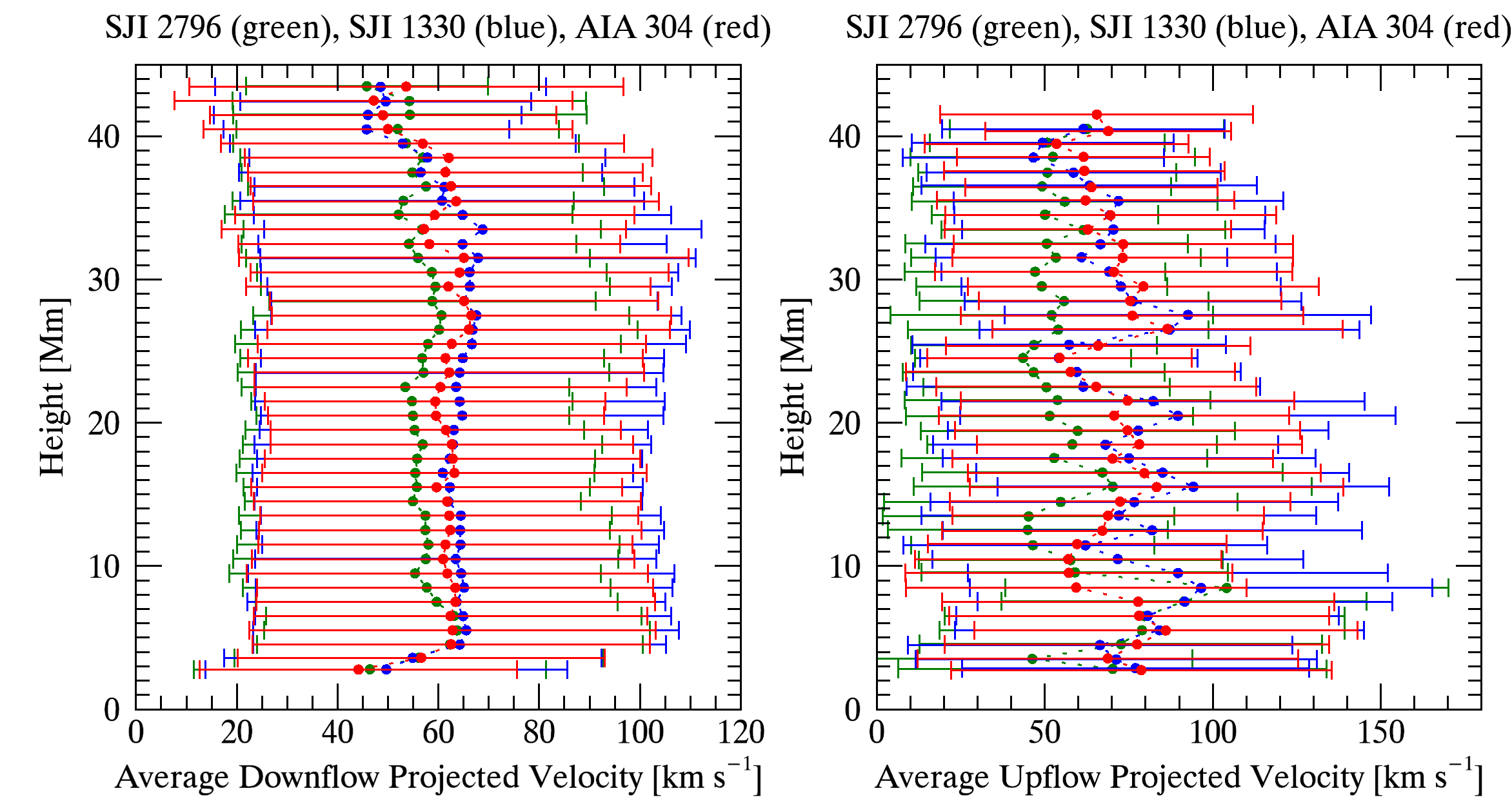}
\caption{The variation of the downflow (left) and upflow (right) motions and their standard deviation at each height bin during the gradual phases. Green, blue and red colors denote SJI~2796~\AA, SJI~1400~\AA, and AIA~304~\AA, respectively. \label{falling_speed_grd}}
\end{figure*}

\begin{figure*}[ht!]
\includegraphics[width=\textwidth]{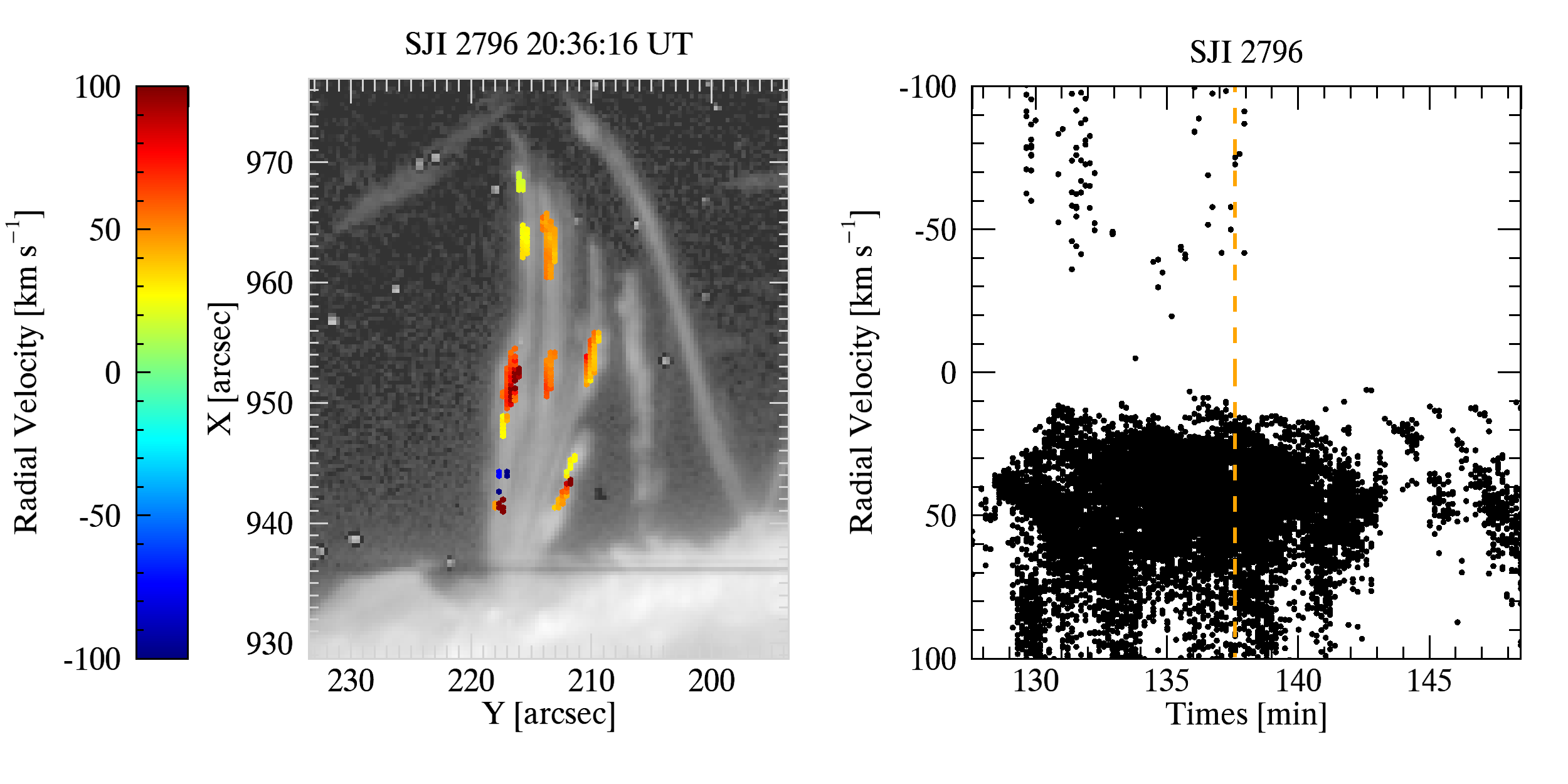}
\caption{Left: One of the shower events observed in the SJI~2796~\AA\ during the gradual phase. The coloured detected pixels show the radial velocity. Here, the positive values correspond to the downflow motions, while the negative values show the upflow motions. Right: The variation of radial velocity with the observational time. The orange dashed line indicates the time. \label{shower1_scatter}}
\end{figure*}

We also examined the behaviour of the average downflow and upflow motions over the projected height during the pre-flare and gradual phases. However, in Figure \ref{falling_speed_grd}, we only show this for the gradual phase since we obtain similar trends during the pre-flare phase. Here, the solar limb is given by zero height. As the rain falls (from roughly 43~Mm to the solar limb), its average downflow projected velocities show a similar trend at all heights, except near the top of the loops (from 43~Mm to 35~Mm), where a clear increase in speeds (roughly from 42~km~s$^{-1}$ to 70~km~s$^{-1}$) is seen. This small acceleration followed by constant downward speed is expected from gas pressure \citep{Oliver2014}. However, it is also likely that projection effects play a role since stronger projection is expected near the loop apex (and therefore smaller speeds). Relative to the downward motion, the upflow projected velocities exhibit more stochastic behaviour. In order to show this behaviour better, we focus on one particular shower event (which we later named shower1 or SH1) in Figure \ref{shower1_scatter}. Here, downflow motions (positive values) are the bulk motions, whereas the upflow motions (negative values) are more localised and sporadic in time. 

\subsubsection{Morphology of Rain Clumps}

\begin{figure*}[ht!]
\includegraphics[width=\textwidth]{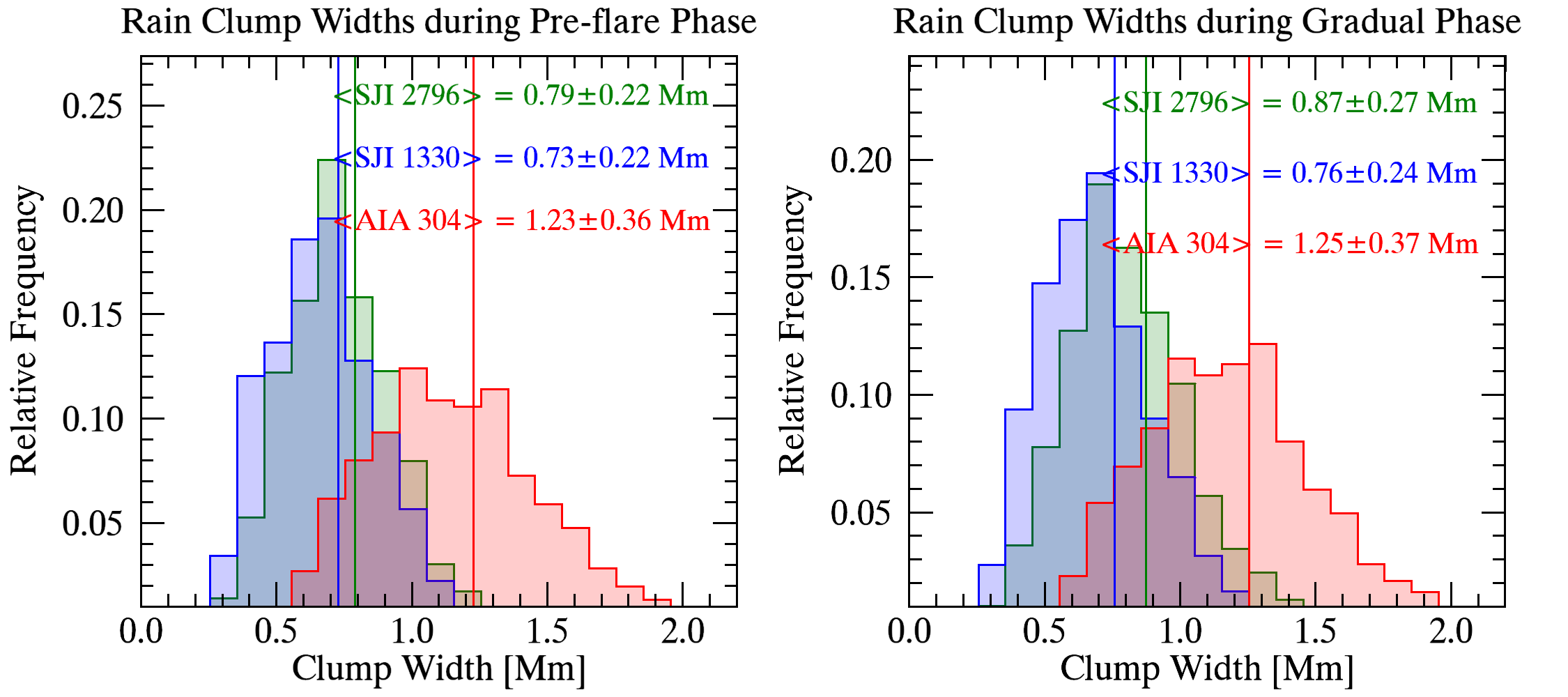} \\
\includegraphics[width=\textwidth]{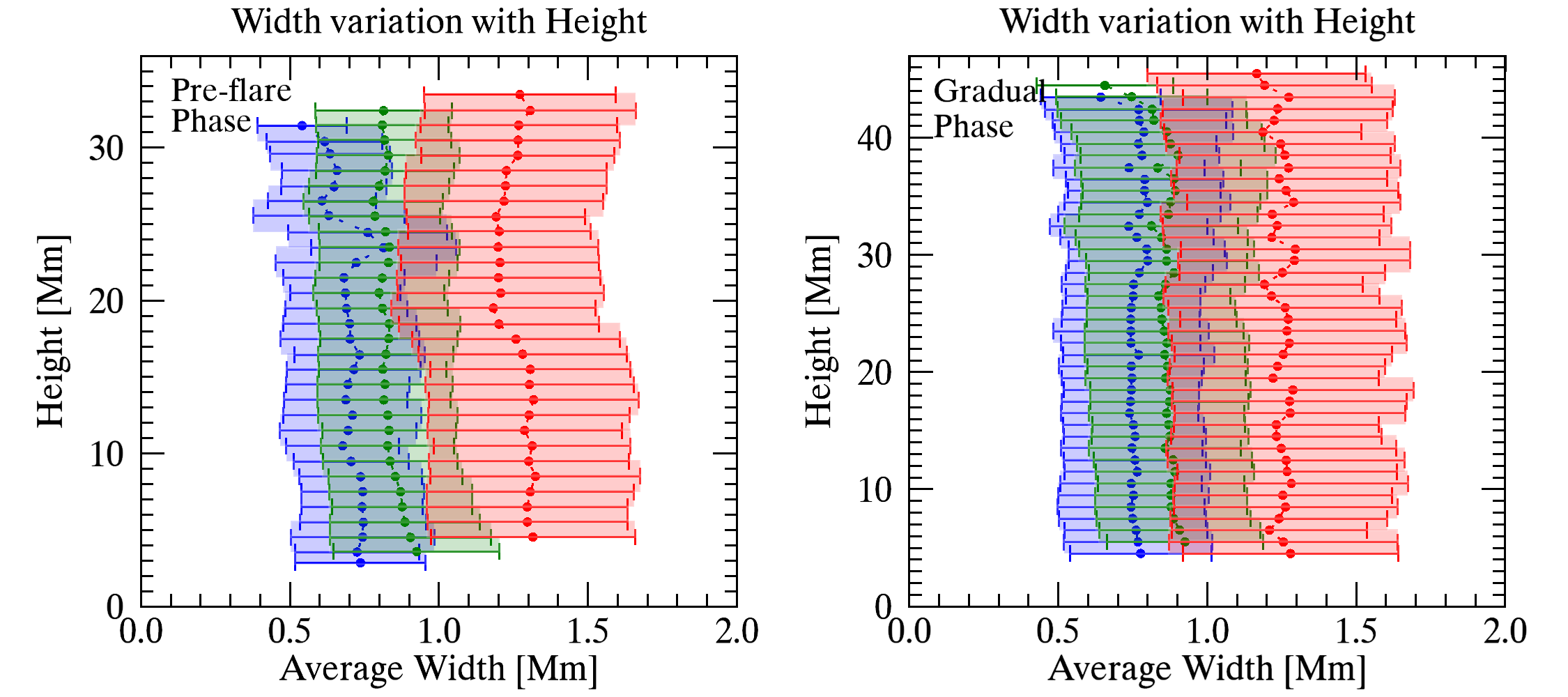} 
\caption{Top: Histograms of rain clump widths during the pre-flare (left) and gradual (right) phases. Green, blue and red denote SJI~2796~\AA, SJI~1330~\AA, and AIA~304~\AA, respectively (see legend). Bottom: The width variation with projected height during the pre-flare (left) and gradual (right) phases. Here, the circles and error bars correspond to the median width and their standard deviation at each height bin. \label{width_morphology_plot}}
\end{figure*}

We also present the statistical analysis of the width and length of individual rain clumps during the pre-flare and gradual phases. The width and length are calculated in the same way as we described in our previous paper by \citet{Sahin2023ApJ...950..171S}. In a nutshell, the width of each rain clump is determined by fitting a Gaussian to the intensity profile along a perpendicular line to the clump's trajectory and calculating the Full Width at Half Maximum (FWHM). To compute the length of the clumps, the center point of the Gaussian fit used for the width calculation is determined, and a straight path is drawn along the same trajectory. The extrema of the clump's length is defined by points where the intensity falls below noise thresholds along the straight path, and the average length is calculated based on these measurements (see  Figure 9 in \citet{Sahin2023ApJ...950..171S}). In Figure \ref{width_morphology_plot}, we show the 1D histogram distribution (top panels) of the rain clump widths in the SJI~2796~\AA\ (green), SJI~1330~\AA\ (blue), and Cool~AIA~304~\AA\ (red) during the pre-flare (left) and gradual phases (right). It is apparent that these distributions are similar in shape across all channels in each phase.

We found that the average widths are 0.79$\pm$0.22~Mm, 0.73$\pm$0.22~Mm, and 1.23$\pm$0.36~Mm for SJI~2796~\AA, SJI~1330~\AA, and Cool~AIA~304~\AA, respectively, in the pre-flare phase. These results are consistent with those shown in our previous paper \citep[e.g.,][]{Sahin2023ApJ...950..171S} where we only focus on the quiescent coronal rain. As discussed in that paper, it is likely that the higher opacity and lower spatial resolution of AIA~304~\AA\ are mainly responsible for the larger widths in that channel. These widths increased by a factor of 1.10, 1.04, and 1.02 in the gradual phase and were found to be 0.87$\pm$0.27~Mm, 0.76$\pm$0.24~Mm, 1.25$\pm$0.37~Mm for the SJI~2796~\AA, SJI~1330~\AA, and AIA~304~\AA, respectively. However, these increases fall within the range of standard deviation (although SJI~2796~\AA\ appears to hold greater relevance), implying that they may lack statistical significance. The bottom plots in the same figure show the width variation with height, where $z=0$ indicates the solar limb. In both pre-flare (left) and gradual (right) phases, the widths remain roughly stable as the rain falls. However, we can distinguish a positive trend in the pre-flare phase in the high-resolution SJI~2796 and 1330 images, with widths increasing 100-200 km on average as they fall from the higher ($\approx$ 32~Mm) to the low ($\approx$ 5~Mm) corona. This trend is not seen in the gradual phase, where widths are largely constant with height. This pre-flare increment could be simply due to the fact that rain appears high up and, even if very rapid, takes some time to grow. In the gradual phase, rain is seen everywhere, particularly in the lower loop portions, reducing this effect. We further note that the widths in the SJI~2796~\AA\ are consistently larger than those in the SJI~1330~\AA\ by $\approx$100~km, for both pre-flare and gradual phase. This may also be due to the opacity difference. These effects are also observed in our previous study (e.g. \citet{Sahin2023ApJ...950..171S}), where we analyzed SJI~2796~\AA\ and SJI~1400~\AA.

\begin{figure*}[ht!]
\includegraphics[width=\textwidth]{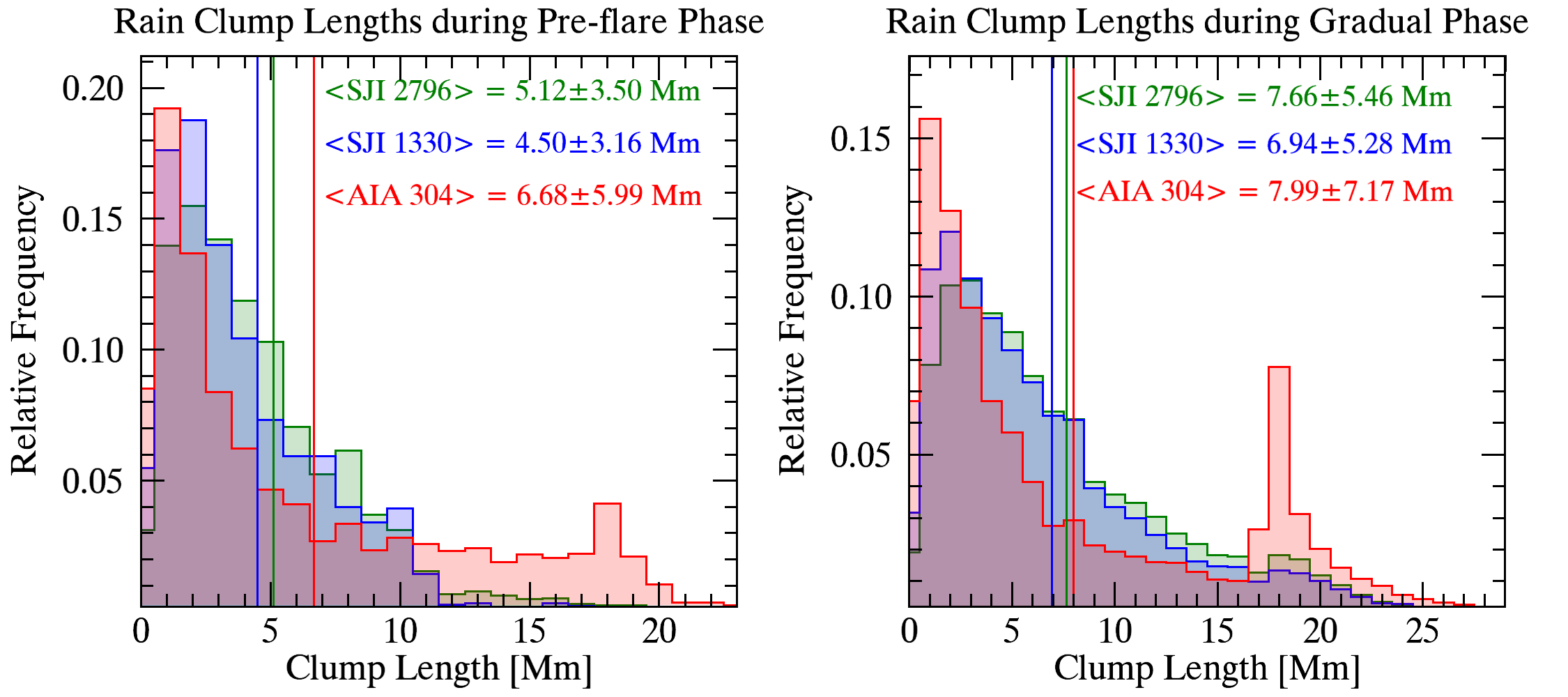} \\
\includegraphics[width=\textwidth]{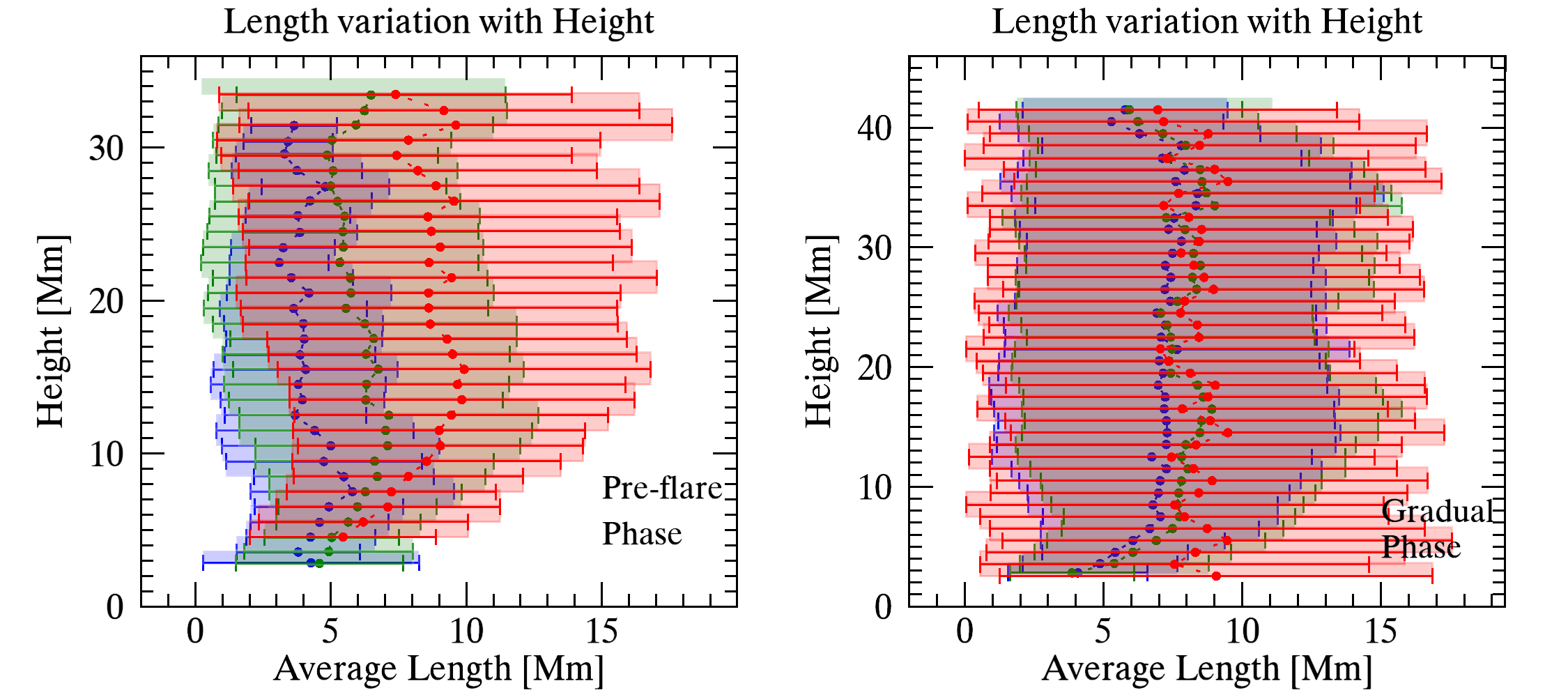} 
\caption{The same plots as shown in Figure \ref{width_morphology_plot} but for the length of rain clumps. \label{length_morphology_plot}}
\end{figure*}

We also show the distribution of the rain clump length in Figure \ref{length_morphology_plot} during the pre-flare (left) and gradual (right) phases. We found that the average lengths are 5.12$\pm$3.50~Mm for the SJI~2796~\AA, 4.50$\pm$3.16~Mm for the SJI~1330~\AA, and 6.68$\pm$5.99~Mm for the Cool AIA~304~\AA. These lengths increased by a factor of 1.50, 1.54, and 1.20 in the gradual phase and were found to be 7.66$\pm$5.46~Mm, 6.94$\pm$5.28~Mm, 7.99$\pm$7.17~Mm for the SJI~2796~\AA, SJI~1330~\AA, and Cool~AIA~304~\AA, respectively. As for the widths, we find a small increase in the lengths in the pre-flare phase of 1000-2000~km with decreasing height for all channels, from 30~Mm down to a height of 13~Mm or so. On the other hand, for the gradual phase, we do not see any trend in any channel, and the lengths become similar, which could simply be attributed to the large quantities of (multi-thermal) rain rapidly occupying the entire loop (clump lengths become roughly 20\% of half of the loop length). Below 13~Mm, we see a strongly decreasing, linear trend of the lengths of all channels for both pre-flare and gradual phases, down to 4000-5000~km at a height of 3000~km. This strong decrease is also obtained for previous datasets \citep{Sahin2023ApJ...950..171S} and can be simply explained by the clumps crossing below the solar limb and the inability of the RHT to continue tracking them. 

We find that the SJI~2796~\AA\ lengths are significantly larger than in the SJI~1330~\AA\ by 1000-2000~km in the pre-flare phase. In turn, the lengths are 2000-3000~km larger in the Cool~AIA~304~\AA\ than in the SJI~2796~\AA (see Figure \ref{length_morphology_plot}). This may be due to the higher opacity and lower resolution of the AIA~304 line. Strong intensity variation along a clump can be interpreted as separate clumps in our algorithm. However, lower resolution reduces these variations, leading to longer clumps with smoother intensities along their lengths.  

\subsection{Chromospheric Evaporation}
\label{subsec:impulsive_phase}

Flare ribbons observed in chromospheric lines like H$\alpha$, directly signify the influence of electron beams on the chromosphere, and mark the initiation of chromospheric evaporation (CE). We start our analysis by investigating the impulsive phase of the solar flare, where we see the flare ribbons and CE (i.e. hot regions). CE occurs between 19:29 and 20:24 UT time range. The diffuse emission signature of CE can be clearly seen in the SJI~1330~\AA\ and (hot) AIA~304~\AA\ channels in Figure \ref{figure1}, as well as in the hotter AIA channels. The SJI~2796~\AA\ and the Cool~AIA~304~\AA\ passbands show an absence of this evaporation. 

\subsubsection{Chromospheric Evaporation Dynamics}

\begin{figure*}[ht]
\includegraphics[width=\textwidth]{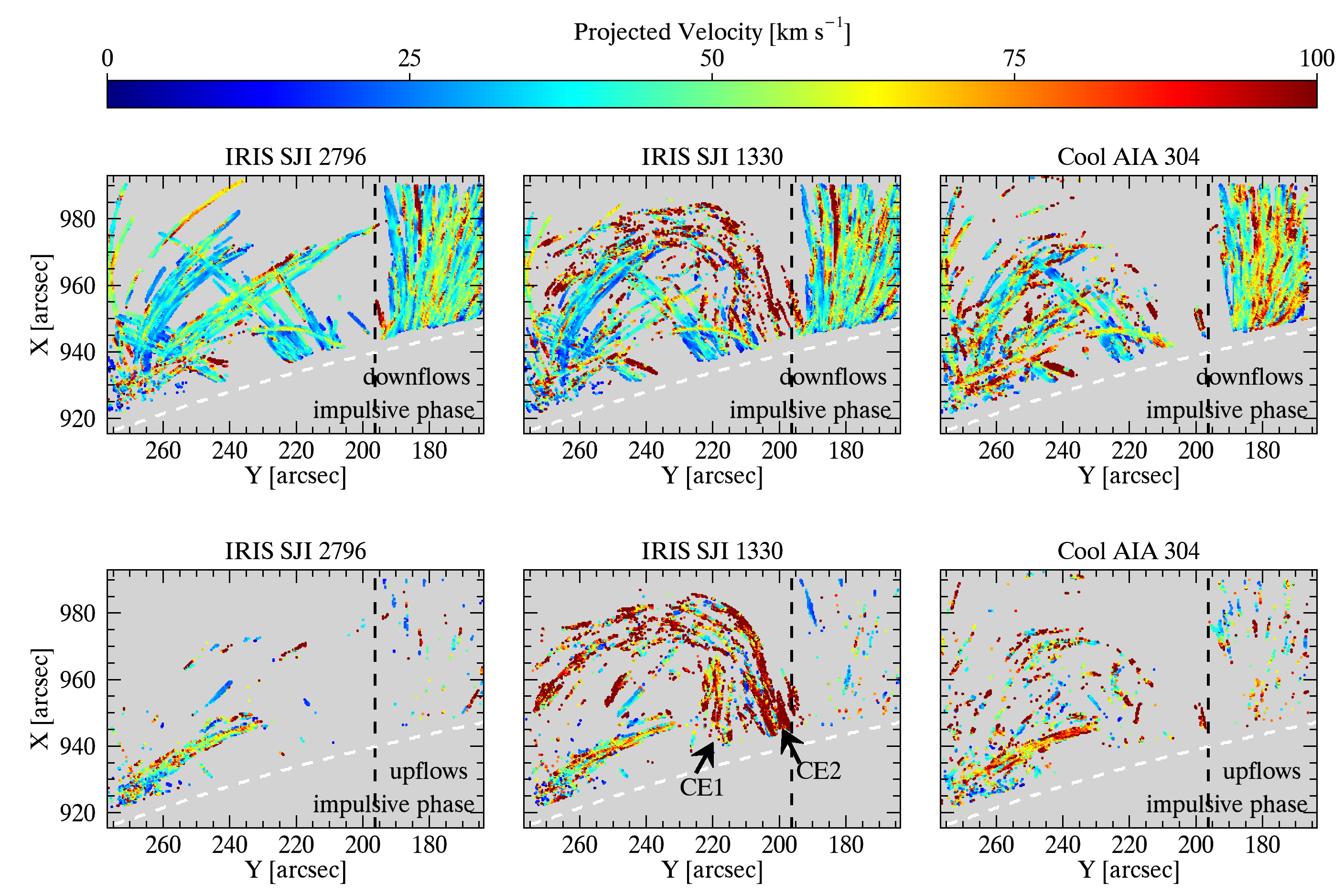}
\caption{The same maps as shown in Figure \ref{pre_rain_dynamics} and Figure \ref{grd_rain_dynamics} but for the impulsive phase.}  \label{imp_ce_dynamics}
\end{figure*}

We also applied the RHT algorithm to SJIs and Cool~AIA data during the impulsive phase and then obtained the projected velocity maps (see Figure \ref{imp_ce_dynamics}) for the downflow (top panels) and upflow (bottom panels) motions. As can be clearly seen in the middle bottom panels, the region filled with red-redish colours in the SJI~1330~\AA\ map corresponds to pixels within the CE region. We would like to note that there are many CE events in this region, and that is difficult to detect all of them because of the very diffuse emission. Therefore, here, we select two CE events in terms of their relative isolation. These two CE events are roughly indicated by small arrows in Figure \ref{imp_ce_dynamics}. As expected, the SJI~2796~\AA\ and Cool~AIA~304~\AA\ channels do not have any signature of these CE events. 

\begin{figure*}[ht!]
\includegraphics[width=\textwidth]{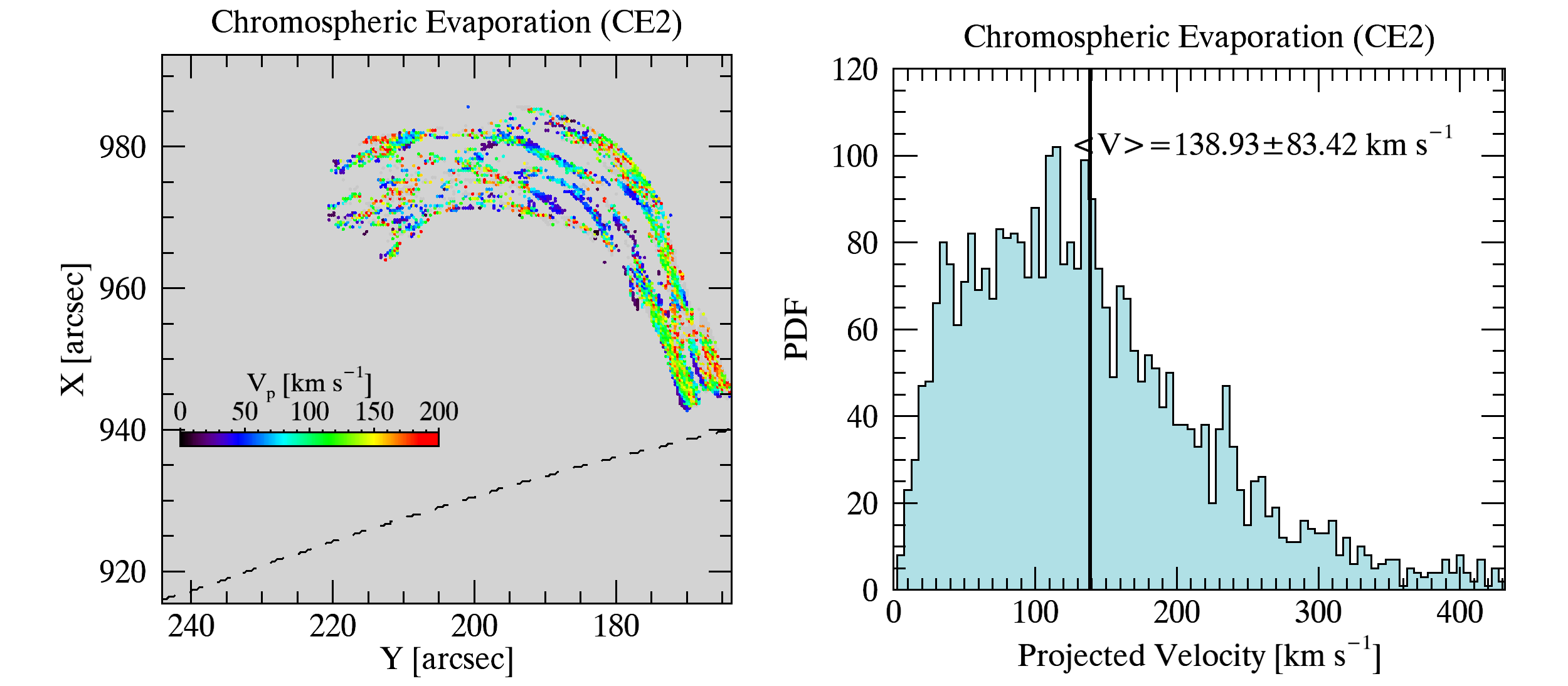}
\caption{Left: Average upflow projected velocities of CE2 in SJI~1330~\AA. Right: 1D velocity histogram distribution of the left panel.}  \label{upflow_chro_dynm}
\end{figure*}

We isolate and classify these pixels showing upflows in SJI~1330~\AA\ during the impulsive phase as potential CE pixels (for the CE2) and plot the corresponding velocity map and histograms in Figure \ref{upflow_chro_dynm}. Note that, in this plot, we released the conditions we used after the RHT routine to capture more material in the CE. We only include pixels for which $\overline{|\theta}_{t}|\le87^\circ$ instead of $\overline{|\theta}_{t}|\le84^\circ$. The maximum velocity extends to just over 400~km~s$^{-1}$. We found that the average speed of CE is 139$\pm$83~km~s$^{-1}$. 

\begin{figure*}[ht!]
\includegraphics[width=\textwidth]{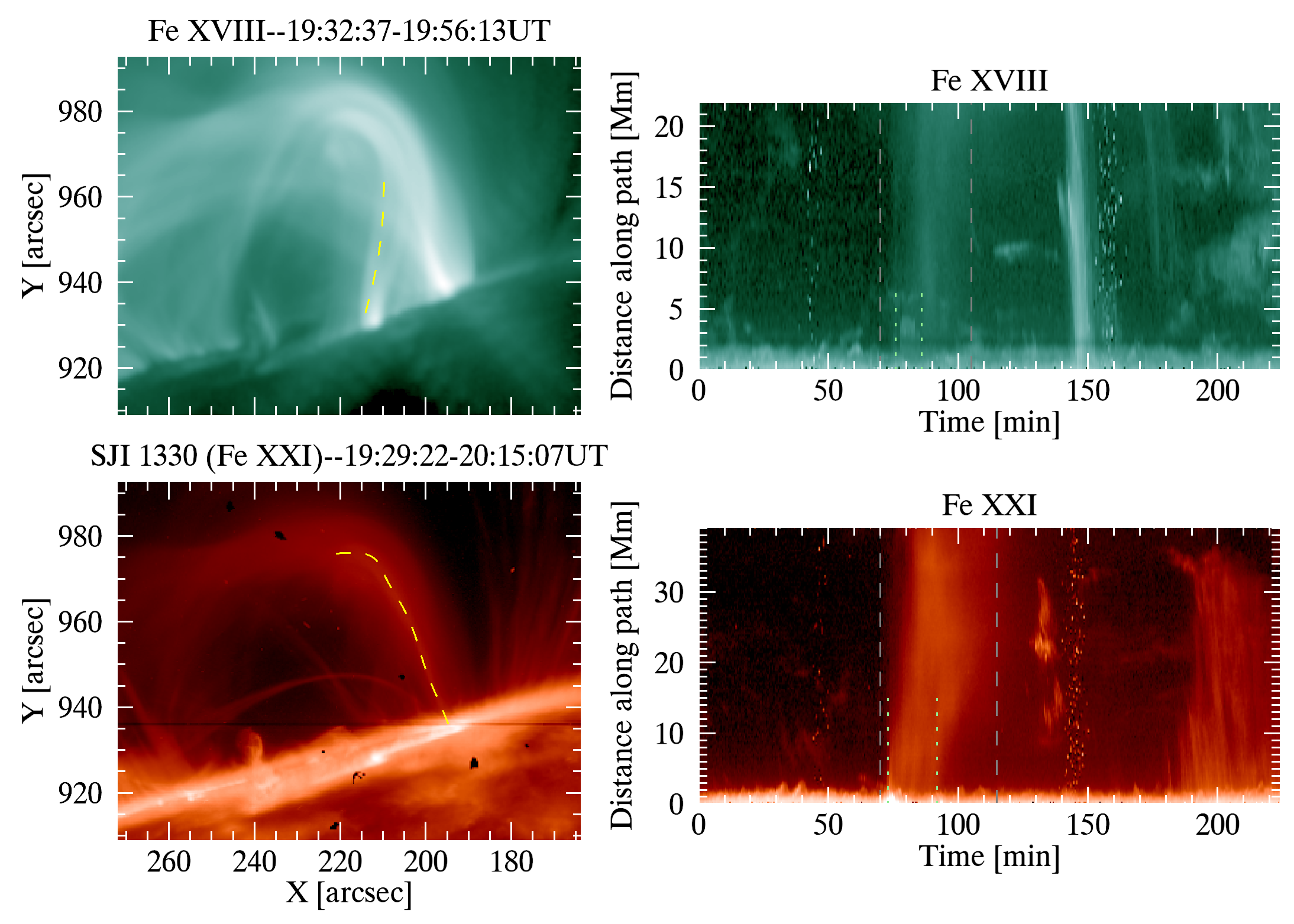} \\
\includegraphics[width=\textwidth]{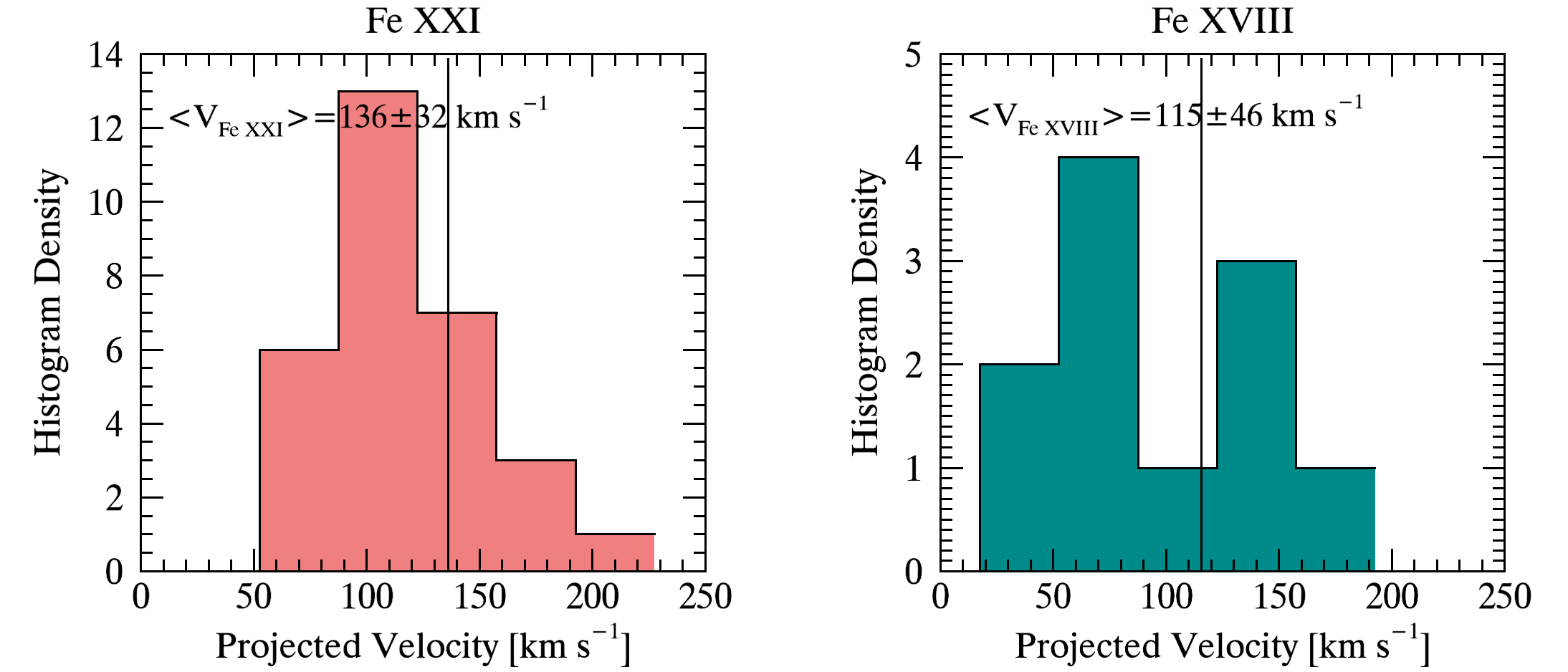} \\
\caption{Observed CE1 (top-left) and CE2 (middle-left) regions, which are obtained by summing the images in the interval 19:32-19:56 UT and  19:29-20:15 UT during the impulsive phase in \ion{Fe}{XVIII} and \ion{Fe}{XXI} (SJI 1330), respectively. Yellow dashed lines in these panels correspond to the chosen and traced paths using the CRISPEX analysis tool. Their time-distance diagrams are shown in the right panels. The grey dashed lines indicate the total time duration of the hot emission in the respective channels, while the green dotted lines indicate the duration of the CE events. The bottom panels show the histograms for the calculated velocities with TANAT for the CE events. Solid vertical lines denote the mean values.}  \label{fig:crispex}
\end{figure*}

We have also tracked the CE events along the loop with the help of the CRisp SPectral EXplorer (CRISPEX) and Timeslice ANAlysis Tool (TANAT), two widget-based tools programmed in the Interactive Data Language (IDL). These tools provide easy browsing of the image or spectral data, the determination of loop trajectory, extraction, and further analysis of time-distance diagrams. Figure \ref{fig:crispex} shows the observed CE events in \ion{Fe}{XVIII} and \ion{Fe}{XXI} with the chosen and analyzed loop path (yellow dashed lines). On the right panels, we show time-distance (x-t diagram) maps obtained from the traced loop path on the left panels. The total time duration (see grey dashed lines in the x-t diagram) for the hot emission that accompanies CE1 and CE2 is roughly 33 and 44~min, respectively. However, this does not directly reflect the duration of each CE event. Indeed, besides CE, the hot emission can be produced by other sources, such as the work performed by the Lorentz force or directly from the reconnection region. To determine the true duration of the CE events we note the presence of bright emission propagating upwards from the footpoint, which last roughly 10~min and 19~min (see green dotted lines) for the CE1 and CE2, respectively. As we will see in section~\ref{sec:ce_spec}, this is corroborated by the SG analysis. We made multiple velocity measurements at different heights in these x-t diagrams, and we found a broad velocity distribution (see histogram plots) between 15 and 200~km~s$^{-1}$, with an average 115$\pm$46~km~s$^{-1}$ for the CE1. We found 136$\pm$32~km~s$^{-1}$ on average for the CE2, which is consistent with the velocity from the RHT result.

%% MOVE TO DISCUSSION --> The velocities related to CE range in value. For instance, an observed velocity range of 126 to 210 km s$^{-1}$ was recently reported by \citet{Li2022ApJ...926..216L} during a solar flare of the C6.7-class.

%\begin{figure*}[ht!]
%\includegraphics[width=\textwidth]{figures/Figure14.png}
%\caption{The evolution of Kelvin-Helmholtz (KH) type vortex observed in the SJI~1330~\AA\ channel (shown by white arrow) in response to the CE}  \label{vortexfigu}
%\end{figure*}

%We also observed a vortex-like structure that forms once the CE reaches the top of the loop, as shown in Figure \ref{vortexfigu}.

%produced by the CE flow, closely resembling the results of \citet{Fang2016ApJ...833...36F} in the 2.5D MHD model. Therefore, it is likely that this is a Kelvin-Helmholtz (KH) vortex produced by shear flow. In the numerical model, the shear flow is produced due to asymmetric heating at both footpoints, leading to asymmetric CE between both footpoints.

\subsubsection{Chromospheric evaporation and rain shower morphology}
\label{sec:ceandsh}

\begin{figure*}[ht!]
\includegraphics[width=\textwidth]{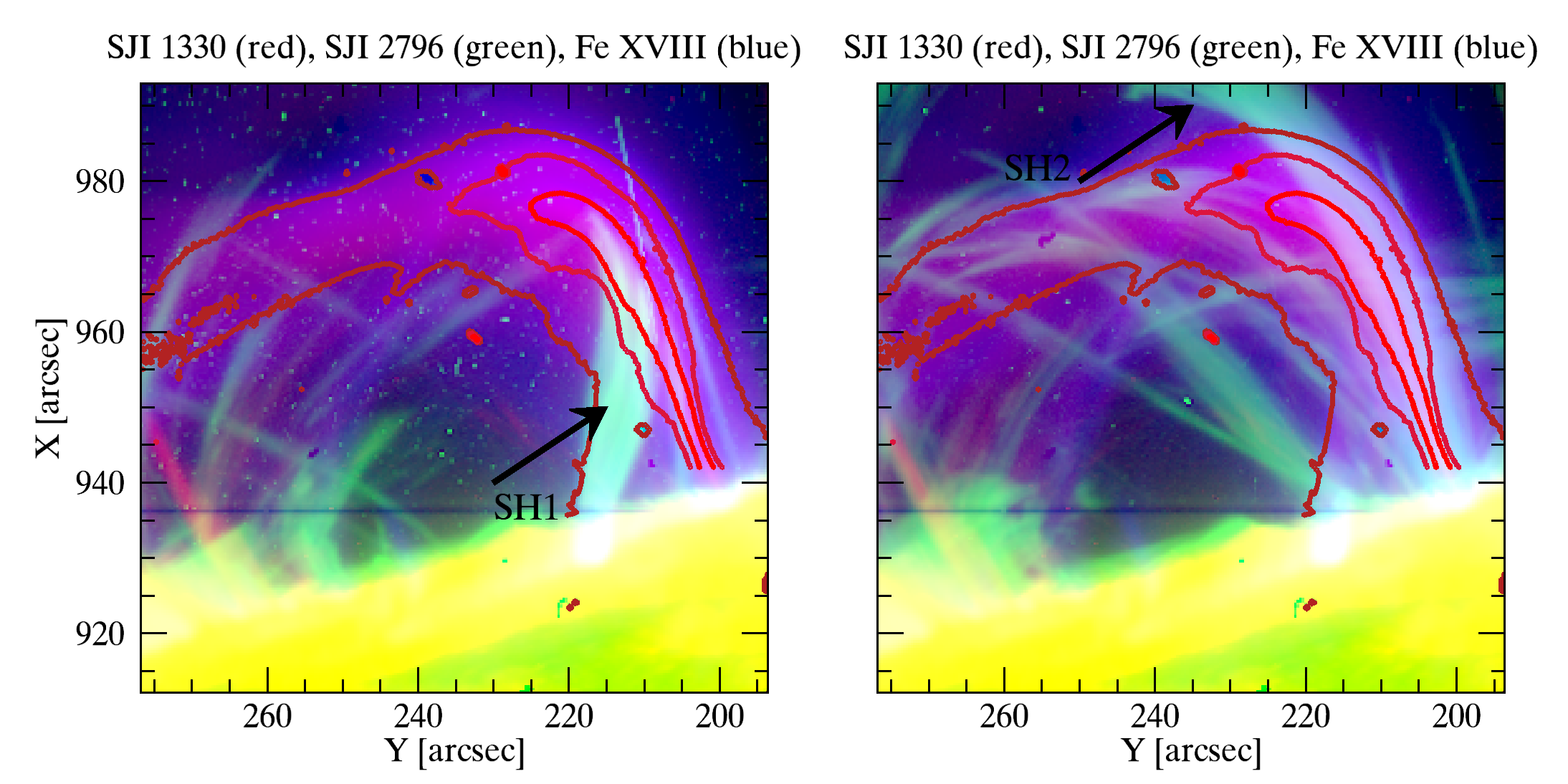}
\caption{Composite image of the SJI~1330\AA\ (red), SJI~2796\AA\ (green), and \ion{Fe}{XVIII} (blue) taken by different observational time range. The SJI~1330\AA\ and \ion{Fe}{XVIII} were obtained by summing the images in the interval 19:30 - 20:25. The SJI~2796\AA\ results from images summed during the 20:25 - 20:37 UT (left) and 21:45 - 21:57 UT (right) interval. The contour lines are derived from the SJI~1330~\AA, emphasizing the CE region based on the different intensity thresholds. \label{alpha_image}}
\end{figure*}

Here, we compare CE and flare-driven coronal rain to determine if the regions where we observe the rain shower align with those of CE. Figure \ref{alpha_image} presents a composite image showing impulsive and gradual phases together. Here, the red and blue corresponds to the SJI~1330~\AA\ and AIA~94~\AA\ images, respectively, taken during the impulsive phase (19:30 - 20:25 UT), while the green is only taken from the SJI~2796~\AA\ images in the time corresponding to the gradual phase (20:25 - 22:02~UT). A very fine intensity structure is seen in the CE, particularly in the SJI~1330~\AA\ channel. This is further shown in the composite image by setting contours in brighter red for higher intensity.

\begin{figure*}[ht!]
\includegraphics[width=\textwidth]{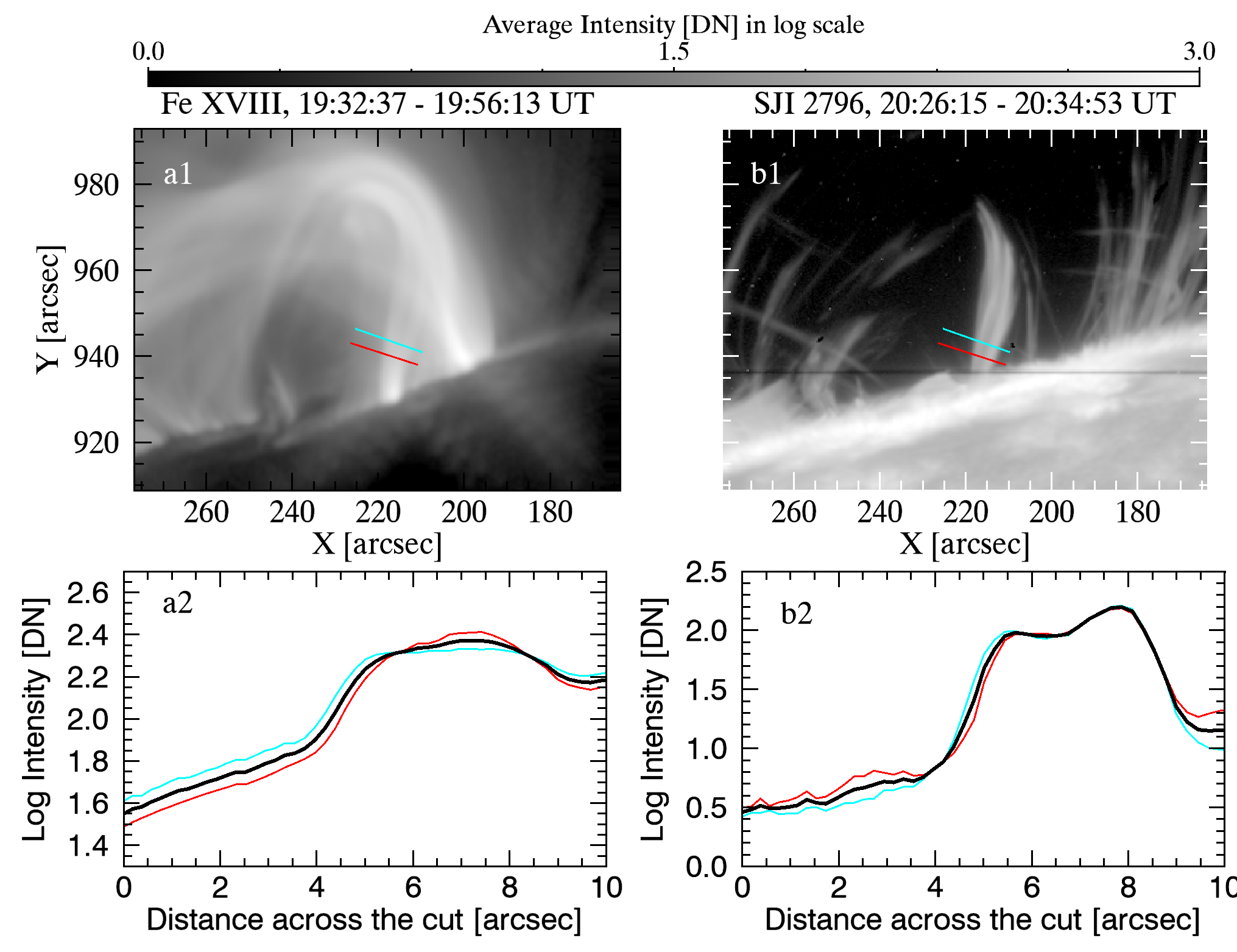}
\caption{Top: Observed CE1 region during the impulsive phase in Fe XVIII (a1) and SH1 during the gradual phase in SJI~2796~\AA\ (b1) in logarithmic scale. The red and cyan colours indicate the perpendicular cuts to the loop trajectory. Bottom: The intensity variation over the distance of these perpendicular cuts for the Fe XVIII (a2) and SJI~2796 (b2). Black curves show the average of these intensity variations.} \label{fig:ce1_sh1}
\end{figure*}

\begin{figure*}[ht!]
\includegraphics[width=\textwidth]{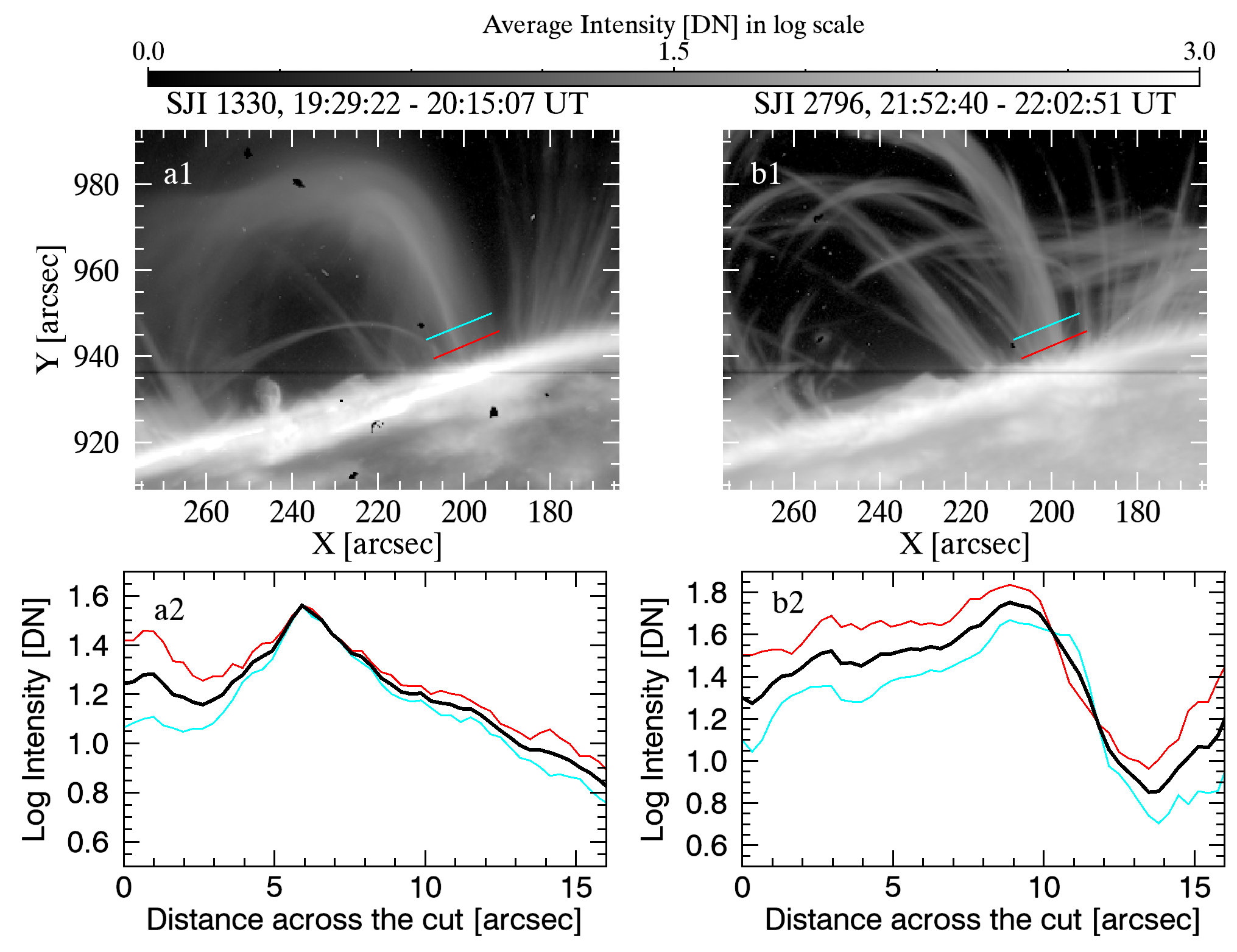}
\caption{Top: Observed CE2 region during the impulsive phase in SJI~1330~\AA\ (a1) and SH2 during the gradual phase in SJI~2796~\AA\ in logarithmic scale. The red and cyan colours indicate the perpendicular cuts to the loop trajectory. Bottom: The intensity variation over the distance of these perpendicular cuts for the SJI~1330 (a2) and SJI~2796 (b2). Black curves show the average of these intensity variations.} \label{fig:ce2_sh2}
\end{figure*}

Similar to the CE events, there are also many more shower events that likely correspond to CE events. We select two main shower events among all showers, namely SH1 and SH2 (see arrows in Figure \ref{alpha_image}), that seem to spatially correspond to the two previously selected CE regions, namely CE1 and CE2. In Figure \ref{fig:ce1_sh1}, we show the CE1 region (left) observed during the impulsive phase and SH1 (right) during the gradual phase. The red and cyan colours show transverse cuts to the CE1 and SH1 events, taken at the same locations. The bottom panels in the same figure provide the time-averaged intensity over the distance of these coloured cuts, where the time intervals are those specified at the beginning of this section. The black line on these plots corresponds to the average intensity of these cuts. Similarly, Figure \ref{fig:ce2_sh2} shows the other CE and SH events: CE2 and SH2. Visual comparison of the overall morphology between CE and SH already reveals a strong similarity. We quantify this similarity by calculating their width. There are essentially two types of widths: an overall width for the entire (overall) CE or shower event (SH) and the strand width. To calculate the overall width, we fit a Gaussian over the intensity profile of the temporally summed image (see top and bottom panels in Figures \ref{fig:ce1_sh1} and \ref{fig:ce2_sh2}, respectively) along each transverse cut (red and cyan) and set the width equal to the Full Width at Half Maximum (FWHM) of the Gaussian fit. In this process, we select the cuts (red and cyan) that are relatively isolated and do not intersect other neighbouring structures. We found that the width of the larger structure (i.e. the overall width) spans a range from [2294$\pm$55, 4813$\pm$208]~km with 3872$\pm$104~km on average for the SH1 and [2771$\pm$83, 5825$\pm$163]~km with 4088$\pm$107~km on average for the CE1. It is [2883$\pm$69,6869$\pm$270]~km with 4427$\pm$157~km on average for the SH2 and [4634$\pm$82,6331$\pm$206]~km with 5526$\pm$155~km on average for the CE2. These calculations are based on the average width measurements over the channel, and the overall average values for each channel (as shown in Figures \ref{appendix1} and \ref{appendix2}) are given in Table \ref{tab:width_CE_SH} for each SH and CE event. Due to the high temperatures, the CE is only visible in certain channels. Also, the SH can be very dim in specific channels. Therefore, we have included only the channels for which the intensity values are larger than a specific noise threshold.

\begin{deluxetable*}{ccccc}[ht!]
\tablenum{4}
\tablecaption{Obtained overall width measurements for the SH and CE events across the channels. \label{tab:width_CE_SH}}
\tablewidth{0pt}
\tablehead{\colhead{} & \colhead{CE1 [km]} & \colhead{CE2 [km]} & \colhead{SH1 [km]} & \colhead{SH2 [km]}}
\startdata
SJI 2796 & - & - & 3377$\pm$63 & 4318$\pm$231 \\ 
SJI 1330 & - & 5847$\pm$206 & 2950$\pm$59 & 6013$\pm$270 \\ 
Cool AIA 304 & - & - & 3313$\pm$59 & 3665$\pm$193   \\ 
AIA 171 & - & - & 3146$\pm$87 & 3584$\pm$143   \\ 
AIA 193 & - & - & 4126$\pm$115 & 4492$\pm$172   \\ 
AIA 211 & - & - & 4804$\pm$153 & 3216$\pm$123   \\ 
AIA 335 & - & 5293$\pm$82 & 4404$\pm$158 & -   \\ 
AIA 94 & 5140$\pm$123 & 6331$\pm$201 & 4279$\pm$93 & 5663$\pm$137   \\ 
AIA 131 & 3036$\pm$90 & 4634$\pm$133 & 4454$\pm$147 & 4559$\pm$120   \\ 
\enddata
\end{deluxetable*}

We also note the sub-structure within each event, which we refer to as CE strands or rain strands, accordingly. Because of the LOS superposition, the width of individual strands can appear much larger than their true value. Hence, for the width calculation, we select the thinnest strand over a particular time. As for the overall width, we fit a Gaussian intensity profile over a path crossing the strand transversely to the local trajectory, and retrieve the FWHM. We found that the average substructure width (i.e. strand width) is 642$\pm$56~km and 1218$\pm$176~km for the SH1 (based on the SJI~2796~\AA) and CE1, respectively, while it is 720$\pm$145~km and 1534$\pm$37~km for the SH2 and CE2, respectively. These results suggest that there is a strong correspondence between shower events and CE in morphology at both sub-structure and larger structure levels.  

\subsection{IRIS Spectra of the chromospheric evaporation}
\label{sec:ce_spec}
\begin{figure*}[ht!]
\includegraphics[width=0.90\textwidth]{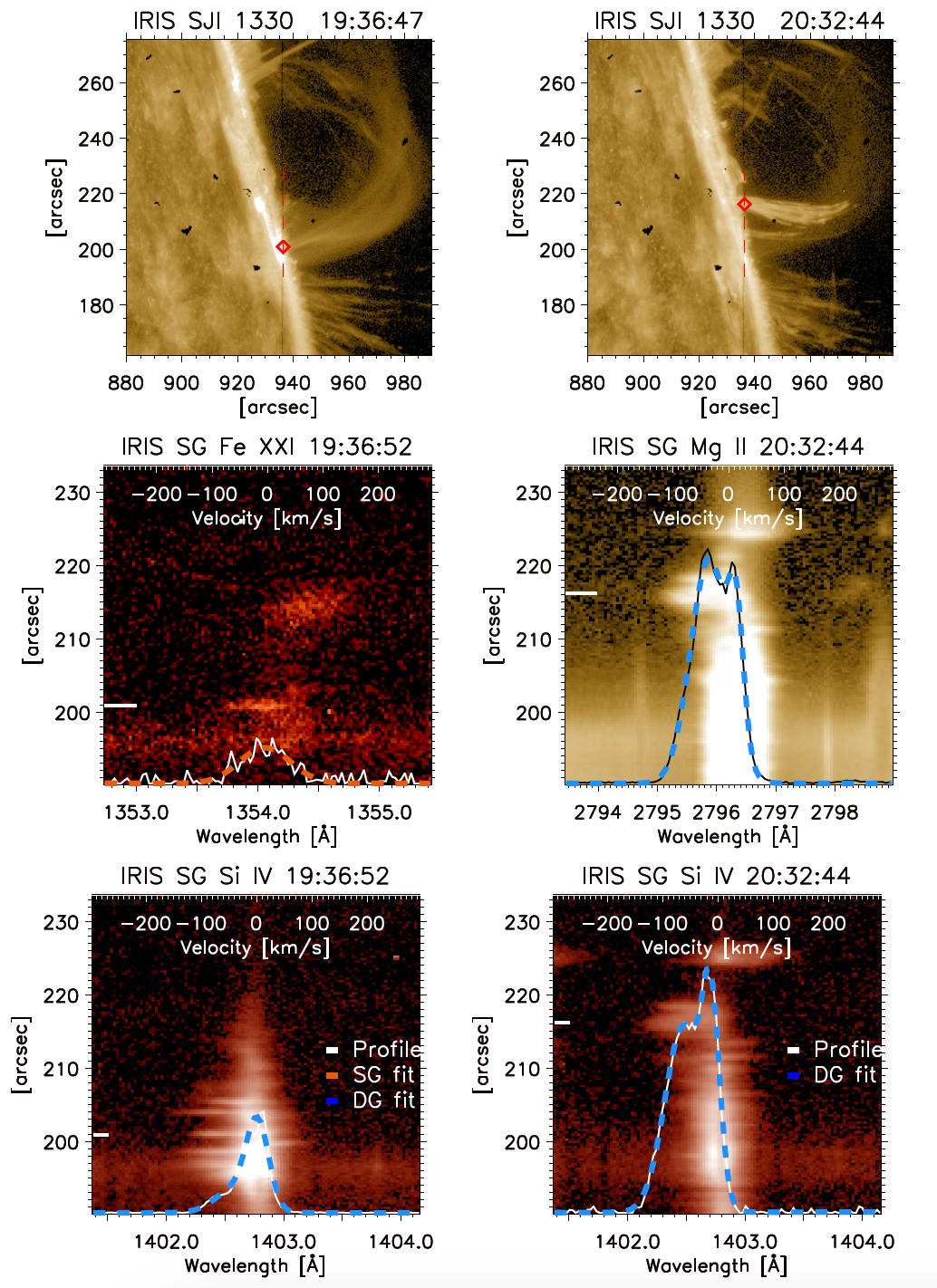}
\caption{IRIS~1330~\AA\ spectra for the CE (including both CE1 and CE2, left panels) and SH1 (right panels) (time is indicated at the top of each panel). The vertical red dashed lines in the IRIS/SJI panels show the portion of the IRIS/SG slit shown in the spectral maps of the \ion{Fe}{XXI} (middle left), \ion{Mg}{II} (middle right) and \ion{Si}{IV} (bottom row) lines. The overlaid line profiles in the maps (white curves) correspond to the positions marked by the red diamonds in the SJI images, corresponding to the location of the slit in the spectral maps indicated by the short horizontal white lines on the left. The spectral profiles are fitted with either a single (`SG', dashed orange curve) or a double (`DG', dashed blue curve) Gaussian.}
\label{spectras}
\end{figure*}

To better constrain the properties of the CE and rain showers, we examine the \ion{Fe}{XXI}, \ion{Mg}{II}~k and \ion{Si}{IV} spectra obtained by IRIS. The left panels in Figure~\ref{spectras} show the spectra in the \ion{Fe}{XXI}~1354.08~\AA\, and \ion{Si}{IV}~1402.77~\AA\, lines of the CE1 and CE2 towards the start (UT19:36:52) along the part of the slit that covers the visible flare loop footpoint (red dashed vertical line in the SJI~1330~\AA\ image of the top panel). In the \ion{Fe}{XXI} spectra, we can see several localised intensity enhancements along the slit, which supports the presence of 10$^{7}$~K plasma in both CEs. The enhanced level of emission across the wavelength range at around 195$\arcsec$ corresponds to the ribbon location. At 200$\arcsec$, we can see a stronger but more localised emission peak (intensity on the order of 10~DN), which corresponds to CE2. The width of the emission is only a few arcsec, which matches very well the CE width seen in the SJI in Figure~\ref{fig:ce2_sh2}. We plot over the spectral maps (wavelength vs slit axis) the spectral profiles at this specific point shown by the red diamond in the SJI figure (see caption for details). We fit with a single Gaussian the \ion{Fe}{XXI} profile, and the best fit from a single or double Gaussian fitting of the \ion{Si}{IV}~1402.77~\AA\, profile. The profile is Doppler shifted towards the blue, with a value of -7$\pm$2.7~km~s$^{-1}$, with a FWHM of 0.494~\AA, which corresponds to 110$\pm$2.7~km~s$^{-1}$. Taking a 10$^{7}$~K peak formation temperature and taking into account the FUV instrumental width, we obtain a non-thermal line width of 25.8$\pm$2.7~km~s$^{-1}$. On the other hand, the \ion{Si}{IV} profile at this point shows a complex and very bright profile ($>$2000~DN) which is best fitted by a double Gaussian. The main component is relatively narrow (FWHM of 48.5$\pm$0.03~km~s$^{-1}$) with almost no Doppler shift (1.69$\pm$0.03~km~s$^{-1}$), and the second, dimmer component is broader (81.4$\pm$0.2~km~s$^{-1}$) with a strong blue shift of -42.1$\pm$0.2~km~s$^{-1}$. Assuming a peak formation temperature of 10$^{4.8}$ for the \ion{Si}{IV} line, we obtain non-thermal line widths of 20~km~s$^{-1}$ and 34~km~s$^{-1}$ for the main and secondary components, respectively, which are similar to those of the \ion{Fe}{XXI} profile. This suggests that the flare heating giving rise to the CE2 is located below the formation height of the \ion{Si}{IV} line (forming in the low TR), and pushes the TR plasma up, heating it to 10$^7$~K. Further up the slit at $\approx215\arcsec$, there is a wider spectral profile that belongs to CE1, about 5$\arcsec$ in width, which has a redshift of 58.7$\pm$3.6~km~s$^{-1}$ with a broad FWHM of 137$\pm$3.6~km~s$^{-1}$, corresponding to a non-thermal line width of 43.7$\pm$3.6~km~s$^{-1}$.

\begin{figure*}
    \includegraphics[width=\textwidth]{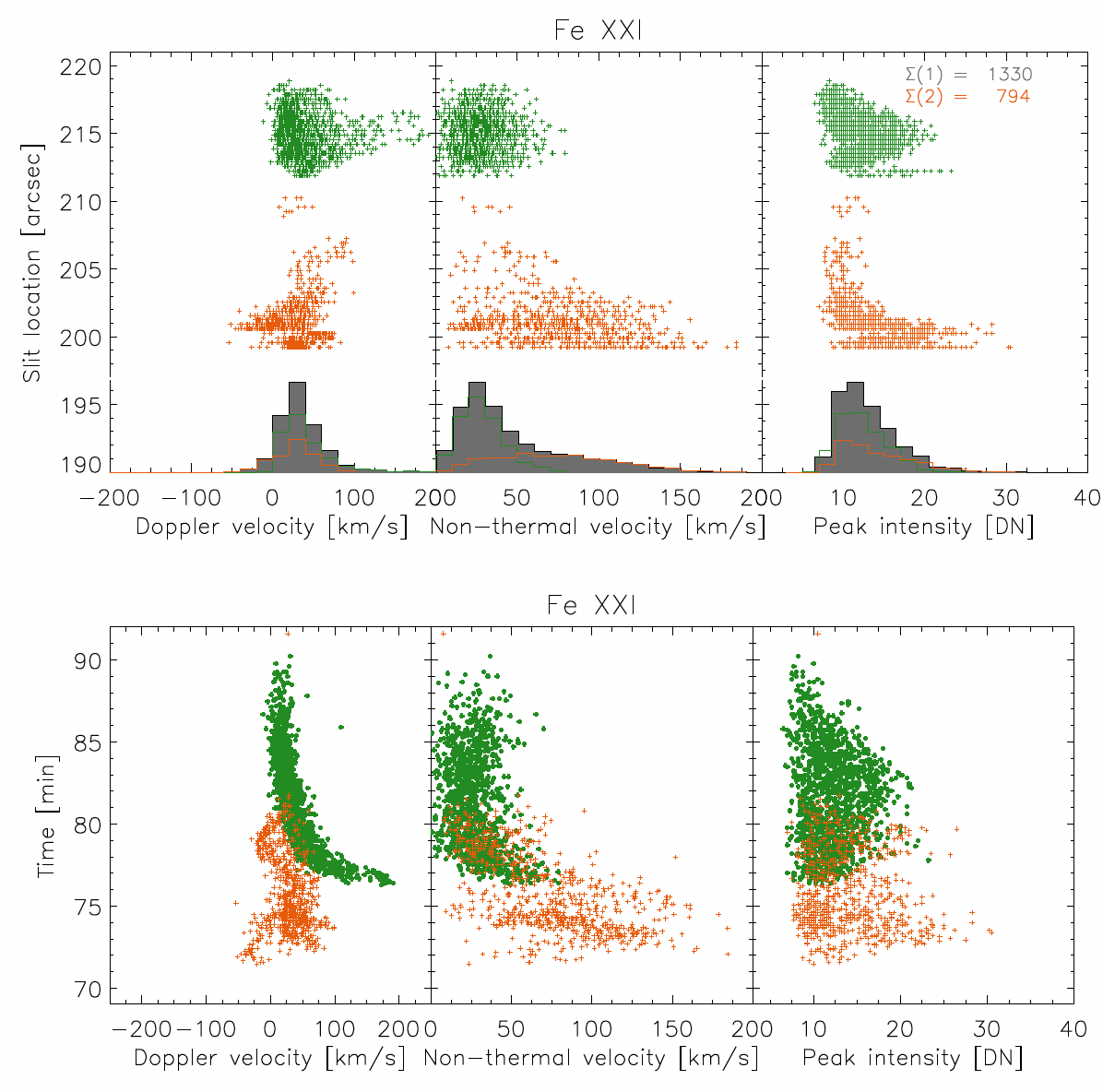}
    \caption{Spatial and temporal distribution of the \ion{Fe}{XXI}~1354.08~\AA\, spectra. We select spectral line profiles satisfying specific requirements (see section \ref{sec:method}) and fit them with a single Gaussian. We restrict the slit range between 199\arcsec and 220\arcsec to focus on CE1 (green points) and CE2 (orange points). From the Gaussian fitting we obtain the Doppler velocity (left panels), non-thermal velocity (middle panels, obtained by subtracting the instrumental line width and the thermal component taking the peak formation temperature for the line), and the peak intensity (right panels). The spatial and temporal distribution are shown in the top and bottom rows, respectively. We also show the histograms of the distribution for each quantity on the top row, with the total in grey, and the green and orange outline histograms for the CE1 and CE2, respectively. The sum of points for each CE is shown on the top right panel with the respective number.}
\label{fig:spec_stats_fe21}
\end{figure*}

In Figure~\ref{fig:spec_stats_fe21}, we show statistics of the spectral fitting results of the \ion{Fe}{XXI} line along the slit (top panel) and over time (bottom panel). We choose to remove the location corresponding to the ribbon to focus on both CEs instead. We also pose a relatively large threshold for the peak intensity of the profile of 5~DN (with the background noise at $\approx 2~$DN). We can identify CE1 between 212\arcsec and 219\arcsec (green points in the figure), whose spectral signatures can be detected for 14~min, and CE2 around $201\arcsec$ up to 205\arcsec or so (red points in the figure), which lasts ~10~min. The widths closely match the overall widths of the CEs seen in SJI, shown in Figures~\ref{fig:ce1_sh1} and \ref{fig:ce2_sh2}. Furthermore, we observe strong variations in speed on the order of $1-2$\arcsec, which matches the widths of the substructures seen in the SJI. The duration for both CEs is significantly shorter than the total duration of the hot emission seen in the SJI, but match the duration over which we see the upward propagating features from the footpoint in the SJI (see x-t diagrams in Figure \ref{fig:crispex}), thereby confirming our previous interpretation that the CEs duration is short. It is nonetheless possible that further CE is happening at these locations, but is too faint to be observed with the SG instrument.  Indeed, we note that both CEs have low \ion{Fe}{XXI} intensities, between 10~DN and 20~DN, and our intensity threshold is set rather high. Regarding the dynamics, CE1 exhibits a strong red shift up to $190~$km~s$^{-1}$, with an average of 40~km~s$^{-1}$ and a standard deviation of 31~km~s$^{-1}$. The non-thermal velocities are relatively small, with an average of 28~km~s$^{-1}$ and a standard deviation of 13~km~s$^{-1}$. On the other hand, CE2 shows a strong, more localised, blue shift at $\approx-60$~km~s$^{-1}$, with an average of 26~km~s$^{-1}$ and a standard deviation of 37~km~s$^{-1}$. The non-thermal velocities of CE2 are significantly larger than those of CE1, with a maximum up to 184~km~s$^{-1}$, an average of 73~km~s$^{-1}$ and a standard deviation of 36~km~s$^{-1}$. The difference in the Doppler shift extrema and average non-thermal velocities between both CEs suggests a difference in the magnetic field topology with respect to the LOS, and also the presence of magnetic shear and unresolved structure at the loop footpoint.

The temporal evolution of CEs shows a rapid decrease in both Doppler and non-thermal velocities for CE1, but only a decrease in non-thermal velocity for CE2, with the fast values lasting less than 2~min for CE1 and $\approx5$~min for CE2. This suggests a transition from explosive to gentle evaporation. The duration of the fast and gentle velocity phases are consistent with previous reports \citep{Fletcher_2013ApJ...771..104F, Sellers_2022ApJ...936...85S}. The decrease in the non-thermal velocity for CE2 occurs without a significant decrease in the Doppler velocity, suggests the presence, and corresponding reduction, of turbulence \citep{Polito_2019ApJ...879L..17P}.

One question is whether the observed POS speeds with SJI are a good measure of the total speeds. The spectra reveals large Doppler and non-thermal velocities but highly localised in space and in time. As mentioned above, this suggests the presence of magnetic shear and unresolved structure at the footpoint. This would suggest that the POS velocities measured higher up are not far from the total velocity values. However, we can calculate an upper bound estimate $v_{tot}=\sqrt{v_{POS}^2+v_{LOS}^2}$ taking as the LOS component $v_{LOS}$ the largest value between the average Doppler velocity and the non-thermal velocity for each CE. This gives total velocities of $122\pm38$~km~s$^{-1}$ and $154\pm33$~km~s$^{-1}$ for CE1 and CE2, respectively.

\begin{figure*}
    \includegraphics[width=\textwidth]{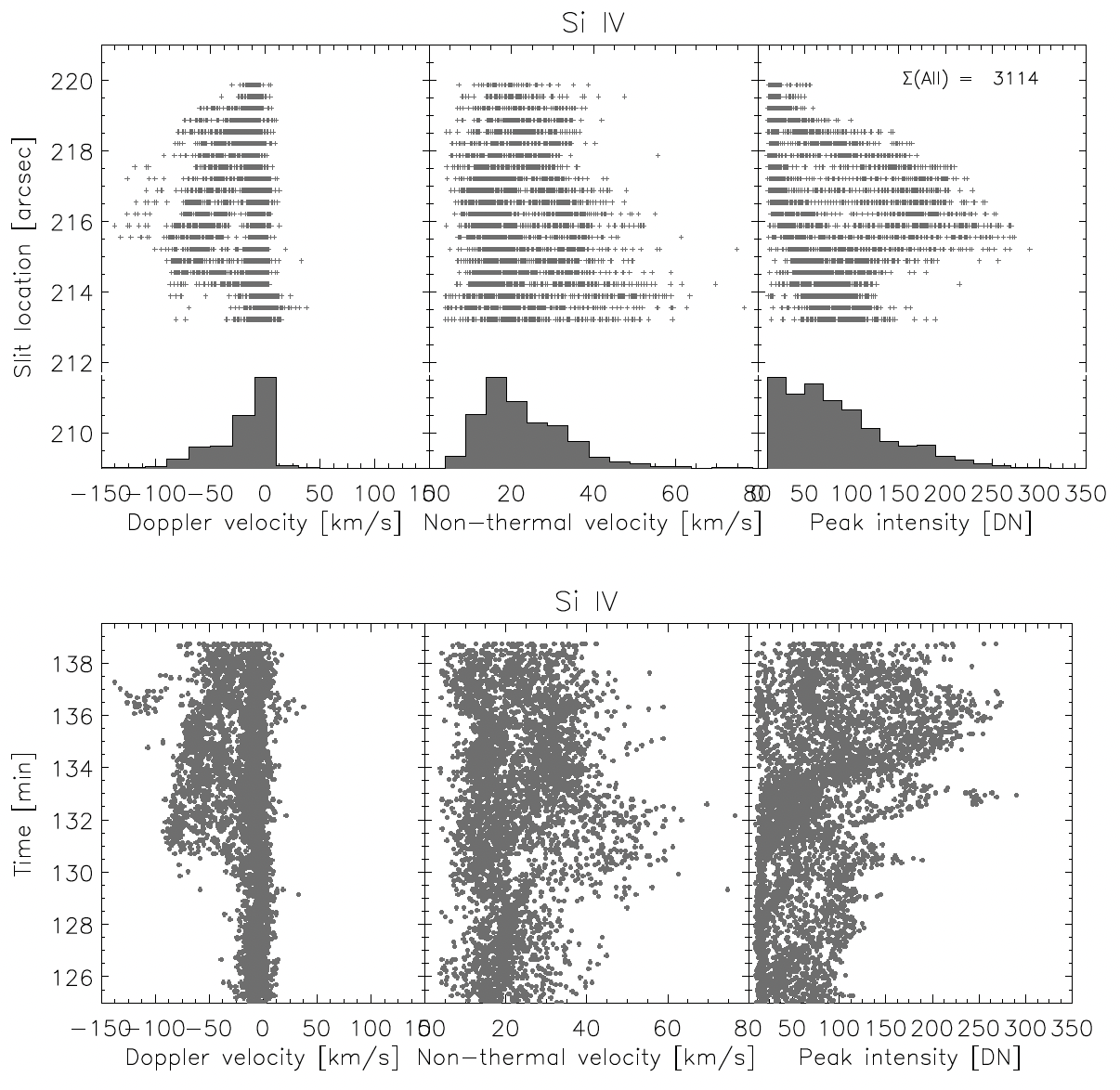}
    \caption{Spatial and temporal distribution of the \ion{Si}{IV}~1402.77~\AA\, spectra. Similar to Figure~\ref{fig:spec_stats_fe21} for the \ion{Si}{IV} line, and we focus only on SH1 by restricting the slit range between 211\arcsec and 220\arcsec (as seen from Figure~\ref{fig:spec_stats_fe21}) and the time range over which SH1 is observed (and avoiding the cosmic ray episode, see Figure~\ref{sp_int_variation}). The best fit is selected from a single or double Gaussian. The sum of all points is shown on the top right panel.}
\label{fig:spec_stats_si4}
\end{figure*}

\subsection{IRIS Spectra of the rain shower}
We now analyse the spectra of the rain showers. Because of the very low location in height of the slit, we can only properly detect the spectroscopic features of SH1. Indeed, we could not find any Doppler signature for SH2, in neither the \ion{Mg}{II} nor \ion{Si}{IV} lines. It is possible that because of the small shift in location introduced by the solar rotation (due to West limb location of event and SH2 occurring 130~min after CE2), the slit at that location has shifted too low in height for the rain to be observed. Since the slit is higher up for SH1, the effect of solar rotation is less pronounced.

In the right panels of Figure~\ref{spectras}, we show the map for the \ion{Mg}{II}~k line (middle panel) and \ion{Si}{IV} line (bottom panel) for a snapshot during SH1. We can see the blue-shifted rain as it falls and we select a point in SH1 (red diamond in the SJI image) and overlay the corresponding profile over the spectral maps. In this case, double Gaussian fits provide the best fits for both \ion{Mg}{II}~k and \ion{Si}{IV} spectra. For \ion{Mg}{II}~k, we obtain a Doppler shift of 13.85$\pm$0.06~km~s$^{-1}$ and -35.05$\pm$0.075~km~s$^{-1}$. The corresponding non-thermal line width (assuming a formation temperature of 10$^4$~K for \ion{Mg}{II}~k and including the instrumental width for the FUV) is 13.20$\pm$0.0075~km~s$^{-1}$ and 33.54$\pm$0.06~km~s$^{-1}$. For \ion{Si}{IV}, both components are blue-shifted, with -12.73$\pm$0.25~km~s$^{-1}$ and -63.13$\pm$0.37~km~s$^{-1}$, and corresponding non-thermal line widths of 15.05$\pm$0.25~km~s$^{-1}$ and 33.20$\pm$0.37~km~s$^{-1}$. We note that the non-thermal line widths are very similar between both lines (as expected for coronal rain), while the Doppler velocities are not. It is likely that the \ion{Mg}{II} profiles suffer from large optical thickness, thereby affecting the measured Doppler velocities. We can therefore assume that the real Doppler velocity of the rain is the most blue-shifted component of \ion{Si}{IV}.

Figure~\ref{fig:spec_stats_si4} shows the spatial (top panel) and temporal (bottom panel) distribution of the spectral line fitting for the \ion{Si}{IV} line. We focus on SH1 (for the reasons explained above) by selecting the slit range determined by the \ion{Fe}{XXI} results, that is, between 211\arcsec and 220\arcsec. We also limit the results to a time range that contains the interval over which the SH1 is seen to cross the slit in the SJI, and also before the occurrence of the cosmic ray episode (see Figure~\ref{sp_int_variation}). The appearance of the rain is clear in the Doppler velocity temporal distribution, where the appearance of a strong blue shifted component is continuously seen from $\approx132$~min. The maximum blue shifted value is -138~km~s$^{-1}$. We obtain an average Doppler velocity for the Doppler shifted component (corresponding to the rain) of -56~km~s$^{-1}$ with a standard deviation of 18~km~s$^{-1}$. The corresponding average non-thermal velocity is 30~km~s$^{-1}$ with a standard deviation of 8~km~s$^{-1}$. We note that the combined dynamics from the average Doppler and non-thermal velocities of the SH1 and CE1 are very similar in magnitude. The fact that the non-thermal velocity of the SH1 is smaller than that of CE1 supports the presence of turbulence and/or unresolved emission at faster speeds for the CE (as observed in the POS component). As done for the CE, we can provide an upper bound for the total average velocity for SH1 taking the average Doppler velocity into account. We obtain $v_{tot}=83\pm16$~km~s$^{-1}$.

\subsection{Mass-flux analysis}
\subsubsection{Energetics and Thermodynamics of Chromospheric Evaporation}

\begin{figure*}[ht!]
\includegraphics[width=\textwidth]{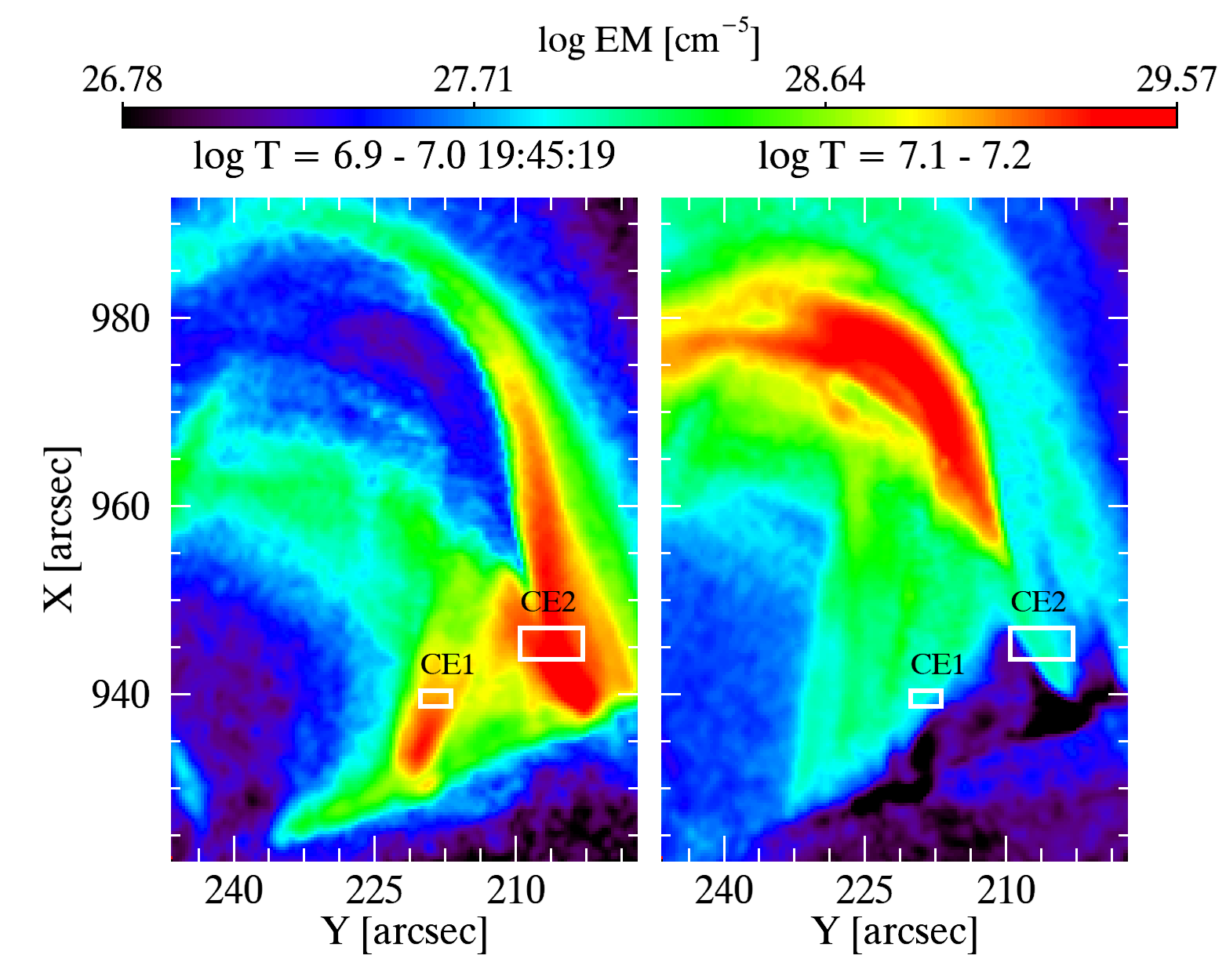} 
\caption{EM maps for the $\log T = 6.9 - 7.0$ (left) and $\log T = 7.1 - 7.2$ (right). The white rectangle areas show the focused region within the CE1 and CE2 for the detailed mass-flux analysis. \label{fig:em_ce_map}}
\end{figure*}

We investigate the mass and energy cycles during the impulsive phase. As we mentioned, we have two main CEs (CE1 and CE2) at different times. We first applied the DEM method based on the ``simreg" technique \citep{Plowman2020ApJ...905...17P} and obtained the emission measure (EM) for each temperature bin, starting from $\log T=5.5$ to 7.2 for each CE. Figure \ref{fig:em_ce_map} displays two EM maps corresponding to the temperature bins of $\log T = 6.9 - 7.0$ (on the left) and $\log T = 7.1 - 7.2$ (on the right). These temperature ranges exclusively encompass instances where CE is observed. We defined sub-regions in the respective temperature bins (see Figure \ref{fig:em_ce_map}) marked by white rectangles for both CE1 and CE2. To define these sub-regions, we intentionally excluded footpoints due to the presence of the solar disc. Our analysis commences by calculating the average EM variations for both temperature bins and for each CE event across distinct time intervals derived from SG analysis. The time interval for the $\log T = 6.9 - 7.0$ and  $\log T = 7.1 - 7.2$ bin extends from 19:35:07 to 19:49:19 ($\Delta t \approx 14$~min). In the case of CE2, these bins encompass the period from 19:35:31 to 19:45:55 ($\Delta t\approx 10$~min).
%over which we see the strongest emission.

We take the temperature of the CE $T_{CE}$ to be the average DEM-weighted temperature ($\overline{T}_{DEM}$), which is calculated using the Equation \ref{average_temp}:
\begin{equation}
\label{average_temp}
%\overline{T}_{DEM}^{CE} = \frac{\int DEM(T)\times TdT}{\int DEM(T)dT} 
\begin{split}
\overline{T}_{DEM} & = \frac{\int (DEM(T) - DEM(T)_{t_0})\times TdT}{\int (DEM(T)-DEM(T)_{t_0}) dT}\\
T_{CE} &=\overline{T}_{DEM},
\end{split}
\end{equation}
where the integral is performed exclusively over the two temperature bins where the CE1 and CE2 are observed. Here, $t_0$ denotes a time immediately prior to the occurrence of the CE. However, $DEM(T)_{t_0}=0$ in this case, since no emission in the hot bins is seen prior to CE. We find $T_{CE}=[8.95\pm3.28]\times 10^{6}$~K and $[8.62\pm3.28]\times 10^{6}$~K for CE1 and CE2, respectively.

Then, the average $EM$ over these temperature bins is given by  $\Delta EM$ from Equation \ref{average_em}. 

\begin{equation}
\label{average_em}
%EM= \int DEM(T)dT.
\Delta EM= \int (DEM(T)-DEM(T)_{t_0}) dT,
\end{equation}
where $DEM(T)_{t_0}=0$ for CE. The average $EM$ is found to be $[1.86\pm1.55]\times 10^{28}$~cm$^{-5}$ for the CE1 and $[6.55\pm3.93]\times10^{28}$~cm$^{-5}$ for the CE2. With the help of these findings, we then calculate the total number density ($n_{CE}$) and mass density ($\rho_{CE}$) of the CE as follows: 

\begin{equation}
\label{eq:ce_density}
\begin{split}
    n_{DEM} & = \sqrt{\frac{1.2\times EM}{w_{overall}}} \\
    n_{CE} &=1.92n_{DEM}\\
    \rho_{CE}& = 1.17 n_{DEM} m_{p},
\end{split}
\end{equation}
where $n_{DEM}$ denotes the electron number density and where we have assumed a 10\% Helium abundance, a fully ionised plasma and $m_{p}$ is the proton mass. Although we observe substructure within the CE, the internal structure (for instance, if composed of individual strands), is unclear. Hence, we take $w_{overall}$ as the observed overall width of the CE. To estimate the density of the CEs we focus on the observed strands. We find a strand mass density of $\rho_{CE}=[1.44\pm0.02]\times10^{-14}$~g~cm$^{-3}$ for the CE1 and $[2.33\pm0.03]\times10^{-14}$~g~cm$^{-3}$ for the CE2. Then, we calculate the mass rate ($R_{tot}$), the mass flux ($F_{tot}$), and the total mass $M_{to}$ for each CE according to:

\begin{equation}
\label{eq:ce_mass_flux_rate}
\begin{split}
    R_{tot} & = \rho A v_{tot} \\
    F_{tot} & = \rho v_{tot} \\
    M_{tot} & = R_{tot} \times \Delta t
\end{split}
\end{equation}

Here, the CE total velocity is given by $v_{tot}$, the cross-sectional area $A$ of the CE is $A_{CE}=\pi r^2$, assuming a cylinder, and the mass density $\rho=\rho_{CE}$ for the CE. We use the upper bound for the total velocity found by combining the SJI and SG measurements, $[122\pm38]$~km~s$^{-1}$ for CE1 and $[154\pm33]$~km~s$^{-1}$ for CE2. The mass rate ($R_{tot}$) and the mass flux ($F_{tot}$) are found to be $[2.31\pm0.01]\times10^{10}$~g~s$^{-1}$ and $[1.76\pm0.01]\times10^{-7}$~g~cm$^{-2}$~s$^{-1}$, respectively, for the CE1. These are $[8.61\pm0.03]\times10^{10}$~g~s$^{-1}$ and $[3.59\pm0.08]\times10^{-7}$~g~cm$^{-2}$~s$^{-1}$, respectively, for the CE2. Given the $\Delta t=14$~min and 10~min duration of CE1 and CE2, the total mass going into the loop is $M_{tot} =[1.94\pm0.01]\times10^{13}$~g and $[5.16\pm0.02]\times10^{13}$~g, respectively. The kinetic energy ($E_{k}$) is given by:

\begin{equation}
\label{eq:kinetic_energy}
\begin{split}
    E_{k} & = \frac{1}{2}\rho v_{tot}^{2} V \\
    V & =Av_{tot}\Delta t.
\end{split}
\end{equation}

For the CE we take the mass density $\rho=\rho_{CE}$, volume $V=V_{CE}$ and cross-sectional area $A=A_{CE}$. We find that the kinetic energies are $[1.44\pm0.02]\times10^{27}$~erg for the CE1 and $[6.12\pm0.05]\times10^{27}$~erg for the CE2. Finally, the thermal energy ($E_{th}$) is given by:

\begin{equation}
\label{eq:thermal_energy}
    E_{th} = \frac{3}{2}nk_B T V,  \\ 
\end{equation}

where for the CE we take the total number density $n=n_{CE}$, temperature $T=T_{CE}$ and volume $V=V_{CE}$, with $k_{B}$ the Boltzmann constant. We find thermal energies ($E_{th}$) of $[3.54\pm0.01]\times10^{28}$~erg and $[9.06\pm0.03]\times10^{28}$~erg for the CE1 and CE2 events, respectively. Then, the total energy ($E_{tot}=E_{k}+E_{th}$) is $[3.68\pm0.02]\times10^{28}$~erg for the CE1 and $[9.67\pm0.03]\times10^{28}$~erg for the CE2. A summary of all these calculations is given in Table \ref{table:mass_energy_calculations}.

In addition to these, we also estimate the total thermal energy of the flare via Equation \ref{eq:thermal_energy}, with $T=\overline{T}_{DEM}=9\times10^6$~K considering the emission measure over the respective duration of the hot emission accompanying each CE event. Also, the volume in this case is that of the flare loop where we see the hot emission. For this measurement, we find the total number density ($n_{DEM}$) to be $1.91\times10^{10}$~cm$^{-3}$. By approximating the loop to a box, we estimate the volume using $V=w^2L$, with $L$ (62.5~Mm) the length of the flare loop, and $w$ (18.3~Mm) the observed width, which we assume roughly equal to the depth. We find that the total thermal energy of the flare is roughly $7.52\times10^{29}$~erg.

\begin{deluxetable*}{ccccc}[ht!]
\tablenum{5}
\tablecaption{Measurements of the CE and SH events. For CE1, CE2, SH1, and SH2 we show the following quantities: duration ($\Delta t$), POS, LOS, and total velocity ($v_{POS}, v_{LOS}, v_{tot}$), strand and overall width ($w_{strand}$ and $w_{overall}$), average emission measurement ($<EM>$), mass rate ($R_{strand}$), mass flux ($F_{strand}$), total mass rate ($R_{tot}$), total mass flux ($F_{tot}$), strand and total mass ($M_{strand}$ and $M_{tot}$), kinetic energy ($E_{k}$), thermal energy ($E_{th}$) and total energy ($E_{tot}=E_{k}+E_{th}$). We also show temperature ($T$), total number density ($n$), and mass density ($\rho$), with the sub-index $CE$ or $rain$ depending on the case. The sub-indexes $cool$, $warm$ denote the cool and warm components of the rain. $N_{strand}$ corresponds to the minimum number of rain strands. \label{table:mass_energy_calculations}}
\tablewidth{0pt}
\tablehead{\colhead{} & \colhead{CE1} & \colhead{CE2} & \colhead{SH1} & \colhead{SH2} }
\startdata
%$\Delta$t [min] & 33$_{6.9-7.0}$ \& 18$_{7.0-7.1}$ & 44$_{6.9-7.0}$ \& 26$_{7.0-7.1}$ & 9 & 28 \\ 
$\Delta t$ [min] & 14  & 10 & 9 & 28 \\ 
$v_{POS}$ [km~s$^{-1}$] & [115$\pm$39] & [136$\pm$32]  & [61$\pm$14] & [76$\pm$12] \\ 
$v_{LOS}$ [km~s$^{-1}$] & [40$\pm$31] & [73$\pm$36]  & [56$\pm$18] & - \\ 
$v_{tot}$ [km~s$^{-1}$] & [122$\pm$38] & [154$\pm$33]  & [83$\pm$16] & [76$\pm$12] \\ 
$w_{strand}$ [km] & [1218$\pm$176] & [1535$\pm$37]  & [642$\pm$56] & [720$\pm$145] \\ 
$w_{overall}$ [km] & [4088$\pm$107] & [5526$\pm$155]  & [3872$\pm$104] & [4427$\pm$157] \\ 
$<EM>$ [cm$^{-5}$] & $[1.86\pm1.55]\times10^{28}$ & $[6.55\pm3.93]\times10^{28}$ & $[1.52\pm0.67]\times10^{27}$ & $[1.29\pm0.38]\times10^{27}$ \\ 
%$<EM>$ [cm$^{-5}$] & [3.66$\pm$3.10]$\times$10$^{28}$ & [9.36$\pm$5.21]$\times$10$^{28}$ & [1.52$\pm$0.67]$\times$10$^{27}$ & [1.29$\pm$0.38]$\times$10$^{27}$ \\ 
%$T_{DEM}$ [K] & [9.47$\pm$3.28]$\times$10$^{6}$ &  [8.59$\pm$3.28]$\times$10$^{6}$ & [4.09$\pm$0.59]$\times$10$^{6}$ & [3.03$\pm$1.01]$\times$10$^{6}$ \\ 
%$T_{DEM}$ [K] & $[8.95\pm3.28]\times10^{6}$ &  $[8.62\pm3.28]\times10^{6}$ & $[4.09\pm0.59]\times10^{6}$ & $[3.03\pm1.01]\times10^{6}$ \\ 
%$n_{DEM}$ [cm$^{-3}$] & $[1.67\pm0.67]\times10^{10}$ &  $[2.69\pm0.81]\times10^{10}$ & $[2.31\pm0.27]\times10^{10}$ & $[2.01\pm0.29]\times10^{10}$ \\
%$\rho_{DEM}$ [g~cm$^{-3}$] & $[3.27\pm1.31]\times10^{-14}$ & $[5.27\pm1.58]\times10^{-14}$ & $[2.36\pm0.53]\times10^{-14}$ & $[2.05\pm0.29]\times10^{-14}$  \\ 
$T_{CE}$ [K] & $[8.95\pm3.28]\times10^{6}$ &  $[8.62\pm3.28]\times10^{6}$ & - & - \\ 
$n_{CE}$ [cm$^{-3}$] & $[1.42\pm0.02]\times10^{10}$ &  $[2.29\pm0.03]\times10^{10}$ & - & -\\
$\rho_{CE}$ [g~cm$^{-3}$] & $[1.44\pm0.02]\times10^{-14}$ & $[2.33\pm0.03]\times10^{-14}$ & - & -  \\ 
$T_{warm}$ [K] & - &  - & $[4.09\pm0.59]\times10^{6}$ & $[3.03\pm1.01]\times10^{6}$ \\ 
$n_{warm}$ [cm$^{-3}$] & - &  - & $[1.02\pm0.04]\times10^{10}$ & $[8.90\pm0.89]\times10^{9}$ \\
$\rho_{warm}$ [g~cm$^{-3}$] & - & - & $[1.04\pm0.04]\times10^{-14}$ & $[9.05\pm0.91]\times10^{-15}$  \\ 
$n_{cool}$ [cm$^{-3}$] & - & - & $[4.18\pm0.03]\times10^{12}$ & $[2.69\pm0.09]\times10^{12}$ \\ 
$\rho_{cool}$ [g~cm$^{-3}$] & - & - & $[4.26\pm0.03]\times10^{-12}$ & $[2.74\pm0.09]\times10^{-12}$ \\ 
$T_{cool}$ [K] & - &  - & $10^{4}$ & $10^{4}$ \\ 
$n_{rain}$ [cm$^{-3}$] & - & - & $[6.98\pm0.12]\times10^{11}$ & $[4.35\pm0.91]\times10^{11}$ \\
$\rho_{rain}$ [g~cm$^{-3}$] & - & - & $[7.11\pm0.18]\times10^{-13}$ & $[4.43\pm0.92]\times10^{-13}$ \\
$T_{rain}$ [K] & - &  - & $[5.9\pm1.5]\times10^4$ & $[6.2\pm0.99]\times10^4$ \\ 
$N_{strand}$ & - & - & 6 & 6 \\ 
$R_{strand}$ [g~s$^{-1}$] & - & - & $[1.91\pm0.02]\times10^{10}$ & $[1.37\pm0.12]\times10^{10}$ \\
$F_{strand}$ [g~cm$^{-2}$ s$^{-1}$] & - & - & $[5.90\pm0.07]\times10^{-6}$ & $[3.37\pm0.11]\times10^{-6}$ \\
$M_{strand}$ [g] & - & - & $[1.03\pm0.01]\times10^{13}$ & $[2.30\pm0.21]\times10^{13}$ \\ 
$R_{tot}$ [g~s$^{-1}$] & $[2.31\pm0.01]\times10^{10}$ & $[8.61\pm0.03]\times10^{10}$ & $[1.15\pm0.01]\times10^{11}$ & $[8.23\pm0.74]\times10^{10}$ \\
$F_{tot}$ [g~cm$^{-2}$~s$^{-1}$] & $[1.76\pm0.01]\times10^{-7}$ & $[3.59\pm0.08]\times10^{-7}$ & $[3.54\pm0.04]\times10^{-5}$ & $[2.02\pm0.07]\times10^{-5}$ \\
$M_{tot}$ [g] & $[1.94\pm0.01]\times10^{13}$ & $[5.16\pm0.02]\times10^{13}$ & $[6.19\pm0.07]\times10^{13}$ & $[1.38\pm0.12]\times10^{14}$ \\ 
$E_{k}$ [erg] & $[1.44\pm0.02]\times10^{27}$ & $[6.12\pm0.05]\times10^{27}$ & $[1.29\pm0.01]\times10^{28}$ & $[2.52\pm0.03]\times10^{28}$ \\ 
$E_{th}$ [erg] & $[3.54\pm0.01]\times10^{28}$ & $[9.06\pm0.03]\times10^{28}$ & $[4.57\pm0.01]\times10^{27}$ & $[1.09\pm0.06]\times10^{28}$ \\
$E_{tot}$ [erg] & $[3.68\pm0.02]\times10^{28}$ & $[9.67\pm0.03]\times10^{28}$ & $[1.75\pm0.01]\times10^{28}$ & $[3.61\pm0.03]\times10^{28}$ \\
\enddata
\end{deluxetable*}

\subsubsection{Energetics and Thermodynamics of Flare-driven Coronal Rain}

\begin{figure*}[ht!]
\includegraphics[width=\textwidth]{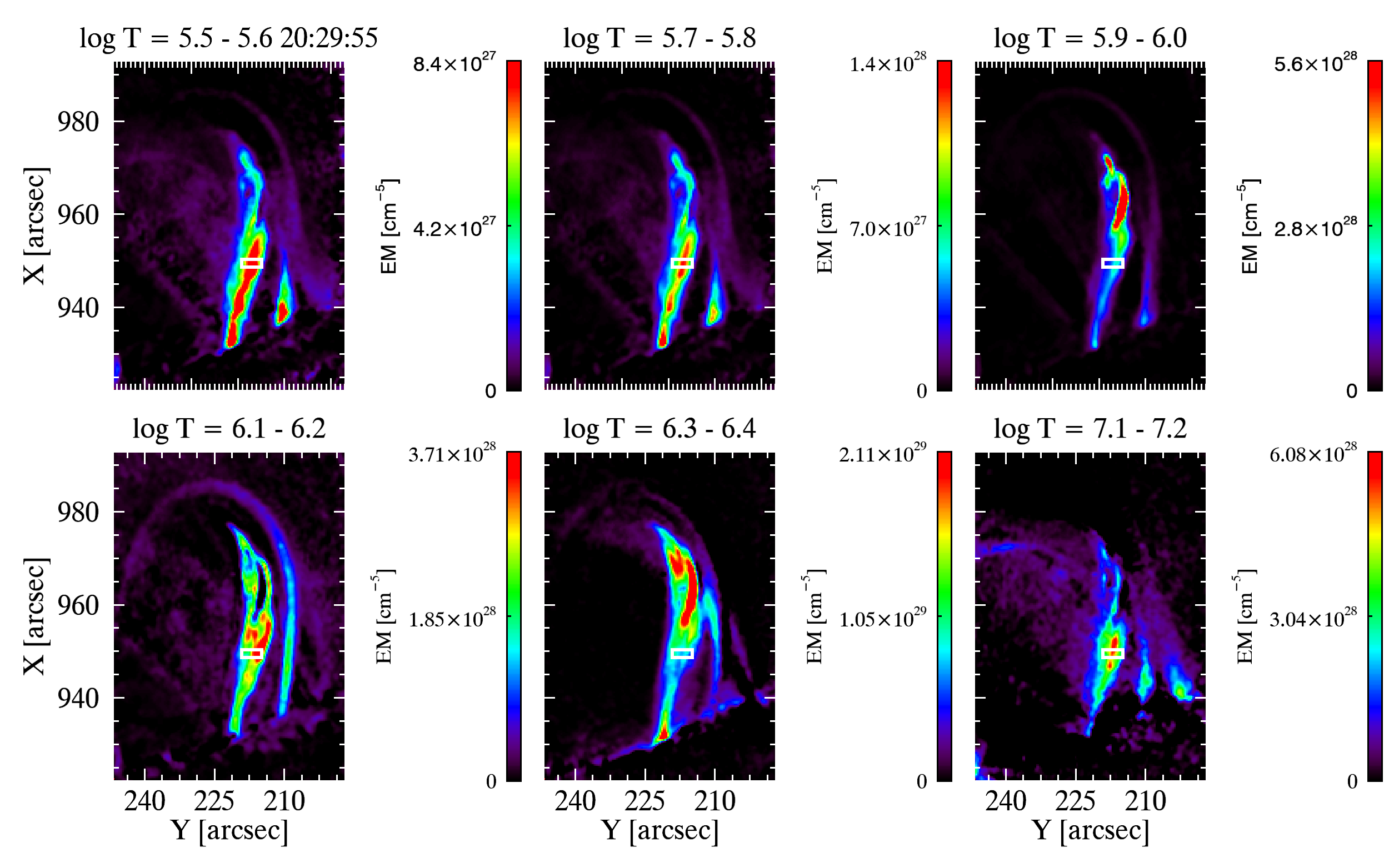} 
\caption{$EM$ maps for some temperature bins indicated in the title. The white rectangle areas show the focused region within the SH1 for the detailed mass-flux analysis. \label{em_maps_sh1}}
\end{figure*}

\begin{figure*}[ht!]
\includegraphics[width=\textwidth]{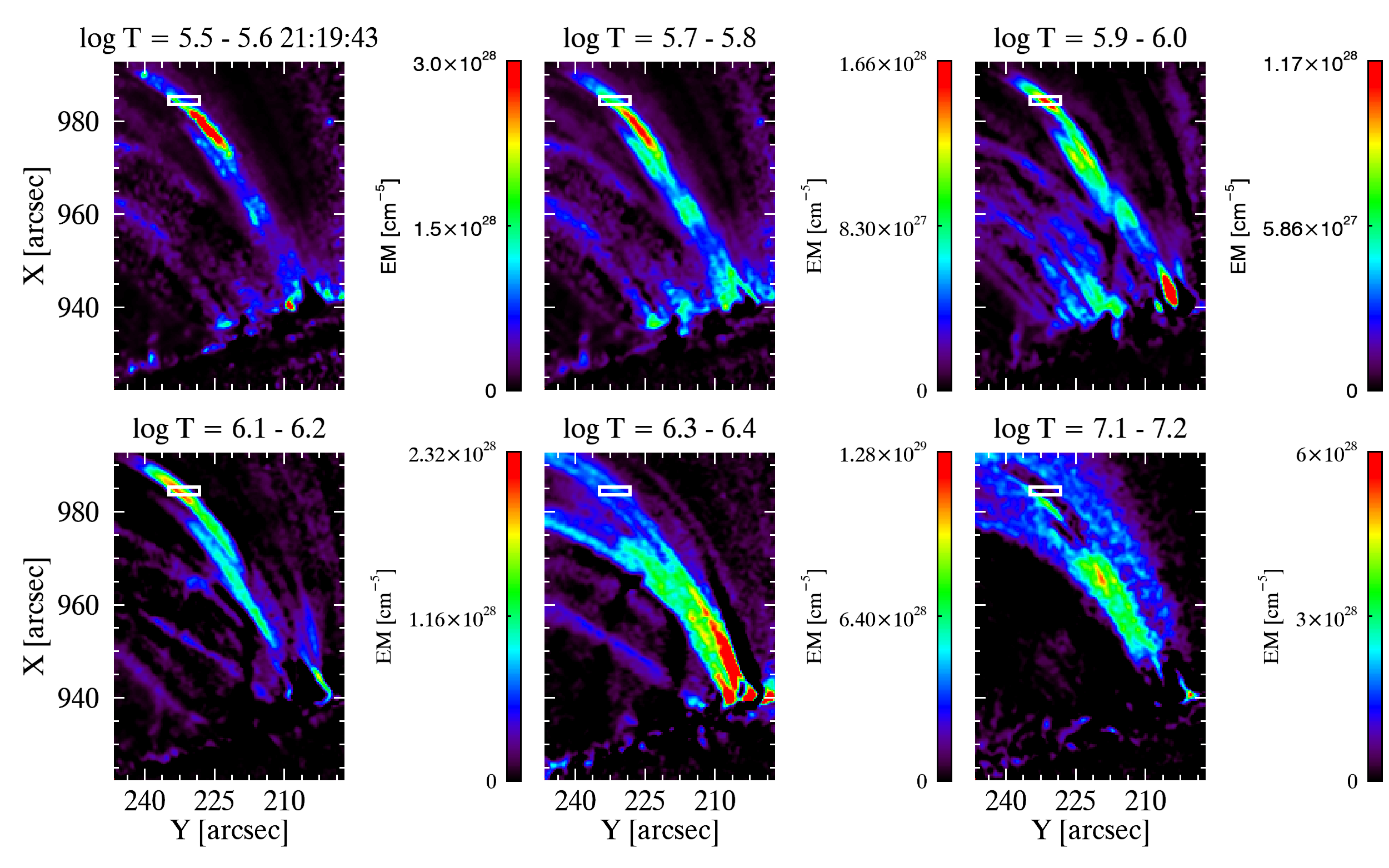} 
\caption{Same as Figure~\ref{em_maps_sh1} but for SH2.} \label{em_maps_sh2}
\end{figure*}

Similar to CE, we also examine the mass and energy flow during two shower events observed at different times, which we previously named SH1 and SH2. SH1 event starts at 20:26:31 UT and ends at 20:35:31 UT, while the SH2 is between 21:18:31 and 21:46:43 UT. We show emission measure (EM) maps of these two shower events for each temperature bin in Figures \ref{em_maps_sh1} and \ref{em_maps_sh2}. Contrary to the CE case, the rain, being multi-thermal, its $EM$ spans many temperature bins. We define as `warm rain' the plasma emitting in the temperature bins above $\log T=5.5$ (where the material is certainly optically thin), and `cool rain' as the material below these temperatures (including the chromospheric range). A further difference with the CE is that prior to the rain occurrence, the $EM$ is significant across the temperature range. Hence, in these maps, we remove the $EM$ value immediately prior to the occurrence of the showers, at $t_0 =$20:26:19 for the SH1 and $t_0 =$21:18:19 for the SH2. This is to remove the strongly emitting background and capture as close as possible the $EM$ from the showers themselves. 

We take the temperature of the warm rain ($T_{warm}$) as the average DEM-weighted temperature:
\begin{equation}\label{eq:te_warm}
    T_{warm}=\overline{T}_{DEM},
\end{equation}
calculated using Equation~\ref{average_temp}. 
%\begin{equation}
%\label{average_temp_2}
%\overline{T}_{DEM, rain} = \frac{\int (DEM(T) - DEM(T)_{t_0})\times TdT}{\int (DEM(T)-DEM(T)_{t_0}) dT} 
%\end{equation}

We find  $T_{warm}=[4.09\pm0.59]\times10^{6}$~K and $[3.03\pm1.01]\times10^{6}$~K for SH1 and SH2, respectively. The average $EM$ is calculated using Equation \ref{average_em} and is found to be $[1.52\pm0.67]\times10^{27}$~cm$^{-5}$ for SH1 and $[1.29\pm0.38]\times10^{27}$~cm$^{-5}$ for SH2.
%\begin{equation}
%\label{average_em_2}
%\Delta EM= \int (DEM(T)-DEM(T)_{t_0}) dT
%\end{equation}
The number density and mass density of the warm plasma is calculated using a similar equation as for the CE:
\begin{equation}
\begin{split}
    n_{DEM} & = \sqrt{\frac{1.2\times EM}{w_{strand}}} \\
    n_{warm} & = 1.92n_{DEM} \\
    \rho_{warm} &=1.17n_{DEM}m_p,    
\end{split}
\end{equation}

where instead of the overall width, we have now taken the strand width $w_{strand}$, since we are able to resolve the rain strands. As previously mentioned, the thinnest rain strand width is found to be $[642\pm56]$~km for the SH1 and $[720\pm145]$~km for the SH2, which we take as representatives of the basic (fundamental) structures that form the rain showers. 

In order to obtain the total number density ($n_{rain}$) and mass density ($\rho_{rain}$) of the cool part of the rain, we assume that the cool and warm parts of the rain are in pressure balance $(p_{rain}=p_{cool}=p_{warm})$, and also that there is only a small magnetic field variation across the rain \citep{Antolin2022A}. The temperature of the cool rain ($T_{cool}$) is taken as the representative temperature value of the SJI~2796 line ($10^4$~K).Then, we have:

\begin{equation}
\label{eq:cool_density}    
\begin{split}
    T_{cool} &=10^4~\mbox{K}\\
    \rho_{cool} &= \frac{\rho_{warm} T_{warm}}{T_{cool}}\\
    n_{cool} &= \frac{\rho_{cool}}{0.61m_{p}}    
\end{split}
\end{equation}

%\begin{equation}
%\label{eq:rain_density}
%\begin{split}
 %   n_{rain} = \frac{\rho_{DEM} T_{DEM}}{1.17m_{p} T_{rain}} \\
  %  \rho_{rain} = n_{rain}\times1.17m_{p}
   % %\rho_{rain} = \frac{\rho_{DEM} T_{DEM}}{T_{rain}}
%\end{split}
%\end{equation}

Similarly to Equation~\ref{eq:ce_density}, we have assumed a 10\% helium abundance in Equation~\ref{eq:cool_density}. We estimate the volume occupied by the cool part of the rain ($V_{cool}$) with Equation~\ref{eq:rain_volume}, based on the thinnest rain strand observed and the minimum number of such strands ($N_{strand}$) needed to cover the SH width in the POS, which we find to be roughly $N_{strand}=6$ for both SH1 and SH2. 

\begin{equation}
\label{eq:rain_volume}    
\begin{split}    
    V_{cool} &=N_{strand}\times \pi \left(\frac{w_{strand}}{2}\right)^2 v_{tot}\Delta t \\
    V_{warm} &= V_{shower}-V_{cool} \\ 
\end{split}
\end{equation}
%Vcool=Shower1=8.7052869e+25 \pm0.14
%Vcool,shower1,rht = 1.0237338e+25
%Vcool=Shower2=3.1190976e+26 \pm3.09

Respectively for SH1 and SH2, we find volumes for the cool part of the rain ($V_{cool}$) of $[8.70\pm0.14]\times10^{25}$~cm$^{3}$ and $[3.12\pm1.96]\times10^{26}$~cm$^{3}$, and shower volumes ($V_{shower}$) of $[5.28\pm2.10]\times10^{26}$~cm$^{3}$ and $[1.96\pm0.32]\times10^{27}$~cm$^{3}$. In an alternative approach, the volume $V_{cool}$ in Equation~\ref{eq:rain_volume} can be calculated using the RHT results as: 

\begin{equation}
    V_{cool}^{RHT} = N_tA_{pixel}w_{strand},
\end{equation}

where $N_t$ is the total number of rain pixels, $A_{pixel}$ is the area of a pixel and where we have assumed that the emitting plasma along the LOS has a thickness equal to the strand width $w_{strand}$. This approach results in a volume $9.21\times10^{25}$~cm$^{3}$, which is roughly the same as that found with Equation~\ref{eq:rain_volume}.

Since we know the volume occupied by the shower ($V_{shower}$), we can then find the volume of the warm rain ($V_{warm}$), the overall total number density ($n_{rain}$), mass density ($\rho_{rain}$) and temperature ($T_{rain}$) of the rain:

\begin{equation}
\label{eq:rain_density}    
\begin{split}    
    n_{rain} &= \frac{(n_{cool}V_{cool}+n_{warm}V_{warm})}{V_{shower}}\\
    \rho_{rain} &= 0.61 n_{rain}m_p \\
    T_{rain} &=\frac{p_{rain}}{k_B n_{rain}}.
\end{split}
\end{equation}

Respectively for SH1 and SH2, we find coronal rain number densities ($n_{rain}$) of $[6.98\pm0.12]\times10^{11}$~cm$^{-3}$ and $[4.35\pm0.91]\times10^{11}$~cm$^{-3}$, and mass densities ($\rho_{rain}$) of $[7.11\pm0.18]\times10^{-13}$~g~cm$^{-3}$ and $[4.43\pm0.92]\times10^{-13}$~g~cm$^{-3}$. Finally, we find the rain temperatures ($T_{rain}$) of $5.90\pm1.50\times10^4$~K and $6.20\pm0.99\times10^4$~K.
 
Taking $v_{tot}$ as the total rain velocity previously estimated, Equation~\ref{eq:ce_mass_flux_rate} gives the mass rate ($R_{strand}$) and the mass flux ($F_{strand}$) as $[1.91\pm0.02]\times10^{10}$~g~s$^{-1}$ and $[5.90\pm0.07]\times10^{-6}$~g~cm$^{-2}$~s$^{-1}$, respectively, per strand for the SH1. For the SH2, these are found to be $[1.37\pm0.12]\times10^{10}$~g~s$^{-1}$ and $[3.37\pm0.11]\times10^{-6}$~g~cm$^{-2}$~s$^{-1}$ per strand, respectively. Given the duration ($\Delta t$) of 9~min for the SH1 and 28~min for the SH2 over all temperature bins from $\log T=5.5$ to $\log T=7.2$, the mass per strand (i.e. $M_{strand}$ = $R_{strand} \times \Delta t$) is found to be $[1.03\pm0.01]\times10^{13}$~g for the SH1 and $[2.30\pm0.21]\times10^{13}$~g for the SH2. To calculate the respective quantities for the entire SH, we multiply by the minimum number of rain strands ($N_{strand}$). Then, we find that the total mass rate ($R_{tot}$) is $[1.15\pm0.01]\times10^{11}$~g~s$^{-1}$ for the SH1 and $[8.23\pm0.74]\times10^{10}$~g~s$^{-1}$ for the SH2. Finally, we find that a total mass $M_{tot}=[6.19\pm0.07]\times10^{13}$~g and $[1.38\pm0.12]\times10^{14}$~g goes down in SH1 and SH2, respectively. 

The kinetic energy ($Ek$) of the SH events is given by Equation~\ref{eq:kinetic_energy}, with $\rho=\rho_{rain}$ and $V=V_{shower}$. We find $E_k=[1.29\pm0.01]\times10^{28}$~erg and  $[2.52\pm0.03]\times10^{28}$~erg, respectively, for SH1 and SH2.

We also calculated the thermal energy of the showers using the Equation~\ref{eq:thermal_energy}, with $n=n_{cool}$, $T=T_{cool}$ and $V=V_{shower}$\footnote{Note that because the pressure of the cool and warm rain material is the same, it is equivalent to take $n=n_{warm}$ and $T=T_{warm}$ in Equation~\ref{eq:thermal_energy}.}, resulting in $[4.57\pm0.01]\times10^{27}$~erg for the SH1 and $[1.09\pm0.06]\times10^{28}$~erg for the SH2. Consequently, the total energies for the SH1 and SH2 are $[1.75\pm0.01]\times10^{28}$~erg and $[3.61\pm0.03]\times10^{28}$~erg, respectively. A summary of all these calculations is given in Table \ref{table:mass_energy_calculations}.

%\begin{equation}
%\label{eq:thermal_energy_sh}
%    E_{th}  = \frac{3}{2}n_{cool}k_{B}T_{cool}V_{shower}
%\end{equation}

Since we obtained density parameters for the flare-driven, we would also like to compare them with their quiescent counterpart. For this, we selected a few clumps from the pre-flare phase and obtained average $EM$ and density measurements using the same equations above. The average $EM$ for the quiescent rain is found to be $[1.60\pm0.45]\times10^{26}$~cm$^{-5}$, which is an order of magnitude lower than that of the flare-driven coronal rain. The average mass density of the warm quiescent rain ($\rho_{warm}$) is $[3.38\pm0.13]\times10^{-15}$~g~cm$^{-3}$. Then, the average total number ($n_{rain}$) and mass densities ($\rho_{rain}$) of quiescent coronal rain are $[9.01\pm0.70]\times10^{10}$~cm$^{-3}$ and $[9.18\pm0.71]\times10^{-14}$~g~cm$^{-3}$, respectively, indicating that the flare-driven rain is $\approx 6$ times denser than the quiescent coronal rain.

\section{Discussion and Conclusions \label{sec:sum}}

In this paper, we present a detailed investigation of the mass and energy flow during a C2.1-class flare, whose start and end are marked by the chromospheric evaporation and coronal rain, respectively. We determine their morphology, mass and energy, and compare the values between the observed inflow into the flaring loop and outflow from the loop. We further present a statistical comparison between the quiescent (pre-flare) and flare-driven coronal rain (gradual). 

We have checked the rain quantity and rain intensity in the pre-flare and gradual phases. We found that the rain quantity increases by a factor of 3.09, 6.15, and 3.89 times from pre-flare to gradual phase for the SJI~2796~\AA, SJI~1330~\AA, and Cool~AIA~304~\AA, respectively. Similarly, the rain intensity also increases by a factor of 3.14, 2.95, and 4.35 for the SJI~2796~\AA, SJI~1330~\AA, and Cool~AIA~304~\AA, respectively, from pre-flare to gradual phases. On the other hand, the average mass density of coronal rain increases by a factor of 6.28 from pre-flare to the gradual phases. This strong increase in density is not reflected by the strong increase in intensity, which may reflect the transition to optically thick radiation of the flare-driven rain. The net increase in rain quantity in terms of area from pre-flare to gradual phase in the SJI~2796~\AA, 1330~\AA, and Cool~AIA~304~\AA\ is comparable, with the range of $[5.26, 5.92]\times10^{12}$~km$^2$ across all channels. This similarity suggests that the plasma is continuously cooling all the way down to the SJI~2796~\AA\ temperatures. The similar trend across the channels also suggests a very fast cooling in the $10^4-10^5$~K range, with basically no appreciable time difference in the intensity peaks. Furthermore, this also suggests that very little material stays at 10$^5$~K temperatures. However, if everything cools to the SJI~2796~\AA\ temperature formation and nothing remains at higher temperatures, a decrease in the amount of rain in the SJI~1330~\AA\ and Cool~AIA~304~\AA\ channels would be observed. This suggests that there is a mechanism, such as the CCTR \citep{antolin2020}, that guarantees strong co-located emission in time and space at hotter temperatures even if there is continuous catastrophic cooling all the way down to the SJI~2796~\AA.

%This is expected due to the flare-driven rain has a higher intensity in the low TR \citep{Foukal1978ApJ...223.1046F, Lacatus2017ApJ...842...15L}. 

We have also investigated the morphology of these two types of coronal rain. We found that the width of rain clumps varies from 0.2~Mm to 2~Mm with roughly 0.7~Mm (for the SJI channels) and 1.2~Mm (for the AIA channel) on average in both pre-flare and gradual phases. These results are in agreement with the current literature in terms of quiescent coronal rain studies \citep[such as][]{antolin2015, Li2022ApJ...926..216L, Sahin2023ApJ...950..171S}. In this study, we have also included flare-driven coronal rain morphology and on average, there is basically no change in the width from pre-flare to gradual phases, with only a maximum increase of 10\% (for the SJI~2796~\AA), which can be explained based on the amount of rain increase in the POS, leading to more LOS superposition of structures. Furthermore, for both the pre-flare and gradual phases, the width of the rain clumps remains fairly uniform as they fall along the loop, suggesting that the factor influencing rain width is unaffected by the varying thermodynamic conditions with height (e.g. progressive cooling or increased compression downstream from the rain). All these results highlight the fundamental and consistent nature of rain widths largely independent of the physical quantities that vary strongly during the flare or with height, such as the field strength, densities and temperatures. This largely invariant nature of the rain widths reflects a more fundamental mechanism setting the widths (contrary to the lengths). An early study by \citet{VanderLinden_1991SoPh..134..247V} demonstrated that the widths could be set by the eigenfunction of the thermal mode, which suggests that rain widths reflect the detailed evolution of the instability of this MHD wave. A recent 2.5D MHD numerical simulation by \citet{Antolin2022A} showed the emergence of a fundamental magnetic field strand generated from TI due to gas pressure loss and flux freezing, suggesting that total pressure balance across the field may define the rain width. This is further supported by the recent 3D MHD numerical modelling from \citet{Ruan2024arXiv240319204R}, where the Lorentz force and the pressure gradient closely balance each other across the flare loop during its evolution. This means that we can use pressure equilibrium purely between the gas pressure in the flare loop $p_{gas}$ and the surrounding magnetic pressure to calculate the average magnetic field strength ($B$) during the impulsive and gradual phases, as $\frac{B^2}{8\pi} = p_{gas}$. Taking the thermodynamic values for each case (see Table~\ref{table:mass_energy_calculations}), we obtain a magnetic field strength of 23.78~G and 10.92~G for the impulsive and gradual phases, respectively, which are consistent with the numerical results of \citet{Ruan2024arXiv240319204R} for a similar C-class flare. The decrease in field strength, which translates into a magnetic total energy of $\approx3.71\times10^{29}$~erg, is consistent with the calculated total thermal energy of the flare ($7.52\times10^{29}$~erg). DKIST, with the capability of measuring the coronal magnetic field strength, should be able to directly assess these theoretical assumptions. 

Contrary to the clump widths, the clump lengths vary from a few Mm to 22~Mm. We found that the average length of flare-driven coronal rain is 1.50, 1.54, and 1.20 times larger than the pre-flare quiescent coronal rain. This can be simply explained by the large increase in rain quantity, with the growth largely occurring along the field for a given clump (while the rain width stays constant). Contrary to the invariance of the rain widths with height, the lengths of rain clumps increase at lower heights up to a threshold set by the falling time of the rain, the length of the loop, and the inability of the RHT method to follow the clumps on disk.

We then examined the rain dynamics during the pre-flare and gradual phases and found that both types of rain exhibit a wide range of velocities spanning from a few km~s$^{-1}$ to $200$~km~s$^{-1}$. These results are in agreement with the previous numerical \citep{fang2015, Li2022ApJ...926..216L} and observational \citep{antolin2012, Sahin2023ApJ...950..171S} studies. Rain clumps do not only move downward but also upward and in varying trajectories. The occurrence of the upward motions and the change in the trajectory of rain clump is first noted by \citet{antolin2010}. Recent studies \citep{Li2022ApJ...926..216L, Sahin2023ApJ...950..171S} have shown how ubiquitous these upward motions of coronal rain are in the solar atmosphere. \citet{Sahin2023ApJ...950..171S} found that upflow velocities can have higher values than downflow velocities. However, these upflow motions correspond to more stochastic and sporadic motions. In other words, contrary to downward motions, upward motions are more localised events and do not correspond to bulk flows. Our findings showed that downflow velocity values are increased by a factor of 1.40 from the pre-flare to the gradual phase, with minimal variation observed across the channels. The upflow speeds show negligible to little increase from pre-flare to gradual phases (with a maximum factor of 1.25 for SJI~1330~\AA). We have also examined the average projected velocity in relation to both downward and upward motions as a function of height. At higher heights, we observe downward accelerations in both pre-flare and gradual phases. However, the rain maintains a consistently steady speed during its falling, potentially owing to a combination of effective gravity and pressure variations \citep{antolin2010, Oliver2014, Martinez-Gomez2020}. In contrast to the downward motion, we observe a more chaotic behaviour in the upward motion concerning the variation in projected velocity with height. The bulk increase in downflow speeds is in agreement with the hydrodynamic effect discussed in \citet{Oliver2014, Martinez-Gomez2020}. It has been shown numerically that, as the mass and density of the clumps increase, the drag due to the gas pressure gradient force leads to an increase in speed given by $v_{max}=2.56\Theta^{0.64}$ \citep{Martinez-Gomez2020}, where  $\Theta=\rho_{rain}$/$\rho_{DEM}$ is the density ratio between the rain clump and the surrounding region \citep{Martinez-Gomez2020}. In this study, we found that the maximum POS velocity for flare-driven coronal rain ($\approx70$~km~s$^{-1}$ on average at a height of 5~Mm, cf. Fig.~\ref{falling_speed_grd}) is 1.4 times larger than the corresponding maximum velocity for quiescent rain (roughly 50~km~s$^{-1}$ on average at a height of 5~Mm, not shown here). We have also calculated the average total number and mass densities of quiescent and flare-driven coronal rain. For quiescent coronal rain, these values were found to be [9.01$\pm$0.70]$\times$10$^{10}$~cm$^{-3}$ and [9.18$\pm$0.71]$\times$10$^{-14}$~g~cm$^{-3}$, respectively, which are consistent with usual quiescent rain densities \citet{Antolin_Froment_10.3389/fspas.2022.820116}. For the flare-driven rain, we have found $[5.77\pm0.55]\times10^{-13}$~g~cm$^{-3}$ for the mass densities, which constitutes an increase of roughly 6 times the quiescent values. Using the previous results, we can now test the formula by \citet{Martinez-Gomez2020}. We can write the formula as follows:

%For quiescent coronal rain, these values were found to be [3.00$\pm$1.21]$\times$10$^{11}$~cm$^{-3}$ and [5.87$\pm$2.02]$\times$10$^{-13}$~g~cm$^{-3}$, respectively, which are consistent with the studies of \citet{antolin2015} and \citet{froment2020}. For the flare-driven rain, we have found $[1.49\pm0.46]\times10^{12}$~cm$^{-3}$ and $[2.92\pm0.90]\times10^{-12}$~g~cm$^{-3}$ for the number and mass densities, respectively, which constitutes an increase of roughly 5 over the quiescent values. Using the previous results, we can now test the formula by \citet{Martinez-Gomez2020}. We can write the formula as follows:

\begin{equation}
    \label{eq:vratio}   
    \frac{v_{rain,fl}}{v_{rain,qs}} = \left[\left(\frac{\rho_{rain,fl}}{\rho_{rain,qs}}\right)\left(\frac{\rho_{DEM,qs}}{\rho_{DEM,fl}}\right)\right]^{0.64}
\end{equation}

Here, $v_{rain,fl}$ and $v_{rain,qs}$ correspond to the maximum velocities of flare-driven and quiescent coronal rain, respectively. By taking the density values of the rain ($\rho_{rain}$) and the surrounding area ($\rho_{DEM}$ from outside the rain), we obtain the expected velocity ratio in Equation~\ref{eq:vratio}, $\frac{v_{rain,fl}}{v_{rain,qs}}=1.65$, closely matching the observational result. This supports the theory that gas pressure drag controls the final speed of the rain \citep{Oliver2014}.

A potential explanation for the small increased upflow velocities could be attributed to the appearance of the rain itself. As it falls, cooling occurs, resulting in increased opacity and further rain appearance along the trajectory, which in turn gives rise to the perception of localised, rapid motions depending on the cooling rate. This effect would be particularly visible at the trail of the rain, leading to apparent upflows, and therefore may better explain the localised (and sporadic) behaviour of the upflows. However, it remains unclear whether this effect can fully account for the sporadic nature of the upflows.  

%for the density calculation since we have calculated the average mass density of quiescent coronal rain. As given previously, $v_{max}$ for the flare-driven coronal rain is 1.43 times larger than $v_{max}$ for the quiescent coronal rain and given this formula, the density ratio between the quiescent and flare-driven coronal rain also should be 1.43. Then, this equation gives the representative mass density for the quiescent coronal rain as 1.66$\times$10$^{-12}$~g~cm$^{-3}$. However, this value is roughly 2 times denser than the value we found for the quiescent coronal rain.

%Another plausible explanation is the existence of occasional small heating events at the loop footpoints, whose localized nature may impact individual clumps but not the overall bulk flow.

%For instance, \citet{Mackay2001SoPh..198..289M} showed that the chromosphere acts as a piston, which effectively halts the downward progression of rain by compressing the region downstream. Recently, \citet{Adrover2021A&A...649A.142A} carried out an analytical examination of critical points within a loop, treating it as a dynamic system arising from line-tying conditions in the lower atmosphere. This study showed that these line-tying conditions trigger oscillations, resulting in upward flows within the system.

We have focused on the impulsive phase and found that the RHT is also able to capture the dynamics of CE. Once again, there are many more CE events; however, we only focused on the best CE events regarding their isolation since it is difficult to analyze all of them due to their diffuse emission. We also analyzed dynamics of both CE events using the CRISPEX/TANAT and found a broad velocity distribution between $15$ and $230$~km~s$^{-1}$, with an average of 125~km~s$^{-1}$. A natural question is whether some of the observed upward propagating features correspond to slow magnetoacoustic waves rather than flows. Indeed, it is usually very difficult to distinguish between a wave and a flow \citep{Moortel2015SoPh..290..399D}. The numerical flare model by \citet{Fang2015ApJ...813...33F} reveals that initial flow motions in flaring loops are often accompanied by slow-mode waves, with the two phenomena initially appearing indistinguishable. In their numerical model, which includes temperatures similar to those reported in our study, both wave and flow propagate at the same speed, which initially is close to 600~km~s$^{-1}$, corresponding to the sound speed for a 10$^{7.1}$~K plasma. However, subsequent analysis reveals distinct characteristics of propagating slow modes, including periodic motion and intensity modulation corresponding to reflections at the loop footpoints. In our study, the speeds we observe are significantly lower than the sound speed. Also, we do not observe a modulation of the intensity (which would correspond to the reflection of the slow mode from the far footpoint). Hence, this suggests that the signature we observe is only the flow. Furthermore, the upward propagating features that we interpret as a flow exists over tens of minutes, long enough for a slow mode to propagate back and forth between the loop (and damp, given the very fast damping times usually reported). We were also able to observe the selected CE events in the \ion{Fe}{XXI}~1354.08~\AA\, line with the SG spectrometer of \textit{IRIS}, thereby providing a measure for the Doppler and non-thermal velocities, as well as additional support for the flow nature and $10^7$~K temperature of the CEs. Combining the LOS and POS average velocities, we were able to give upper bounds for the total velocities of the CEs. We estimated that the average total upward velocities are $122\pm38$~km~s$^{-1}$ and $154\pm33$~km~s$^{-1}$ for CE1 and CE2, respectively. 

We then presented a comparative study of these two main CE events with two associated SH events. Despite the diffuse emission associated with CE, substructures within it suggest the presence of strands with a minimum width of $1376\pm106$~km on average, seen with both the SJI and SG instruments. The observed minimum sub-structure for the SH events, on the other hand, is $681\pm100$~km on average. \citet{Jing2016NatSR...624319J} found a strong similarity between the widths of flare ribbons and the widths of rain during an M-class flare, with values in the range of 80-200~km. They interpret their result as a fundamental scale of mass and energy transport during flares. Since CE is directly associated with the flare ribbons, we can compare our results with those of \citet{Jing2016NatSR...624319J}. The CE and rain widths that we find are a factor of $5-10$ larger, which could be attributed to the $5-10$ times coarser spatial resolution (particularly in hot lines) between \textit{GST} and \textit{IRIS}, the diffuse coronal emission (which introduces stronger LOS superposition effects), and the different opacity between H$\alpha$ and the \textit{IRIS} spectral lines. Furthermore, our measurements of the CE widths are performed at (low) coronal heights, and it is likely that the loop area expansion also implies an increase in the width of the CE strands. These factors could also be responsible for the factor of two discrepancy that we find between the CE and the rain widths. On the other hand, the overall average widths of the CE are found to be $4807\pm131$~km with the SJI (matched with SG), which are very similar to the overall average widths of the SHs of $4149\pm130$~km. This suggests that spatial resolution and the diffuse coronal emission leading to LOS superposition may be the dominant factors between the factor of two discrepancy. These results align with those of \citet{Jing2016NatSR...624319J} in that the rain structure can reflect the spatial scales of energy transport during flares. We showed that this association can be further extended to the amount of mass and energy in the evaporated material.

It is important, however, to distinguish the strand widths in CE with those that are found for EUV strands during rain events. As shown in \citet{Antolin2022A}, the CCTR can strongly emit in the EUV, leading to strands of the same width as the rain. A very recent study by \citet{Antolin2023A&A...676A.112A} with Solar Orbiter/HRIEUV in the 174~\AA\ channel, with the highest resolution ever achieved in the EUV of the solar corona, also showed that the observed rain widths are similar to those of EUV strands. However, they also noted that these EUV strands primarily become evident just before the appearance of rain. This implies that the EUV strand widths may be governed by the cooling mechanism (i.e., TI) rather than the heating length scales. In essence, both heating and cooling length scales can result in similar widths. %\sss{We found that the number of strands is 4 and 6 for the CE and SH events, respectively. These numbers are compatible. Even though we kept them different in this analysis,  we also considered it to take an equal number of strands for both CE and SH events. One plausible explanation could be attributed to the lower resolution in the hot lines and the more diffuse nature of the hot emission, which can be responsible for the differences in width strands between CE and shower. However, in any case, the widths of coronal strands are found to be similar to the widths of rain at high resolution \citep{antolin2015, Williams2020ApJ...892..134W}.}

The spectral results revealed a significant difference in Doppler and non-thermal velocities between the CEs, suggesting the presence of magnetic shear close to the footpoints of the loop. Furthermore, we found an interesting difference between the dynamics of SH1 and CE1. While CE1 is less Doppler shifted but with a larger non-thermal line width, SH1 is more Doppler shifted (with opposite sign to CE) but with a smaller non-thermal line width, but both SH1 and CE1 have very similar summed Doppler and non-thermal velocities. Given the strong morphological similarity between both, this indicates the presence of unresolved emission and/or turbulence at faster speeds for the CE (supported by the POS measurements with the SJI). With the SG results we were able to properly determine the duration of the CE events. We noticed a rapid decrease in speed on a timescale of $2-5$~min, which suggested a transition from explosive to gentle evaporation. The observed timescales and speeds are consistent with previous reports \citep{Fletcher_2013ApJ...771..104F, Sellers_2022ApJ...936...85S}. The decrease in nonthermal velocity for CE2 was not accompanied by a decrease in Doppler velocity, indicating an initially present turbulence that decreases over time \citep{Polito_2019ApJ...879L..17P}.

Finally, we examined the mass and energy cycle of the flare, where the CE marks the upflow into the flare loop during the impulsive phase, and the SH marks the downflow out of the loop during the gradual phase. Taking an average total number density of $[1.85\pm0.02]\times10^{10}$~cm$^{-3}$ for the CE events, an average overall width of $4807\pm131$~km, we obtain an average upflow mass flux for CE events of $[2.67\pm0.04]\times10^{-7}$~g~cm$^{-2}$~s$^{-1}$. For the SH events, using the corresponding average values (see Table \ref{table:mass_energy_calculations}), we found a value of $[2.78\pm0.05]\times10^{-5}$~g~cm$^{-2}$~s$^{-1}$. We then estimated the mass going up to be roughly $[3.55\pm0.01]\times10^{13}$~g on average for the CE events. On the other hand, the estimated total mass going down is $[9.99\pm0.65]\times10^{13}$~g on average for the SH events, and is, therefore, roughly three times larger than the mass pushed up. Regarding the energy budget, we found the average kinetic and thermal energies to be $[3.78\pm0.03]\times10^{27}$~erg and $[6.30\pm0.02]\times10^{28}$~erg, respectively, for the CE events. For SH events, we found average kinetic and thermal energies of $[1.90\pm0.02]\times10^{28}$~erg and  $[7.73\pm0.31]\times10^{27}$~erg, respectively. In conclusion, the average total energies are $[6.67\pm0.02]\times10^{28}$~erg and $[2.68\pm0.02]\times10^{28}$~erg for the CE and SH events, respectively. This result shows that the rain energy corresponds to roughly half of the CE energy, with the other half probably converted into heat. We note that, in terms of rain energy, only 15 showers would account for the estimated total flare energy. A number of factors can account for the difference in mass between upflow through CE and downflow through SH. First, it is likely that most of the material in the loop is becoming thermally unstable (the loop becomes fully evacuated), with the resulting coronal rain flowing mainly towards the observed footpoint, as seen in AIA~304~\AA. Second, it is highly probable that CE occurs at the other footpoint of the loop (which remains beyond the IRIS FOV), and we would expect an additional source for the CE to contribute roughly in equal amounts. Third, it is possible that the CE lasts for longer times but remains undetected. This would be the case if the CE occurs at hotter temperature regimes outside those of AIA and \textit{IRIS}, or remains in a gentle state but is too faint. The next-generation spectrometers, such as \textit{MUSE} \citep{DePontieu_2022ApJ...926...52D, Cheung_2022ApJ...926...53C} and \textit{EUVST} \citep{Imada_2024}, which probe a wider temperature and velocity range at higher sensitivity, will be able to elucidate this problem. These findings support the noteworthy role of coronal rain in the mass and energy exchange between the chromosphere and the corona during a flare.

\begin{acknowledgments}
The authors would like to thank the anonymous referee for the constructive comments during the reviewing process of this manuscript. P.A. acknowledges funding from his STFC Ernest Rutherford Fellowship (No. ST/R004285/2). This research was supported by the International Space Science Institute (ISSI) in Bern, through ISSI International Team project $\#545$ (``Observe Local Think Global: What Solar Observations can Teach us about Multiphase Plasmas across Physical Scales"). IRIS is a NASA small explorer mission developed and operated by LMSAL, with mission operations executed at NASA Ames Research Center and major contributions to downlink communications funded by ESA and the Norwegian Space Centre. SDO is a mission for NASA’s Living With a Star (LWS) program. AIA is an instrument onboard the Solar Dynamics Observatory. All SDO data used in this work are available from the Joint Science Operations Center (http://jsoc.stanford.edu) without restriction.
\end{acknowledgments}

\appendix
%\restartappendixnumbering
\renewcommand\thefigure{\thesection A.\arabic{figure}} 
%\section{Rain shower morphology}
\setcounter{figure}{0}    

%As explained in Section \ref{sec:ceandsh}, we obtained the overall width of the SH1  event. 

\begin{figure*}[ht!]
\includegraphics[width=\textwidth]{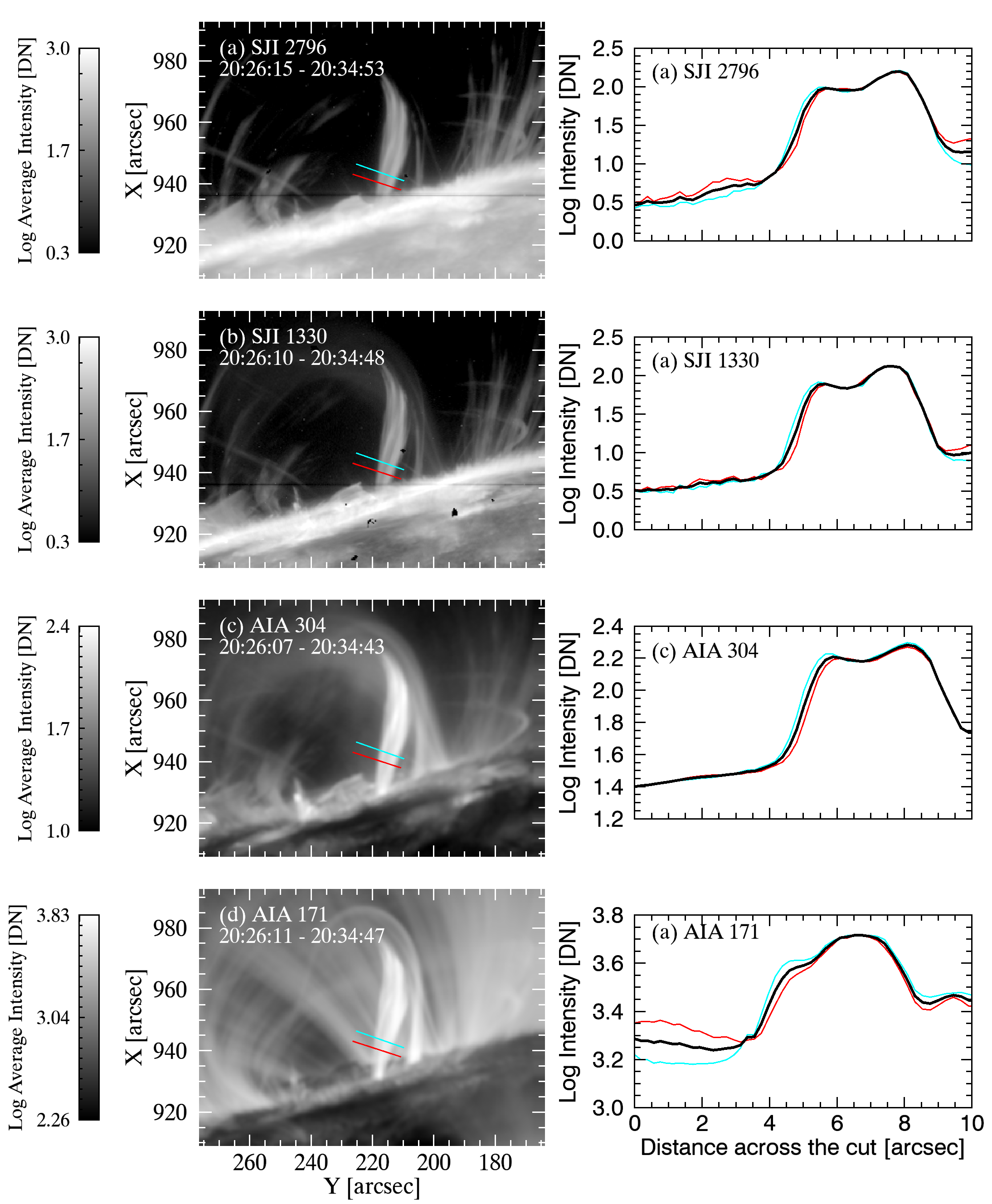}
\caption{Left: Observed SH1 event during the gradual phase in SJI~2796~\AA, SJI~1330~\AA, AIA~304~\AA, and AIA~171~\AA. The red and cyan colours indicate the perpendicular cuts to the loop trajectory. Right: The intensity variation over the distance of these perpendicular cuts. Black curves show the average of these intensity variations.}
\label{appendix1}
\end{figure*}

\begin{figure*}[ht!]
\includegraphics[width=\textwidth]{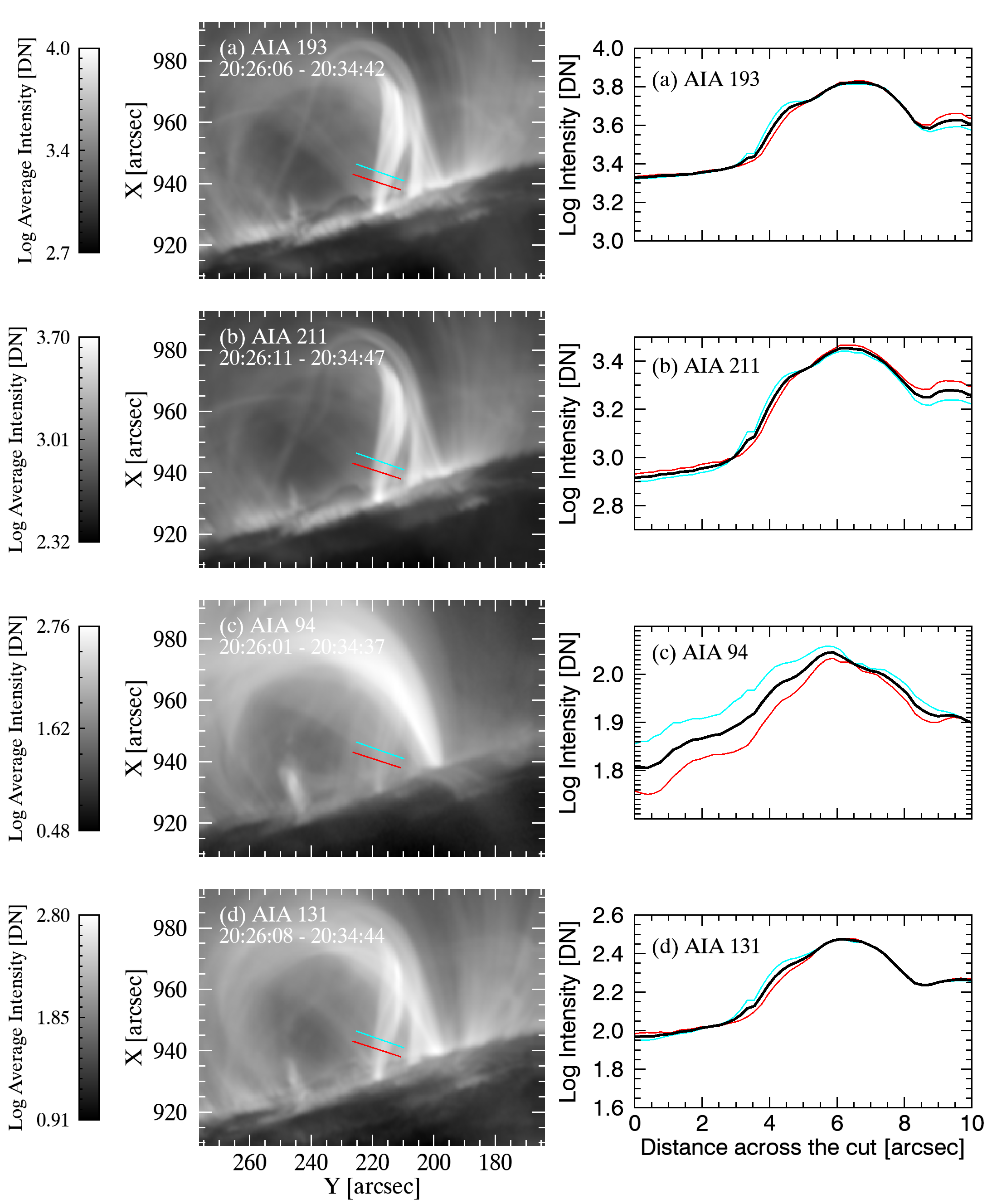}
\caption{Left: Observed SH1 event during the gradual phase in AIA~193~\AA, 211~\AA, 94~\AA, and 131~\AA. The red and cyan colours indicate the perpendicular cuts to the loop trajectory. Right: The intensity variation over the distance of these perpendicular cuts. Black curves show the average of these intensity variations.}
\label{appendix2}
\end{figure*}

\bibliography{sample631}{}

\begin{thebibliography}{}
\expandafter\ifx\csname natexlab\endcsname\relax\def\natexlab#1{#1}\fi
\providecommand{\url}[1]{\href{#1}{#1}}
\providecommand{\dodoi}[1]{doi:~\href{http://doi.org/#1}{\nolinkurl{#1}}}
\providecommand{\doeprint}[1]{\href{http://ascl.net/#1}{\nolinkurl{http://ascl.net/#1}}}
\providecommand{\doarXiv}[1]{\href{https://arxiv.org/abs/#1}{\nolinkurl{https://arxiv.org/abs/#1}}}

\bibitem[{{Ahn} {et~al.}(2014){Ahn}, {Chae}, {Cho}, {Song}, {Yang}, {Goode},
  {Cao}, {Park}, {Nah}, {Jang}, \& {Park}}]{Ahn2014SoPh..289.4117A}
{Ahn}, K., {Chae}, J., {Cho}, K.-S., {et~al.} 2014, \solphys, 289, 4117,
  \dodoi{10.1007/s11207-014-0559-x}

\bibitem[{{Antiochos} {et~al.}(1999){Antiochos}, {MacNeice}, {Spicer}, \&
  {Klimchuk}}]{Antiochos_1999ApJ...512..985A}
{Antiochos}, S.~K., {MacNeice}, P.~J., {Spicer}, D.~S., \& {Klimchuk}, J.~A.
  1999, \apj, 512, 985, \dodoi{10.1086/306804}

\bibitem[{{Antolin}(2020)}]{antolin2020}
{Antolin}, P. 2020, Plasma Physics and Controlled Fusion, 62, 014016,
  \dodoi{10.1088/1361-6587/ab5406}

\bibitem[{{Antolin} {et~al.}(2024){Antolin}, {Auchère}, {Winch}, {Soubrié},
  \& {Oliver}}]{Antolin2024}
{Antolin}, P., {Auchère}, F., {Winch}, E., {Soubrié}, E., \& {Oliver}, R.
  2024, Living Reviews in Solar Physics, PREPRINT (Version 1) available at
  Research Square, \dodoi{10.21203/rs.3.rs-3953676/v1}

\bibitem[{Antolin \& Froment(2022)}]{Antolin_Froment_10.3389/fspas.2022.820116}
Antolin, P., \& Froment, C. 2022, Frontiers in Astronomy and Space Sciences, 9,
  \dodoi{10.3389/fspas.2022.820116}

\bibitem[{{Antolin} {et~al.}(2022){Antolin}, {Mart{\'\i}nez-Sykora}, \&
  {{\c{S}}ahin}}]{Antolin2022A}
{Antolin}, P., {Mart{\'\i}nez-Sykora}, J., \& {{\c{S}}ahin}, S. 2022, \apjl,
  926, L29, \dodoi{10.3847/2041-8213/ac51dd}

\bibitem[{{Antolin} \& {Rouppe van der Voort}(2012)}]{antolin2012}
{Antolin}, P., \& {Rouppe van der Voort}, L. 2012, \apj, 745, 152,
  \dodoi{10.1088/0004-637X/745/2/152}

\bibitem[{{Antolin} {et~al.}(2010){Antolin}, {Shibata}, \&
  {Vissers}}]{antolin2010}
{Antolin}, P., {Shibata}, K., \& {Vissers}, G. 2010, \apj, 716, 154,
  \dodoi{10.1088/0004-637X/716/1/154}

\bibitem[{{Antolin} {et~al.}(2015){Antolin}, {Vissers}, {Pereira}, {Rouppe van
  der Voort}, \& {Scullion}}]{antolin2015}
{Antolin}, P., {Vissers}, G., {Pereira}, T.~M.~D., {Rouppe van der Voort}, L.,
  \& {Scullion}, E. 2015, \apj, 806, 81, \dodoi{10.1088/0004-637X/806/1/81}

\bibitem[{{Antolin} {et~al.}(2023){Antolin}, {Dolliou}, {Auch{\`e}re},
  {Chitta}, {Parenti}, {Berghmans}, {Aznar Cuadrado}, {Barczynski}, {Gissot},
  {Harra}, {Huang}, {Janvier}, {Kraaikamp}, {Long}, {Mandal}, {Peter},
  {Rodriguez}, {Sch{\"u}hle}, {Smith}, {Solanki}, {Stegen}, {Teriaca},
  {Verbeeck}, {West}, {Zhukov}, {Appourchaux}, {Aulanier}, {Buchlin},
  {Delmotte}, {Gilles}, {Haberreiter}, {Halain}, {Heerlein}, {Hochedez}, {Gyo},
  {Poedts}, \& {Rochus}}]{Antolin2023A&A...676A.112A}
{Antolin}, P., {Dolliou}, A., {Auch{\`e}re}, F., {et~al.} 2023, \aap, 676,
  A112, \dodoi{10.1051/0004-6361/202346016}

\bibitem[{{Chen} {et~al.}(2022){Chen}, {Tian}, {Li}, {Peter}, {Chitta}, \&
  {Hou}}]{ChenH_2022AA...659A.107C}
{Chen}, H., {Tian}, H., {Li}, L., {et~al.} 2022, \aap, 659, A107,
  \dodoi{10.1051/0004-6361/202142093}

\bibitem[{{Cheung} {et~al.}(2022){Cheung}, {Mart{\'\i}nez-Sykora}, {Testa}, {De
  Pontieu}, {Chintzoglou}, {Rempel}, {Polito}, {Kerr}, {Reeves}, {Fletcher},
  {Jin}, {N{\'o}brega-Siverio}, {Danilovic}, {Antolin}, {Allred}, {Hansteen},
  {Ugarte-Urra}, {DeLuca}, {Longcope}, {Takasao}, {DeRosa}, {Boerner},
  {Jaeggli}, {Nitta}, {Daw}, {Carlsson}, {Golub}, \&
  {The}}]{Cheung_2022ApJ...926...53C}
{Cheung}, M. C.~M., {Mart{\'\i}nez-Sykora}, J., {Testa}, P., {et~al.} 2022,
  \apj, 926, 53, \dodoi{10.3847/1538-4357/ac4223}

\bibitem[{{Chitta} {et~al.}(2016){Chitta}, {Peter}, \&
  {Young}}]{Chitta2016A&A...587A..20C}
{Chitta}, L.~P., {Peter}, H., \& {Young}, P.~R. 2016, \aap, 587, A20,
  \dodoi{10.1051/0004-6361/201527340}

\bibitem[{{{\c{S}}ahin} \& {Antolin}(2022)}]{Sahin2022ApJ...931L..27S}
{{\c{S}}ahin}, S., \& {Antolin}, P. 2022, \apjl, 931, L27,
  \dodoi{10.3847/2041-8213/ac6fe9}

\bibitem[{{{\c{S}}ahin} {et~al.}(2023){{\c{S}}ahin}, {Antolin}, {Froment}, \&
  {Schad}}]{Sahin2023ApJ...950..171S}
{{\c{S}}ahin}, S., {Antolin}, P., {Froment}, C., \& {Schad}, T.~A. 2023, \apj,
  950, 171, \dodoi{10.3847/1538-4357/acd44b}

\bibitem[{{Culhane} {et~al.}(1970){Culhane}, {Vesecky}, \&
  {Phillips}}]{Culhane1970SoPh...15..394C}
{Culhane}, J.~L., {Vesecky}, J.~F., \& {Phillips}, K.~J.~H. 1970, \solphys, 15,
  394, \dodoi{10.1007/BF00151847}

\bibitem[{{De Groof} {et~al.}(2004){De Groof}, {Berghmans}, {van
  Driel-Gesztelyi}, \& {Poedts}}]{DeGroof2004}
{De Groof}, A., {Berghmans}, D., {van Driel-Gesztelyi}, L., \& {Poedts}, S.
  2004, \aap, 415, 1141, \dodoi{10.1051/0004-6361:20034252}

\bibitem[{{De Moortel} {et~al.}(2015){De Moortel}, {Antolin}, \& {Van
  Doorsselaere}}]{Moortel2015SoPh..290..399D}
{De Moortel}, I., {Antolin}, P., \& {Van Doorsselaere}, T. 2015, \solphys, 290,
  399, \dodoi{10.1007/s11207-014-0610-y}

\bibitem[{{De Pontieu} {et~al.}(2014){De Pontieu}, {Title}, {Lemen}, {Kushner},
  {Akin}, {Allard}, {Berger}, {Boerner}, {Cheung}, {Chou}, {Drake}, {Duncan},
  {Freeland}, {Heyman}, {Hoffman}, {Hurlburt}, {Lindgren}, {Mathur}, {Rehse},
  {Sabolish}, {Seguin}, {Schrijver}, {Tarbell}, {W{\"u}lser}, {Wolfson},
  {Yanari}, {Mudge}, {Nguyen-Phuc}, {Timmons}, {van Bezooijen}, {Weingrod},
  {Brookner}, {Butcher}, {Dougherty}, {Eder}, {Knagenhjelm}, {Larsen},
  {Mansir}, {Phan}, {Boyle}, {Cheimets}, {DeLuca}, {Golub}, {Gates}, {Hertz},
  {McKillop}, {Park}, {Perry}, {Podgorski}, {Reeves}, {Saar}, {Testa}, {Tian},
  {Weber}, {Dunn}, {Eccles}, {Jaeggli}, {Kankelborg}, {Mashburn}, {Pust},
  {Springer}, {Carvalho}, {Kleint}, {Marmie}, {Mazmanian}, {Pereira}, {Sawyer},
  {Strong}, {Worden}, {Carlsson}, {Hansteen}, {Leenaarts}, {Wiesmann},
  {Aloise}, {Chu}, {Bush}, {Scherrer}, {Brekke}, {Martinez-Sykora}, {Lites},
  {McIntosh}, {Uitenbroek}, {Okamoto}, {Gummin}, {Auker}, {Jerram}, {Pool}, \&
  {Waltham}}]{depontieu2014}
{De Pontieu}, B., {Title}, A.~M., {Lemen}, J.~R., {et~al.} 2014, \solphys, 289,
  2733, \dodoi{10.1007/s11207-014-0485-y}

\bibitem[{{De Pontieu} {et~al.}(2022){De Pontieu}, {Testa},
  {Mart{\'\i}nez-Sykora}, {Antolin}, {Karampelas}, {Hansteen}, {Rempel},
  {Cheung}, {Reale}, {Danilovic}, {Pagano}, {Polito}, {De Moortel},
  {N{\'o}brega-Siverio}, {Van Doorsselaere}, {Petralia}, {Asgari-Targhi},
  {Boerner}, {Carlsson}, {Chintzoglou}, {Daw}, {DeLuca}, {Golub}, {Matsumoto},
  {Ugarte-Urra}, {McIntosh}, \& {the MUSE
  team}}]{DePontieu_2022ApJ...926...52D}
{De Pontieu}, B., {Testa}, P., {Mart{\'\i}nez-Sykora}, J., {et~al.} 2022, \apj,
  926, 52, \dodoi{10.3847/1538-4357/ac4222}

\bibitem[{{Del Zanna}(2013)}]{DelZanna2013A&A...558A..73D}
{Del Zanna}, G. 2013, \aap, 558, A73, \dodoi{10.1051/0004-6361/201321653}

\bibitem[{{Del Zanna} {et~al.}(2021){Del Zanna}, {Dere}, {Young}, \&
  {Landi}}]{Del_Zanna_2021}
{Del Zanna}, G., {Dere}, K.~P., {Young}, P.~R., \& {Landi}, E. 2021, The
  Astrophysical Journal, 909, 38, \dodoi{10.3847/1538-4357/abd8ce}

\bibitem[{{Doschek} {et~al.}(1994){Doschek}, {Mariska}, {Strong}, {Bentley},
  {Brown}, {Culhane}, {Lang}, {Sterling}, \&
  {Watanabe}}]{Doschek1994ApJ...431..888D}
{Doschek}, G.~A., {Mariska}, J.~T., {Strong}, K.~T., {et~al.} 1994, \apj, 431,
  888, \dodoi{10.1086/174540}

\bibitem[{{Dud{\'\i}k} {et~al.}(2016){Dud{\'\i}k}, {Polito}, {Janvier},
  {Mulay}, {Karlick{\'y}}, {Aulanier}, {Del Zanna}, {Dzif{\v{c}}{\'a}kov{\'a}},
  {Mason}, \& {Schmieder}}]{Dudik2016ApJ...823...41D}
{Dud{\'\i}k}, J., {Polito}, V., {Janvier}, M., {et~al.} 2016, \apj, 823, 41,
  \dodoi{10.3847/0004-637X/823/1/41}

\bibitem[{{Fang} {et~al.}(2013){Fang}, {Xia}, \& {Keppens}}]{fang2013}
{Fang}, X., {Xia}, C., \& {Keppens}, R. 2013, \apjl, 771, L29,
  \dodoi{10.1088/2041-8205/771/2/L29}

\bibitem[{{Fang} {et~al.}(2015{\natexlab{a}}){Fang}, {Xia}, {Keppens}, \& {Van
  Doorsselaere}}]{fang2015}
{Fang}, X., {Xia}, C., {Keppens}, R., \& {Van Doorsselaere}, T.
  2015{\natexlab{a}}, \apj, 807, 142, \dodoi{10.1088/0004-637X/807/2/142}

\bibitem[{{Fang} {et~al.}(2015{\natexlab{b}}){Fang}, {Yuan}, {Van
  Doorsselaere}, {Keppens}, \& {Xia}}]{Fang2015ApJ...813...33F}
{Fang}, X., {Yuan}, D., {Van Doorsselaere}, T., {Keppens}, R., \& {Xia}, C.
  2015{\natexlab{b}}, \apj, 813, 33, \dodoi{10.1088/0004-637X/813/1/33}

\bibitem[{{Fisher} {et~al.}(1985){Fisher}, {Canfield}, \&
  {McClymont}}]{Fisher1985ApJ...289..414F}
{Fisher}, G.~H., {Canfield}, R.~C., \& {McClymont}, A.~N. 1985, \apj, 289, 414,
  \dodoi{10.1086/162901}

\bibitem[{{Fletcher} {et~al.}(2013){Fletcher}, {Hannah}, {Hudson}, \&
  {Innes}}]{Fletcher_2013ApJ...771..104F}
{Fletcher}, L., {Hannah}, I.~G., {Hudson}, H.~S., \& {Innes}, D.~E. 2013, \apj,
  771, 104, \dodoi{10.1088/0004-637X/771/2/104}

\bibitem[{{Fletcher} {et~al.}(2011){Fletcher}, {Dennis}, {Hudson}, {Krucker},
  {Phillips}, {Veronig}, {Battaglia}, {Bone}, {Caspi}, {Chen}, {Gallagher},
  {Grigis}, {Ji}, {Liu}, {Milligan}, \& {Temmer}}]{Fletcher2011SSRv..159...19F}
{Fletcher}, L., {Dennis}, B.~R., {Hudson}, H.~S., {et~al.} 2011, \ssr, 159, 19,
  \dodoi{10.1007/s11214-010-9701-8}

\bibitem[{{Foukal}(1978)}]{Foukal1978ApJ...223.1046F}
{Foukal}, P. 1978, \apj, 223, 1046, \dodoi{10.1086/156338}

\bibitem[{{Froment} {et~al.}(2020){Froment}, {Antolin}, {Henriques},
  {Kohutova}, \& {Rouppe van der Voort}}]{froment2020}
{Froment}, C., {Antolin}, P., {Henriques}, V.~M.~J., {Kohutova}, P., \& {Rouppe
  van der Voort}, L.~H.~M. 2020, \aap, 633, A11,
  \dodoi{10.1051/0004-6361/201936717}

\bibitem[{{Golding} {et~al.}(2017){Golding}, {Leenaarts}, \&
  {Carlsson}}]{Golding2017A&A...597A.102G}
{Golding}, T.~P., {Leenaarts}, J., \& {Carlsson}, M. 2017, \aap, 597, A102,
  \dodoi{10.1051/0004-6361/201629462}

\bibitem[{{Golub} \& {Pasachoff}(2009)}]{Golub2009soco.bookG}
{Golub}, L., \& {Pasachoff}, J.~M. 2009, {The Solar Corona}

\bibitem[{{Heinzel} \& {Shibata}(2018)}]{Heinzel2018ApJ...859..143H}
{Heinzel}, P., \& {Shibata}, K. 2018, \apj, 859, 143,
  \dodoi{10.3847/1538-4357/aabe78}

\bibitem[{Hough(1962)}]{hough1962method}
Hough, P.~V. 1962, Method and means for recognizing complex patterns,  US
  Patent 3,069,654

\bibitem[{{Imada} {et~al.}(2024){Imada}, {Ugarte-Urra}, {De Pontieu},
  {Shimizu}, {Warren}, {Hansteen}, {Antolin}, {Iijima}, {Martinez-Sykora},
  {Shoda}, {Reep}, {Toriumi}, {Katsukawa}, {Hara}, {Kawate}, {Watanabe},
  {Yokoyama}, {Barczynski}, {Dominique}, {Pereira}, {Froment}, {Brooks},
  {Asai}, {Masuda}, {Savage}, {Matthews}, {Cheung}, {Tei}, {Oba}, {Shortt},
  {Andretta}, {Auchere}, {Teriaca}, \& {Harra}}]{Imada_2024}
{Imada}, S., {Ugarte-Urra}, I., {De Pontieu}, M., {et~al.} 2024, Submitted to
  \pasj

\bibitem[{{Jing} {et~al.}(2016){Jing}, {Xu}, {Cao}, {Liu}, {Gary}, \&
  {Wang}}]{Jing2016NatSR...624319J}
{Jing}, J., {Xu}, Y., {Cao}, W., {et~al.} 2016, Scientific Reports, 6, 24319,
  \dodoi{10.1038/srep24319}

\bibitem[{{Kleint} {et~al.}(2014){Kleint}, {Antolin}, {Tian}, {Judge}, {Testa},
  {De Pontieu}, {Mart{\'\i}nez-Sykora}, {Reeves}, {Wuelser}, {McKillop},
  {Saar}, {Carlsson}, {Boerner}, {Hurlburt}, {Lemen}, {Tarbell}, {Title},
  {Golub}, {Hansteen}, {Jaeggli}, \& {Kankelborg}}]{Kleint2014ApJ...789L..42K}
{Kleint}, L., {Antolin}, P., {Tian}, H., {et~al.} 2014, \apjl, 789, L42,
  \dodoi{10.1088/2041-8205/789/2/L42}

\bibitem[{{Klimchuk}(2019)}]{Klimchuk_2019SoPh..294..173K}
{Klimchuk}, J.~A. 2019, \solphys, 294, 173, \dodoi{10.1007/s11207-019-1562-z}

\bibitem[{{Klimchuk} \& {Luna}(2019)}]{Klimchuk_2019ApJ...884...68K}
{Klimchuk}, J.~A., \& {Luna}, M. 2019, \apj, 884, 68,
  \dodoi{10.3847/1538-4357/ab41f4}

\bibitem[{{Kohutova} {et~al.}(2019){Kohutova}, {Verwichte}, \&
  {Froment}}]{Kohutova2019A&A...630A.123K}
{Kohutova}, P., {Verwichte}, E., \& {Froment}, C. 2019, \aap, 630, A123,
  \dodoi{10.1051/0004-6361/201936253}

\bibitem[{{Landi} {et~al.}(2012){Landi}, {Del Zanna}, {Young}, {Dere}, \&
  {Mason}}]{Landi2012ApJ...744...99L}
{Landi}, E., {Del Zanna}, G., {Young}, P.~R., {Dere}, K.~P., \& {Mason}, H.~E.
  2012, \apj, 744, 99, \dodoi{10.1088/0004-637X/744/2/99}

\bibitem[{{Leenaarts} {et~al.}(2013){Leenaarts}, {Pereira}, {Carlsson},
  {Uitenbroek}, \& {De Pontieu}}]{Leenaarts2013ApJ...772...89L}
{Leenaarts}, J., {Pereira}, T.~M.~D., {Carlsson}, M., {Uitenbroek}, H., \& {De
  Pontieu}, B. 2013, \apj, 772, 89, \dodoi{10.1088/0004-637X/772/2/89}

\bibitem[{{Lemen} {et~al.}(2012){Lemen}, {Title}, {Akin}, {Boerner}, {Chou},
  {Drake}, {Duncan}, {Edwards}, {Friedlaender}, {Heyman}, {Hurlburt}, {Katz},
  {Kushner}, {Levay}, {Lindgren}, {Mathur}, {McFeaters}, {Mitchell}, {Rehse},
  {Schrijver}, {Springer}, {Stern}, {Tarbell}, {Wuelser}, {Wolfson}, {Yanari},
  {Bookbinder}, {Cheimets}, {Caldwell}, {Deluca}, {Gates}, {Golub}, {Park},
  {Podgorski}, {Bush}, {Scherrer}, {Gummin}, {Smith}, {Auker}, {Jerram},
  {Pool}, {Soufli}, {Windt}, {Beardsley}, {Clapp}, {Lang}, \&
  {Waltham}}]{lemen2012}
{Lemen}, J.~R., {Title}, A.~M., {Akin}, D.~J., {et~al.} 2012, \solphys, 275,
  17, \dodoi{10.1007/s11207-011-9776-8}

\bibitem[{{Li} {et~al.}(2018){Li}, {Zhang}, {Peter}, {Chitta}, {Su}, {Xia},
  {Song}, \& {Hou}}]{LiL_2018ApJ...864L...4L}
{Li}, L., {Zhang}, J., {Peter}, H., {et~al.} 2018, \apjl, 864, L4,
  \dodoi{10.3847/2041-8213/aad90a}

\bibitem[{{Li} {et~al.}(2022){Li}, {Keppens}, \&
  {Zhou}}]{Li2022ApJ...926..216L}
{Li}, X., {Keppens}, R., \& {Zhou}, Y. 2022, \apj, 926, 216,
  \dodoi{10.3847/1538-4357/ac41cd}

\bibitem[{{Liu} {et~al.}(2016){Liu}, {Antolin}, \&
  {Sun}}]{Liu_2016SPD....47.0402L}
{Liu}, W., {Antolin}, P., \& {Sun}, X. 2016, in AAS/Solar Physics Division
  Meeting, Vol.~47, AAS/Solar Physics Division Abstracts \#47, 4.02

\bibitem[{{Mart{\'\i}nez-G{\'o}mez} {et~al.}(2020){Mart{\'\i}nez-G{\'o}mez},
  {Oliver}, {Khomenko}, \& {Collados}}]{Martinez-Gomez2020}
{Mart{\'\i}nez-G{\'o}mez}, D., {Oliver}, R., {Khomenko}, E., \& {Collados}, M.
  2020, \aap, 634, A36, \dodoi{10.1051/0004-6361/201937078}

\bibitem[{{Mason} {et~al.}(2019){Mason}, {Antiochos}, \&
  {Viall}}]{Mason2019ApJ...874L..33M}
{Mason}, E.~I., {Antiochos}, S.~K., \& {Viall}, N.~M. 2019, \apjl, 874, L33,
  \dodoi{10.3847/2041-8213/ab0c5d}

\bibitem[{{Milligan} \& {Dennis}(2009)}]{Milligan2009ApJ...699..968M}
{Milligan}, R.~O., \& {Dennis}, B.~R. 2009, \apj, 699, 968,
  \dodoi{10.1088/0004-637X/699/2/968}

\bibitem[{{M{\"u}ller} {et~al.}(2003){M{\"u}ller}, {Hansteen}, \&
  {Peter}}]{muller2003}
{M{\"u}ller}, D.~A.~N., {Hansteen}, V.~H., \& {Peter}, H. 2003, \aap, 411, 605,
  \dodoi{10.1051/0004-6361:20031328}

\bibitem[{{M{\"u}ller} {et~al.}(2004){M{\"u}ller}, {Peter}, \&
  {Hansteen}}]{muller2004}
{M{\"u}ller}, D.~A.~N., {Peter}, H., \& {Hansteen}, V.~H. 2004, \aap, 424, 289,
  \dodoi{10.1051/0004-6361:20040403}

\bibitem[{{Nelson} {et~al.}(2020){Nelson}, {Krishna Prasad}, \&
  {Mathioudakis}}]{Nelson2020A&A...636A..35N}
{Nelson}, C.~J., {Krishna Prasad}, S., \& {Mathioudakis}, M. 2020, \aap, 636,
  A35, \dodoi{10.1051/0004-6361/201937357}

\bibitem[{{Oliver} {et~al.}(2014){Oliver}, {Soler}, {Terradas}, {Zaqarashvili},
  \& {Khodachenko}}]{Oliver2014}
{Oliver}, R., {Soler}, R., {Terradas}, J., {Zaqarashvili}, T.~V., \&
  {Khodachenko}, M.~L. 2014, \apj, 784, 21, \dodoi{10.1088/0004-637X/784/1/21}

\bibitem[{{Pesnell} {et~al.}(2012){Pesnell}, {Thompson}, \&
  {Chamberlin}}]{pesnell2012}
{Pesnell}, W.~D., {Thompson}, B.~J., \& {Chamberlin}, P.~C. 2012, \solphys,
  275, 3, \dodoi{10.1007/s11207-011-9841-3}

\bibitem[{{Plowman} \& {Caspi}(2020)}]{Plowman2020ApJ...905...17P}
{Plowman}, J., \& {Caspi}, A. 2020, \apj, 905, 17,
  \dodoi{10.3847/1538-4357/abc260}

\bibitem[{{Polito} {et~al.}(2019){Polito}, {Testa}, \& {De
  Pontieu}}]{Polito_2019ApJ...879L..17P}
{Polito}, V., {Testa}, P., \& {De Pontieu}, B. 2019, \apjl, 879, L17,
  \dodoi{10.3847/2041-8213/ab290b}

\bibitem[{{Rathore} \& {Carlsson}(2015)}]{Rathore2015ApJ...811...80R}
{Rathore}, B., \& {Carlsson}, M. 2015, \apj, 811, 80,
  \dodoi{10.1088/0004-637X/811/2/80}

\bibitem[{{Reep} {et~al.}(2020){Reep}, {Antolin}, \&
  {Bradshaw}}]{Reep2020ApJ...890..100R}
{Reep}, J.~W., {Antolin}, P., \& {Bradshaw}, S.~J. 2020, \apj, 890, 100,
  \dodoi{10.3847/1538-4357/ab6bdc}

\bibitem[{{Ruan} {et~al.}(2024){Ruan}, {Keppens}, {Yan}, \&
  {Antolin}}]{Ruan2024arXiv240319204R}
{Ruan}, W., {Keppens}, R., {Yan}, L., \& {Antolin}, P. 2024, arXiv e-prints,
  arXiv:2403.19204, \dodoi{10.48550/arXiv.2403.19204}

\bibitem[{{Schad}(2017)}]{schad2017}
{Schad}, T. 2017, \solphys, 292, 132, \dodoi{10.1007/s11207-017-1153-9}

\bibitem[{{Schad}(2018)}]{Schad2018ApJ...865...31S}
{Schad}, T.~A. 2018, \apj, 865, 31, \dodoi{10.3847/1538-4357/aad962}

\bibitem[{{Schad} {et~al.}(2016){Schad}, {Penn}, {Lin}, \&
  {Judge}}]{Schad2016ApJ...833....5S}
{Schad}, T.~A., {Penn}, M.~J., {Lin}, H., \& {Judge}, P.~G. 2016, \apj, 833, 5,
  \dodoi{10.3847/0004-637X/833/1/5}

\bibitem[{{Schrijver}(2001)}]{schrijver2001}
{Schrijver}, C.~J. 2001, \solphys, 198, 325, \dodoi{10.1023/A:1005211925515}

\bibitem[{{Scullion} {et~al.}(2016){Scullion}, {Rouppe van der Voort},
  {Antolin}, {Wedemeyer}, {Vissers}, {Kontar}, \& {Gallagher}}]{scullion2016}
{Scullion}, E., {Rouppe van der Voort}, L., {Antolin}, P., {et~al.} 2016, \apj,
  833, 184, \dodoi{10.3847/1538-4357/833/2/184}

\bibitem[{{Sellers} {et~al.}(2022){Sellers}, {Milligan}, \&
  {McAteer}}]{Sellers_2022ApJ...936...85S}
{Sellers}, S.~G., {Milligan}, R.~O., \& {McAteer}, R.~T.~J. 2022, \apj, 936,
  85, \dodoi{10.3847/1538-4357/ac87a9}

\bibitem[{{Shibata} \& {Magara}(2011)}]{Shibata2011LRSP....8....6S}
{Shibata}, K., \& {Magara}, T. 2011, Living Reviews in Solar Physics, 8, 6,
  \dodoi{10.12942/lrsp-2011-6}

\bibitem[{{Sweet}(1958)}]{Sweet1958IAUS....6..123S}
{Sweet}, P.~A. 1958, in Electromagnetic Phenomena in Cosmical Physics, ed.
  B.~{Lehnert}, Vol.~6, 123

\bibitem[{{Testa} \& {Reale}(2012)}]{Testa2012ApJ...750L..10T}
{Testa}, P., \& {Reale}, F. 2012, \apjl, 750, L10,
  \dodoi{10.1088/2041-8205/750/1/L10}

\bibitem[{{Tian} \& {Chen}(2018)}]{Tian2018ApJ...856...34T}
{Tian}, H., \& {Chen}, N.~H. 2018, \apj, 856, 34,
  \dodoi{10.3847/1538-4357/aab15a}

\bibitem[{{Ugarte-Urra} \& {Warren}(2014)}]{Ugarte-Urra2014ApJ...783...12U}
{Ugarte-Urra}, I., \& {Warren}, H.~P. 2014, \apj, 783, 12,
  \dodoi{10.1088/0004-637X/783/1/12}

\bibitem[{{van der Linden} \&
  {Goossens}(1991)}]{VanderLinden_1991SoPh..134..247V}
{van der Linden}, R.~A.~M., \& {Goossens}, M. 1991, \solphys, 134, 247,
  \dodoi{10.1007/BF00152647}

\bibitem[{{Vashalomidze} {et~al.}(2015){Vashalomidze}, {Kukhianidze},
  {Zaqarashvili}, {Oliver}, {Shergelashvili}, {Ramishvili}, {Poedts}, \& {De
  Causmaecker}}]{Vashalomidze_2015AA...577A.136V}
{Vashalomidze}, Z., {Kukhianidze}, V., {Zaqarashvili}, T.~V., {et~al.} 2015,
  \aap, 577, A136, \dodoi{10.1051/0004-6361/201424101}

\bibitem[{{Verwichte} {et~al.}(2017){Verwichte}, {Antolin}, {Rowlands},
  {Kohutova}, \& {Neukirch}}]{Verwichte2017A&A...598A..57V}
{Verwichte}, E., {Antolin}, P., {Rowlands}, G., {Kohutova}, P., \& {Neukirch},
  T. 2017, \aap, 598, A57, \dodoi{10.1051/0004-6361/201629634}

\bibitem[{{Young} {et~al.}(2015){Young}, {Tian}, \&
  {Jaeggli}}]{Young2015ApJ...799..218Y}
{Young}, P.~R., {Tian}, H., \& {Jaeggli}, S. 2015, \apj, 799, 218,
  \dodoi{10.1088/0004-637X/799/2/218}

\end{thebibliography}
\bibliographystyle{aasjournal}

\end{document}